\documentclass[longauth]{aa}
\usepackage[utf8]{inputenc}
\usepackage[varg]{txfonts}
\usepackage[usenames, dvipsnames]{color}
\usepackage{amsmath}
\usepackage{graphicx}
\graphicspath{{./fig/}}
\usepackage[english]{babel}
\usepackage{siunitx}
\usepackage{float}
\usepackage{url}
\usepackage{longtable}
\usepackage{fancyhdr}
\usepackage{natbib}
\usepackage[normalem]{ulem}
\usepackage[version=4]{mhchem}

\makeatletter

\def\instrefs#1{{\def\scsep{\def\scsep{,}}\@for\w:=#1\do{\scsep\ref{inst:\w}}}}

\renewcommand{\inst}[1]{\unskip$^{\instrefs{#1}}$}

\let\orgautoref\autoref
\renewcommand{\autoref}
        {\def\equationautorefname{Eq.}
         \def\figureautorefname{Fig.}
         \def\sectionautorefname{Sect.}
         \def\subsectionautorefname{Sect.}
         \def\subsubsectionautorefname{Sect.}
         \orgautoref}

\renewcommand*\aa@pageof{, page \thepage{} of \pageref*{LastPage}}

\makeatother

\begin{document}

\title{TOI-1468: A system of two transiting planets, a super-Earth and a mini-Neptune, on opposite sides of the radius valley} 

\titlerunning{A super-Earth and a mini-Neptune orbiting TOI-1468}

\author{
P.\,Chaturvedi \inst{tls}
\and P.\,Bluhm \inst{lsw} 
\and E.\,Nagel \inst{tls,hs} 
\and A.\,P.\,Hatzes \inst{tls}
\and G.\,Morello \inst{iac,ull}
\and M.\,Brady \inst{chicago}
\and J.\,Korth \inst{cha}
\and K.\,Molaverdikhani \inst{lmu,eo,lsw}
\and D.\,Kossakowski \inst{mpia}
\and J.\,A.\,Caballero \inst{cabesac}
\and E.\,W.\,Guenther \inst{tls}
\and E.\,Pall\'e \inst{iac,ull}
\and N.\,Espinoza \inst{stsci}
\and A.\, Seifahrt \inst{chicago}
\and N.\,Lodieu \inst{iac,ull}
\and C.\,Cifuentes \inst{cabesac}
\and E.\,Furlan \inst{nexsi}
\and P.\,J.\,Amado \inst{iaa}
\and T.\,Barclay \inst{maryland,goddard}
\and J.\,Bean \inst{chicago}
\and V.\,J.\,S.\,B\'ejar \inst{iac,ull}
\and G.\,Bergond \inst{caha}
\and A.\,W.\,Boyle \inst{nexsi}
\and D.\,Ciardi \inst{nexsi}
\and K.\,A.\,Collins \inst{cfa}
\and K.\,I.\,Collins \inst{gmu}
\and E.\,Esparza-Borges \inst{iac,ull}
\and A.\,Fukui \inst{tokyo,iac} 
\and C.\,L.\,Gnilka \inst{ames}
\and R.\,Goeke \inst{kavli}
\and P.\,Guerra \inst{albanya}
\and Th.\,Henning \inst{mpia}
\and E.\,Herrero \inst{ice,ieec}
\and S.\,B.\,Howell \inst{ames}
\and S.\,V.\,Jeffers \inst{mpias}
\and J.\,M.\,Jenkins \inst{ames}
\and E.\,L.\,N.\,Jensen \inst{swarthmore}
\and D.\,Kasper \inst{chicago}
\and T.\,Kodama \inst{tokyo}
\and D.\,W.\,Latham \inst{cfa}
\and M.\,J.\,L\'opez-Gonz\'alez \inst{iaa}
\and R.\,Luque\inst{chicago}
\and D.\,Montes \inst{ucm}
\and J.\,C.\,Morales \inst{ice,ieec}
\and M.\,Mori \inst{utokyo}
\and F.\,Murgas \inst{iac}
\and N.\,Narita \inst{tokyo,ACtokyo,iac}
\and G.\,Nowak\inst{iac,ull}
\and H.\,Parviainen \inst{iac,ull}
\and V.\,M.\,Passegger \inst{hs,uok}
\and A.\,Quirrenbach \inst{lsw}
\and S.\,Reffert \inst{lsw}
\and A.\,Reiners \inst{iag}
\and I.\,Ribas \inst{ice,ieec}
\and G.\,R.\,Ricker \inst{kavli}
\and E.\,Rodr\'iguez \inst{iaa}
\and C.\,Rodr\'iguez-L\'opez \inst{iaa}
\and M.\,Schlecker \inst{arizona,mpia}
\and R.\,P.\,Schwarz \inst{patashnick}
\and A.\,Schweitzer \inst{hs}
\and S.\,Seager \inst{kavli}
\and G.\,Stef\'ansson \inst{princeton}
\and C.\,Stockdale \inst{hazelwood}
\and L.\,Tal-Or \inst{ariel,iag}
\and J.\,D.\,Twicken \inst{ames,seti} 
\and S.\,Vanaverbeke \inst{brugge,leuven,iris}
\and G.\,Wang \inst{tsinghua}
\and D.\,Watanabe \inst{planet}
\and J.\,N.\,Winn \inst{princeton}
\and M.\,Zechmeister \inst{iag}
}

\institute{
\label{inst:tls}Th\"uringer Landessternwarte Tautenburg, Sternwarte 5, 07778 Tautenburg, Germany \\
\email{priyanka@tls-tautenburg.de}
\and
\label{inst:lsw}Landessternwarte, Zentrum f\"ur Astronomie der Universit\"at Heidelberg, K\"onigstuhl 12, 69117 Heidelberg, Germany \\
\email{pbluhm@lsw.uni-heidelberg.de}
\and
\label{inst:hs}Hamburger Sternwarte, Universit\"at Hamburg, Gojenbergsweg 112, 21029 Hamburg, Germany
\and
\label{inst:iac}Instituto de Astrof\'isica de Canarias, 38205 La Laguna, Tenerife, Spain
\and
\label{inst:ull}Departamento de Astrof\'isica, Universidad de La Laguna, 38206 La Laguna, Tenerife, Spain
\and
\label{inst:chicago}Department of Astronomy and Astrophysics, University of Chicago, 5640 S. Ellis Avenue, Chicago, IL 60637, USA
\and
\label{inst:cha}Department of Space, Earth and Environment, Chalmers University of Technology, Onsala Space Observatory, 439 92 Onsala, Sweden
\and 
\label{inst:lmu}Universit\"ats-Sternwarte, Ludwig-Maximilians-Universit\"at
M\"unchen, Scheinerstrasse 1, 81679 M\"unchen, Germany
\and
\label{inst:eo}Exzellenzcluster Origins, Boltzmannstrasse 2, 85748 Garching, Germany
\and
\label{inst:mpia}Max-Planck-Institut f\"ur Astronomie, K\"onigstuhl 17, 69117 Heidelberg, Germany
\and
\label{inst:cabesac}Centro de Astrobiolog\'ia (CSIC-INTA), ESAC, Camino bajo del castillo s/n, 28692 Villanueva de la Ca\~nada, Madrid, Spain
\and 
\label{inst:stsci}Space Telescope Science Institute, 3700 San Martin Drive, Baltimore, MD 21218, USA
\and
\label{inst:nexsi}IPAC, Mail Code 314-6, Caltech, 1200 E. California Blvd., Pasadena, CA 91125, USA
\and
\label{inst:iaa}Instituto de Astrof\'isica de Andaluc\'ia (CSIC), Glorieta de la Astronom\'ia s/n, 18008 Granada, Spain
\and
\label{inst:maryland}University of Maryland, Baltimore, MD 21250, USA
\and
\label{inst:goddard}NASA Goddard Space Flight Center, Greenbelt, MD 20771, USA
\and
\label{inst:caha}Centro Astron\'omico Hispano en Andaluc\'ia, Observatorio de Calar Alto, Sierra de los Filabres, 04550 G\'ergal, Spain
\and
\label{inst:cfa}Center for Astrophysics \textbar \ Harvard \& Smithsonian, 60 Garden Street, Cambridge, MA 02138, USA
\and
\label{inst:gmu}George Mason University, 4400 University Drive, Fairfax, VA 22030, USA
\and
\label{inst:tokyo}Komaba Institute for Science, The University of Tokyo, 3-8-1 Komaba, Meguro, Tokyo 153-8902, Japan
\and
\label{inst:ames}NASA Ames Research Center, Moffett Field, CA 94035, USA
\and
\label{inst:kavli}Kavli Institute for Astrophysics and Space Research, Massachusetts Institute of Technology, Cambridge, MA 02139, USA
\and 
\label{inst:albanya}Observatori Astron\`omic Albany\`a, Cam\'i de Bassegoda s/n, Albany\'a 17733, Girona, Spain
\and 
\label{inst:ice}Institut de Ci\`encies de l’Espai (ICE, CSIC), Campus UAB, Can Magrans s/n, 08193 Bellaterra, Spain
\and 
\label{inst:ieec}Institut d’Estudis Espacials de Catalunya (IEEC), 08034 Barcelona, Spain
\and
\label{inst:mpias}Max-Planck-Institut f\"ur Sonnensystemforschung,
Justus-von-Liebig-weg 3, 37077 G{\"o}ttingen, Germany
\and
\label{inst:swarthmore}Department of Physics and Astronomy, Swarthmore College, Swarthmore, PA 19081, USA
\and
\label{inst:ucm}Departamento de F\'{i}sica de la Tierra y Astrof\'{i}sica and IPARCOS-UCM (Instituto de F\'{i}sica de Part\'{i}culas y del Cosmos de la UCM), Facultad de Ciencias F\'{i}sicas, Universidad Complutense de Madrid, 28040, Madrid, Spain
\and
\label{inst:utokyo}Department of Astronomy, Graduate School of Science, The University of Tokyo, 7-3-1 Hongo, Bunkyo-ku, Tokyo 113-0033, Japan
\and
\label{inst:ACtokyo}Astrobiology Center, 2-21-1 Osawa, Mitaka, Tokyo 181-8588, Japan
\and
\label{inst:uok}Homer L. Dodge Department of Physics and Astronomy, University of Oklahoma, 440 West Brooks Street, Norman, OK 73019, USA
\and
\label{inst:iag}
Institut f\"ur Astrophysik and Geophysik, Georg-August-Universit\"at, Friedrich-Hund-Platz 1, 37077 G\"ottingen, Germany
\and
\label{inst:arizona}Department of Astronomy/Steward Observatory, The University of 
Arizona, 933 North Cherry Avenue, Tucson, AZ 85721, USA
\and
\label{inst:patashnick}Patashnick Voorheesville Observatory, Voorheesville, NY 12186, USA
\and
\label{inst:princeton}Department of Astrophysical Sciences, Princeton University, 4 Ivy Lane, Princeton, NJ 08540, USA
\and
\label{inst:hazelwood}Hazelwood Observatory DO3-32, VIC, Australia
\and
\label{inst:ariel}Department of Physics, Ariel University, Ariel 40700, Israel
\and
\label{inst:seti}SETI Institute, Mountain View, CA  94043, USA
\and
\label{inst:brugge}Vereniging Voor Sterrenkunde, Brieversweg 147, 8310, Brugge, Belgium
\and
\label{inst:leuven}Centre for Mathematical Plasma-Astrophysics, Department of Mathematics, KU Leuven, Celestijnenlaan 200B, 3001 Heverlee, Belgium
\and
\label{inst:iris}AstroLAB IRIS, Provinciaal Domein ``De Palingbeek'', Verbrandemolenstraat 5, 8902 Zillebeke, Ieper, Belgium
\and
\label{inst:tsinghua}Tsinghua International School, Beijing 100084, China
\and
\label{inst:planet}Planetary Discoveries, Fredericksburg, VA 22405, USA
}

\date{Received 19 May 2022 / Accepted dd Month 2022}

\abstract{
We report the discovery and characterization of two small transiting planets orbiting the bright M3.0\,V star TOI-1468 (LSPM J0106+1913), whose transit signals were detected in the photometric time series in three sectors of the {\em TESS} mission. We confirm the planetary nature of both of them using precise radial velocity measurements from the CARMENES and MAROON-X spectrographs, and supplement them with ground-based transit photometry. 
A joint analysis of all these data reveals that the shorter-period planet, TOI-1468\,b  ($P_{\rm b}$ = 1.88\,d), has a planetary mass of $M_{\rm b} = 3.21\pm0.24$\,$M_{\oplus}$ and a radius of $R_{\rm b} =1.280^{+0.038}_{-0.039}\,R_{\oplus}$, resulting in a density of $\rho_{\rm b} = 8.39^{+ 1.05}_{- 0.92}$\,g\,cm$^{-3}$, which is consistent with a mostly rocky composition. 
For the outer planet, TOI-1468\,c ($P_{\rm c} = 15.53$\,d), we derive a mass of $M_{\rm c} = 6.64^{+ 0.67}_{- 0.68}$\,$M_{\oplus}$, a radius of $R_{\rm c} = 2.06\pm0.04\,R_{\oplus}$, and a bulk density of $\rho_{c} = 2.00^{+ 0.21}_{- 0.19}$\,g\,cm$^{-3}$, which corresponds to a rocky core composition with a H/He gas envelope. 
These planets are located on opposite sides of the radius valley, making our system an interesting discovery as there are only a handful of other systems with the same properties. This discovery can further help determine a more precise location of the radius valley for small planets around M dwarfs and, therefore, shed more light on planet formation and evolution scenarios.}


\maketitle\  

\section{Introduction}\label{sec:intro}

A number of space-based transit surveys such as \textit{CoRoT} \citep{2006cosp...36.3749B}, \textit{Kepler} \citep{Borucki2010}, and now {\em TESS} \citep{Ricker2015}, have been able to determine precise orbital periods and radii of several thousands of exoplanets. 
Combining the transit light curves with radial-velocity (RV) measurements yields the planet density, as well as a complete set of orbital parameters. 
Currently, the total number of confirmed exoplanets is more than 5000\footnote{\url{tps://exoplanetarchive.ipac.caltech.edu}, accessed on 4 May 2022.}, resulting in a broad range of measured planet bulk densities, giving us the first hints about the internal composition of planets, which is a crucial element for our understanding of their formation. 
One of the most important results from these discoveries is the large amount of planets with radii smaller than the radius of Neptune but larger than that of the Earth ($R_{\rm p} \approx$ 1--3.9\,$R{_\oplus}$) \citep{2013ApJS..204...24B}. 
However, until the advent of {\em TESS}, most of these exoplanets had been detected around solar-type stars, while a complete picture of the process of planet formation requires an understanding of the architecture around all types of stars. 

Solar-type stars have been the prime targets of many transit searches. Some examples of initial RV surveys that focused on later stars, down to M dwarf spectral types, were the survey of high-metallicity stars (N2K; \citealt{Fischer2005}) and the California planet survey \citep{Howard2010}. 
With advancements in space-based transit missions and higher-precision RV measurements with a broader wavelength coverage, especially toward the red end of the spectrum, such as with the CARMENES \citep{Quirrenbach2014} and the MAROON-X \citep{maroonx2018,Seifahrt2020} spectrographs, we are starting to shift the focus toward M dwarfs, the most abundant stars in our galaxy \citep{Chabrier2003,Henry2018,Reyle2021}. 

One of the main advantages of late-type dwarfs over  solar-type stars are the relative sizes and masses between the host-stars and their planets, which make these systems more detectable via transit and RV techniques. 
This fact has been exploited by surveys that have exclusively focused on searches for planets around M dwarfs, such as the SPIRou Legacy Survey \citep{Cloutier2018}, the M dwarfs in the Multiples survey with Subaru \citep{Ward-Duong2015}, and similar such surveys with UVES, HARPS, CARMENES, and other instruments \citep{Kuerster2003, Charbonneau2008, Zechmeister2009, Bonfils2013, Reiners18}.

Transiting planet discoveries have shown that planetary interiors can be quite diverse, ranging from completely rocky cores to gas-dominated planets. They have also indicated a higher frequency for low-mass planets ($1\,M_{\oplus} \lesssim M_{\rm p} \sin{i} \lesssim 10\,M_{\oplus}$) around low-mass stars ($M_{\odot} \lesssim 0.6\,M_{\oplus}$) in orbits less than 100\,d, compared to solar-type stars \citep{Howard2012, Dressing2013, Hsu2020}. 
In a recent study, \cite{Sabotta2021} found an occurrence rate of $1.32^{+0.33}_{-0.31}$ low-mass planets for low-mass stars in periods up to 100\,d. 
Detailed studies of several of these planets occurring around solar-type stars have revealed a bimodal distribution of planets peaking at 1.3\,$R_{\oplus}$ and 2.6\,$R_{\oplus}$, and consequently a relative paucity of planets between 1.5\,$R_{\oplus}$ and 1.8\,$R_{\oplus}$, also known as the radius valley \citep{Fulton17, Zeng2017, vaneylen2018, Berger2018}.

One of the explanations for this bimodal distribution is the formation of planets in a gas-poor environment. In this scenario, the inner disk, where the planets form, is clear of H gas \citep{Owen2013}. 
Thus, irrespective of the mass of the planet, and in absence of such gases, the close-in planets cannot accrete H and He. However, a planet that formed further away from its host star and subsequently migrated inward may then be able to keep its H+He envelope. Nevertheless, not all systems can be explained in this way. Systems such as \object{K2--3} \citep{Damasso2018} and \object{TOI-1266} \citep{Stefansson2020}, where the inner planets are larger than the outer planets, defy these assumptions. 
These systems could instead be explained by assuming that the outer planets had a richer water ice composition \citep{Owen2020}.

Another explanation is that all planets are formed with an H atmosphere but they lose it during the course of evolution, mainly in the first 100\,Ma after formation \citep{Lammer2014, Linsky2015}. Here, the accreted H+He envelope is removed due to the  extreme ultraviolet (XUV) radiation of the host star \citep{Sanz-Forcada2011,Owen2013, Lopez2013}. The erosion depends on the surface gravity of the planet, its separation from the host star, and the amount of XUV radiation that the planet has received during its lifetime.
The outer atmospheres for planets with masses less than 10\,$M_{\oplus}$, or orbiting very close to the host stars, can be easily eroded by the XUV radiation coming from the host star, especially if the host star is active. 
The activity of M stars increases toward the latest spectral type \citep{Reiners2012} and, since the lifetimes of these stars are also long, they can be in a relatively high-activity phase for a long time. 
As a result, planet atmosphere losses due to XUV erosion can be particularly high. 
Alternatively, the atmospheric losses would be less if the star was relatively inactive when it was young. In a third scenario, where atmospheric losses are driven by the energy release from the formation process \citep{Ginzburg+2018, Gupta2019, gupta2020}, the H+He envelope is removed because young, rocky planets are very hot. 
The removal of the envelope typically takes place on the order of 1\,Ga. Since the planet is the driving force, this loss mechanism should also be relevant for planets orbiting at large separations from their host stars. 
Additionally, \cite{Ginzburg+2018} and \cite{Gupta2019} also predicted that the
location of the radius valley should decrease with orbital period as $R_{\rm valley} \sim P^{-0.13}$. 
It is possible that all these processes are relevant for the evolution of the planets. However, one process could be more relevant for planets orbiting a specific type of star than for another. 

In this paper, we present the discovery of a multi-planetary system with at least two transiting planets around an early-to-mid-type M dwarf, LSPM~J0106+1913 \citep{Lepine2005}, recently cataloged as TOI-1468. The paper is organized as follows. In Sect.\,\ref{sec:phot-TESS}, we describe the space-based photometry from {\em TESS}. Section\,\ref{sec:ground-based} comprises all the ground-based observations including additional photometry, high-resolution imaging, and CARMENES high-resolution spectroscopy. In Sect.\,\ref{sec:star}, we discuss the host star by listing its stellar properties and investigating the rotational period of the star. In Sect.\,\ref{sec:results}, we discuss the detailed modeling of the RV and transit data, and the obtained results. We finally interpret our results in Sect.\,\ref{sec:discussion} and present a brief summary in Sect.\,\ref{sec:summary}.  

\section{{\em TESS} photometry} \label{sec:phot-TESS}

\subsection{Transit search}

\begin{figure*}
    \centering
    \includegraphics[width=\textwidth]{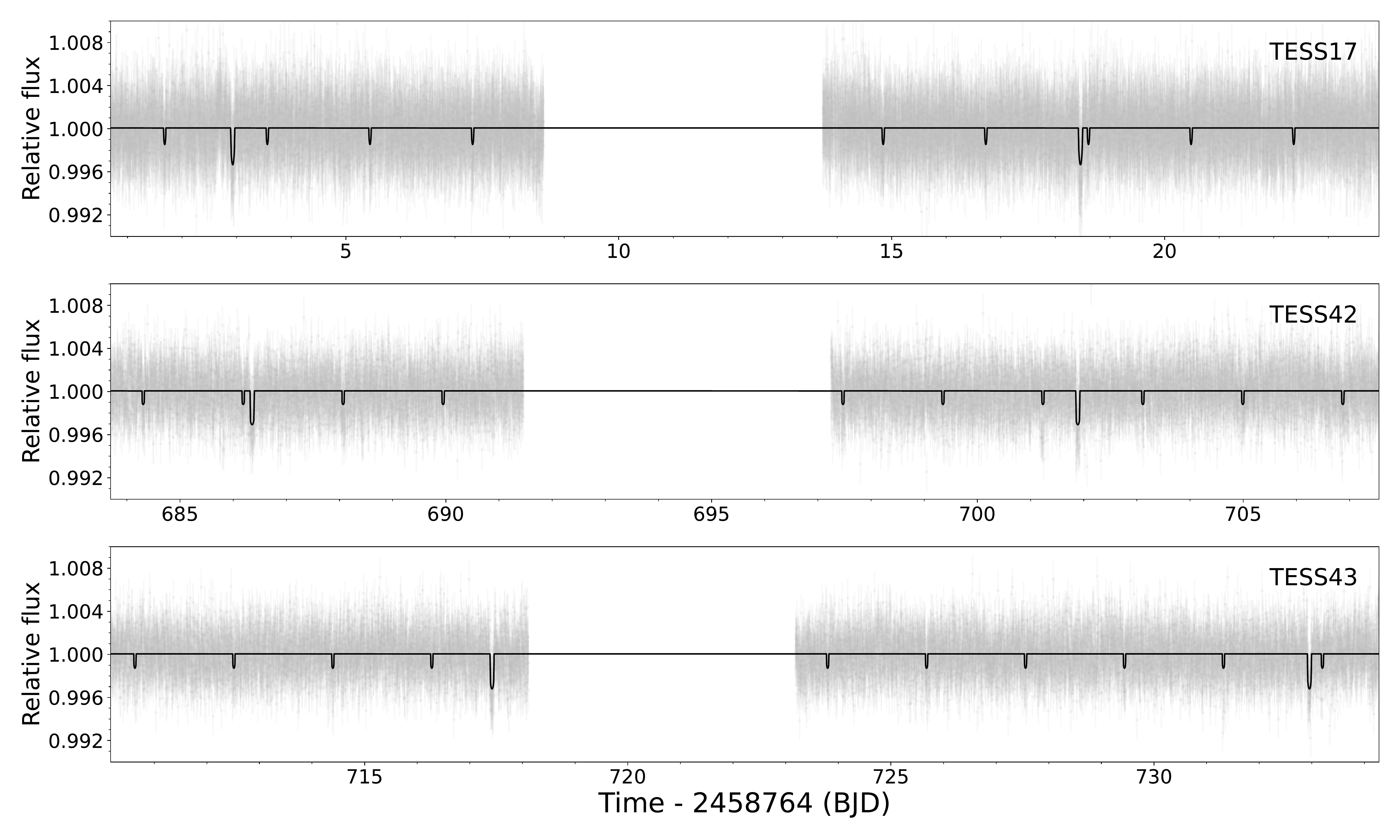}
    \caption{{\em TESS} PDCSAP light curve for TOI-1468 (gray points) for three sectors: 17, 42, and 43, overplotted with the two-transiting-planet model in black.}
    \label{fig:photo-TESS}
\end{figure*}

\begin{figure*}
    \centering
    \includegraphics[width=\textwidth]{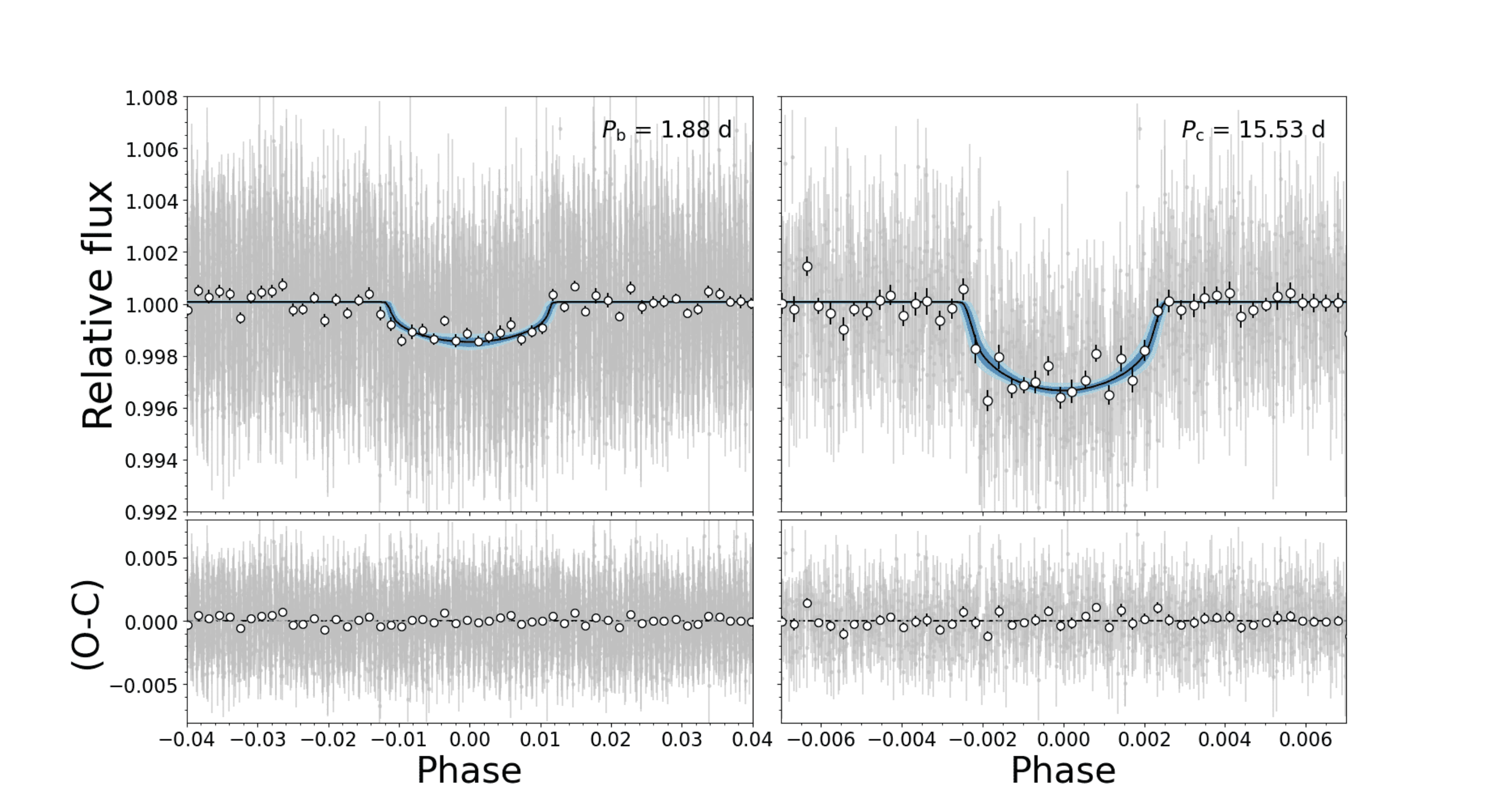}
    \caption{Phase-folded {\em TESS} transit light curves for TOI-1468\,b at 1.88\,d ({\em left}) and TOI-1468\,c at 15.53\,d ({\em right}).
    Gray points are 2\,min (and 20\,s) cadence data, and open circles are binned data  (shown only for reference; data used to fit the model were the unbinned points).
    The best-fit\,\texttt{juliet} model (black line; see Sect.\,\ref{sec:results}) is overplotted for both TOI-1468\,b and TOI-1468\,c, along with shaded regions, light blue for $95\,\%$ and dark blue for $68\,\%$ confidence intervals.}
    \label{fig:photo-TESS_phase}
\end{figure*}

The {\em TESS} mission was designed to perform an all-sky survey to detect transiting planets using its four cameras, each having a field of view of $24\, \times\,24$\,deg$^2$ outfitted with four 2k\,$\times$\,2k CCDs (charge-coupled devices). The light curves are archived in raw and processed format in the Mikulski Archive for Space Telescopes\footnote{\url{https://mast.stsci.edu}, \url{https://archive.stsci.edu/}}. 
TIC 243185500 (discovery name: LSPM J0106+1913)
was observed at 2\,min short-cadence integrations in sector 17. 
The data validation report \citep{Twicken:DVdiagnostics2018, Li2019} produced by the {\em TESS} Science Processing
Operations Center \citep[SPOC;][]{SPOC}) identified transit signals with orbital periods of 1.88\,d and 15.53\,d.
The target star was subsequently promoted to {\em TESS} Object of Interest (TOI) status as TOI-1468 by the {\em TESS} Science Office; the associated planet candidates were designated as TOI-1468.01 (15.53\,d) and TOI-1468.02 (1.88\,d) \citep{Guerrero2021}. 
Finally, TOI 1468 was observed at 2\,min (and 20\,s) cadence in extended mission sectors 42 and 43 (see Table\,\ref{tab:phot_TESS} for details). 
The transit depths for the inner and outer planets are 1.66\,mmag and 3.73\,mmag, respectively. 

We show the {\em TESS} SPOC pre-search data conditioning simple aperture photometry (PDCSAP) \citep{Smith2012, Stumpe2012, Stumpe2014} for sectors 17, 42, and 43 observed for both transiting planets in Fig.\,\ref{fig:photo-TESS}. Phase-folded and best-fit models for both planets are shown in Fig.\,\ref{fig:photo-TESS_phase} (see Sect.\,\ref{subsec:orbit-model} for detailed analysis.)

\begin{table}
\centering
\small
\caption{{\em TESS} observations of TOI-1468.} 
\label{tab:phot_TESS}
\begin{tabular}{cccll}
\hline\hline
\noalign{\smallskip}
Sector   &  Camera   & Cycle   & Start date & End date\\
\noalign{\smallskip}
\hline
\noalign{\smallskip}
17  &  1  &   2 &  07 October 2019   & 02 November 2019\\
42  &  3  &   4 &  20 August 2021    & 16 September 2021\\
43  &  1  &   4 &  16 September 2021 & 12 October 2021\\
\noalign{\smallskip}         
\hline
\end{tabular}
\end{table}

\subsection{Limits on photometric contamination} \label{subsec:contamination}

\begin{figure*}
    \centering
    \includegraphics[width=\textwidth]{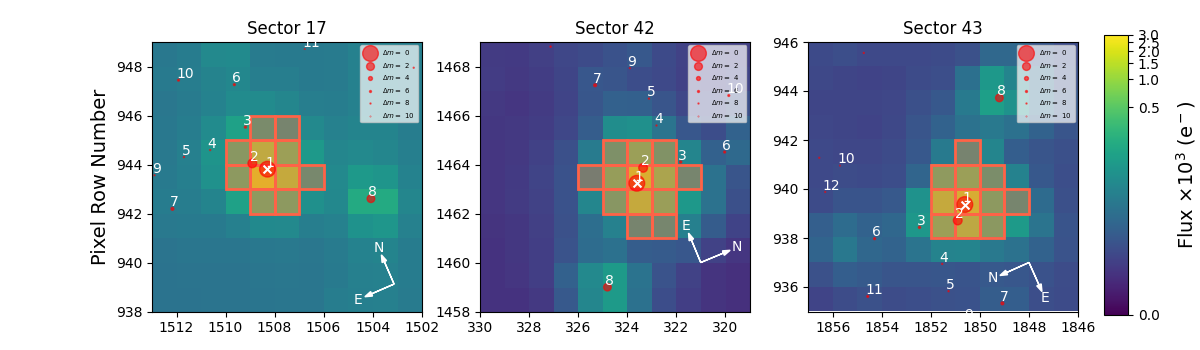}
         \caption{Target pixel files 
         of TOI-1468 in {\em TESS} sectors 17, 42, and 43. The electron counts are color-coded. The red-bordered pixels are used in the simple aperture photometry. The size of the red circles indicates the {\em TESS} magnitudes of all nearby stars and TOI-1468 (circle $\#$1 marked with \guillemotleft$\times$\guillemotright).}
        \label{fig:TPF}
\end{figure*}

The large {\em TESS} pixel size of $\sim21$\,arcsec increases the likelihood of contamination by nearby stars. In Fig.\,\ref{fig:TPF}, we plot all the {\em Gaia} sources within the field of view of the {\em TESS} aperture with the help of \texttt{tpfplotter}\footnote{\url{https://github.com/jlillo/tpfplotter}} \citep{2020A&A...635A.128A}. 
The advantage of this comparison is that both the {\em Gaia} $G_{RP}$ band (630--1050\,nm) and the {\em TESS} $T$ band (600--1000\,nm) have a similar wavelength coverage.
The SPOC crowding metric for TOI-1468 in the three {\em TESS} sectors was $\sim$0.91. This means that according to SPOC modeling after background removal, $\sim91\,\%$ of the flux in the photometric aperture was attributable to the target star, and $\sim9\,\%$ to other sources, 
especially to source \#2 (TIC 243185499, \textit{Gaia} EDR3 2785466581298775552), which is separated from TOI-1468 by $\sim$14\,arcsec and is 1.7\,mag fainter in the $G$ band.
The PDCSAP flux level was reduced to account for contamination by other sources, as described in the SPOC PDCSAP references. 
The high-resolution imaging data for ascertaining any resolved close multiplicity of TOI-1468 is described in Sect.~\ref{subsec:imaging}.

\section{Ground-based observations}\label{sec:ground-based}

\subsection{Ground-based photometry}\label{subsec:ground-phot}

\begin{figure*}
    \centering
    \includegraphics[width=\textwidth]{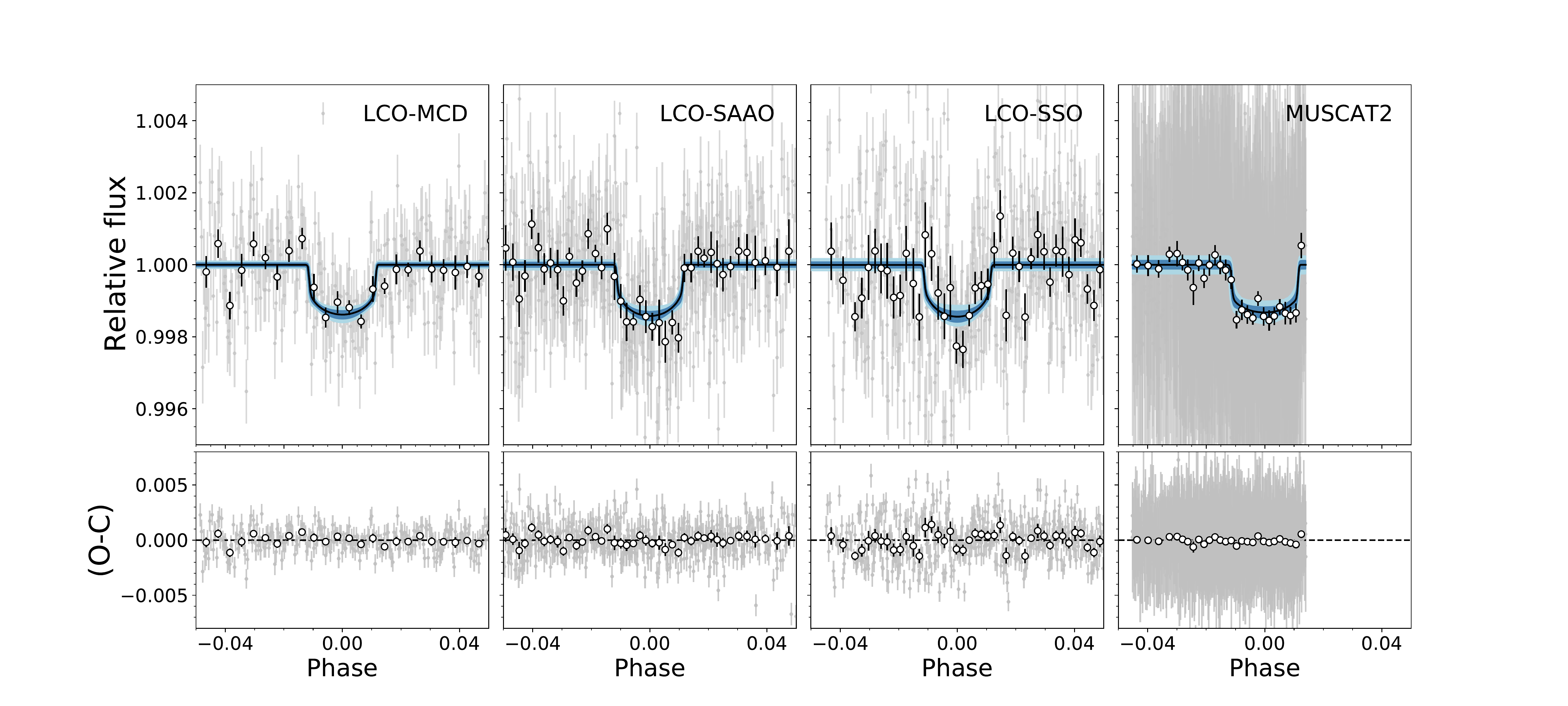}
             \caption{Ground-based photometry data for TOI-1468\,b was observed from various facilities, namely, LCO-McD, LCO-SAAO, LCO-SSO, and MUSCAT2. The normalized flux data are plotted in gray points, with the binned data points highlighted by circles, along with their error bars. The number of data points per bin was ten for LCO-McD, LCO-SAAO, LCO-SS0, and 30 for MUSCAT2. The \texttt{juliet} best-fit model for each set is plotted as a solid black line, along with shaded regions: light blue for the $95\%$ confidence interval, and dark blue for the $68\%$ confidence interval. Details can be found in Sect.\,\ref{sec:results}. The residuals are plotted in the bottom part of each of the panels.}
        \label{fig:phot-ground-1}
\end{figure*}

\begin{figure*}
    \centering
    \includegraphics[width=\textwidth]{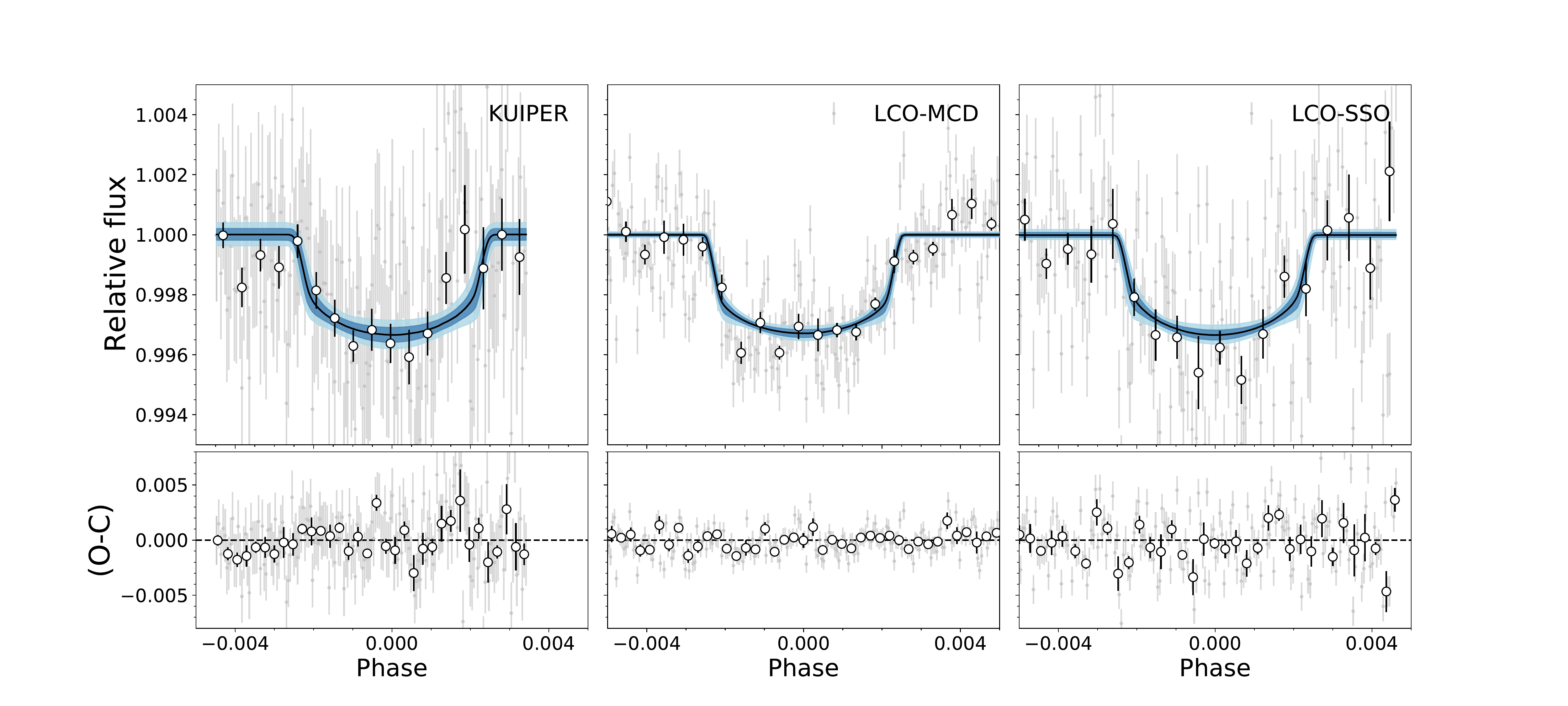}
             \caption{Ground-based photometry data for TOI-1468\,c observed from various facilities, namely, SO-KUIPER, LCO-McD, and LCO-SSO. The normalized flux data are plotted in gray, with the binned data in circles, along with their error bars. The number of data points per bin was ten. The \texttt{juliet} best-fit model for each set is plotted as a solid black line, along with shaded regions: light blue for the $95\%$ confidence interval, and dark blue for the $68\%$ confidence interval. The residuals are plotted in the bottom part of each of the panels.}
        \label{fig:phot-ground-2}
\end{figure*}

Several targeted observations of TOI-1468 were scheduled to monitor the transits for both planetary candidates with various ground-based facilities. 
The summary description of all the observed transits is given in Table\,\ref{tab:ground-based-transits}. 
We further examined archival time-series photometry data of TOI-1468 and listed these observations in Table\,\ref{tab:phot-archive}. 
The photometric data of each facility, phase-folded to the best-fit model (Sect.~\ref{sec:results}), are shown in Fig.\,\ref{fig:phot-ground-1} for TOI-1468\,b and in Fig.\,\ref{fig:phot-ground-2} for TOI-1468\,c. 
All the data sets were modeled with the \texttt{juliet} package and the best-fit model is overplotted in each of the panels (see Sect.\,\ref{subsec:orbit-model} for details). 
In the following paragraphs, we describe the eventually used photometric ground-based photometric data for TOI-1468.
Unused data sets, either from archival or follow-up observations (i.e., MEarth, TRAPPIST, FLWO, GMU), did not have enough quality for the relatively shallow transits of TOI-1468\,b and\,c.

\begin{table*}
\centering
\tiny
\caption{Ground-based observations of TOI-1468 transits.}
\label{tab:ground-based-transits}
\label{tab:phot}
\begin{tabular}{l lll ccc cc l}
\hline
\hline
\noalign{\smallskip}
Planet   &  Telescope   & Camera or  & Filter & Pixel scale & PSF$^a$ & Aperture & Date & Duration & Used \\
~ & ~ & instrument & ~ & (arcsec) & (arcsec) & radius (pixel) & (UT) & (min) & data set$^b$ \\
\noalign{\smallskip}
\hline
\noalign{\smallskip}
\noalign{\smallskip}

b &     MEarth-N (0.4\,m)       & Apogee U42 & $RG715$ &0.75 &  3.2     &  8.0 & 2019-12-12& 384.0 & ...  \\

b & MEarth-Nx7 (0.4\,m) &Apogee U42     & $RG715$ & 0.76& 6.9   &  9.0&2019-12-12& 385.0 & ... \\

b & MEarth-S (0.4\,m)   & Apogee U/F230 &$RG715$ & 0.84& 2.2& 4.2       & 2019-12-12 & 206.0 & ... \\

b &     TCS (1.52\,m) & MuSCAT2 & $g$, $r$, $i$, $z_s$ & 0.44 & 11.8$^c$ & ... & 2019-12-13      & 174.6 & ... \\

b &     TRAPPIST-N (0.60\,m) & Andor IKON-L BEX2-DD & $z$ &0.60 & 3.0 & 6.01 &2019-12-13& 210.0 & ... \\

c &     MEarth-S (0.4\,m) &Apogee U/F230 &$RG715$ &     0.84 & 2.2 & 6.0 & 2019-12-27 & 137.0 & ... \\

c &     MEarth-Sx6 (0.4\,m)     & Apogee U/F230&$RG715$& 0.84 & 5.1& 9.9        & 2019-12-27 & 140.0 & ... \\

c &     SO-Kuiper (1.5\,m) & Mont4k     & $B$ & 0.42    & ... & ... &   2020-01-27 &       176.0 & Yes \\

c &     FLWO (1.2\,m)   & KeplerCam     & $i$ & 0.672 & 2.2     & 6.0   & 2020-01-27 &    148.0 & ... \\

b &     LCOGT-SAAO (1.0\,m) & Sinistro &        $g_p$ & 0.389 & 2.73 &  12.0 & 2020-07-19 &  229.0 & Yes \\

b &     LCOGT-SAAO (1.0\,m) & Sinistro &        $g_p$ &0.389    & 1.81  & 10.0 &  2020-08-19 &    256.0 & Yes \\

b &     LCOGT-SSO (1.0\,m) & Sinistro & $z_s$ & 0.39 &  1.93 &  15.0 &  2020-08-27      & 252.0 & Yes \\

c &     LCOGT-SSO (1.0\,m) &    Sinistro &      $z_s$ & 0.39 &  2.71 &  13.0 & 2020-10-01 &  281.0 & Yes \\

b &     GMU (0.8\,m)    & SBIG STX-16803+FW-7 & $R$  & 0.36 &   5.34 &  15.0 &       2020-10-06 & 194.0 & ... \\

b &     LCOGT-SSO (1.0\,m) & Sinistro & $i_p$ & 0.389 & 4.58 &  17.0 & 2020-10-15 &       277.0 & Yes \\

c &     LCOGT-McD (1.0\,m) &    Sinistro &      $i_p$ & 0.39 &  2.71 &  11.0 &       2020-10-17 & 317.0 & Yes \\

b &     LCOGT-SAAO (1.0\,m)     & Sinistro &    $i_p$   & 0.39 &        4.33 &       14.0 &  2020-10-24      & 337.0 & Yes \\
        
b &     LCOGT-McD (1.0\,m) & Sinistro & $i_p$ & 0.389 & 7.47 &  21.0 &  2020-11-22& 303.0 & Yes \\

b &     TCS (1.52\,m) & MuSCAT2 & $g$, $i$, $z{_s}$ &   0.44 & 11.8$^c$ & ... & 2021-07-14        & 116.0 & Yes   \\

b &     TCS (1.52\,m) & MuSCAT2 & $g$, $i$, $z{_s}$ &   0.44 & 11.8$^c$ & ... & 2021-08-30        & 161.0 & Yes \\
\noalign{\smallskip}         
        \hline
    \end{tabular}
   \tablefoot{
   \tablefoottext{$^a$}{Estimated point spread function.}
   \tablefoottext{$^b$}{Data sets included in the fit.}
   \tablefoottext{$^c$}{Defocused MuSCAT2/TCS.}
   }
\end{table*}

\paragraph{LCOGT.}
Las Cumbres Global Telescope Network \citep[LCOGT; ][]{Brown2013} is a network of 0.4\,-meter, 1-meter, and 2-meter fully automated robotic telescopes spread across the globe. 
We recorded eight transits of TOI-1468 with three 1-meter telescopes of the LCOGT network and three different filters ($g_p$, $z_s$, $i_p$).
In particular, we observed six transits of TOI-1468\,b and two transits of TOI-1468\,c at the LCOGT South African Astronomical Observatory (SAAO) in South Africa on 19 July 2020, 19 August 2020, and 24 October 2020, 
at the LCOGT Siding Spring Observatory (SSO) in Australia) on 27 August 2020, 01 October 2020, and 15 October 2020, 
and at the LCOGT McDonald (McD) Observatory, in the USA on 17 October 2020 and 22 November 2020.
Observation durations varied between 229\,min and 337\,min, significantly longer than the transit durations ($\sim${73--108}\,min).

The data were reduced with the automated Python-based \texttt{BANZAI} pipeline \citep{McCully:2018}. 
The pipeline performs the standard process of data reduction, including the removal of bad pixels, bias subtraction, dark subtraction, flat-field correction, source extraction photometry (with Python and C libraries), and astrometric calibration (with {\tt astrometry.net}). 
Aperture photometry radii varied depending on the local seeing, between 4.3\,arcsec and 6.6\,arcsec.
The transit data were further analyzed using the \texttt{AstroImageJ} (\texttt{AIJ}) software \citep{Collins:2017} and airmass detrended for all the datsets. 

\paragraph{SO-Kuiper.}
The 61-inch Kuiper telescope is operated by the Steward Observatory and is located at 2500\,m at Mt.~Bigelow in the Catalina Mountains north of Tucson, Arizona (USA). 
The 4k\,$\times$\,4k Mont4K CCD was used for the imaging to monitor a single transit of TOI-1468\,c with a $B$ filter on 27 January 2020. The target star was observed for a duration of 4.5\,h with an average seeing of $\sim$\,3\,arcsec. 
The SO-Kuiper data reduction was done with publically available Python pipeline \citep{2018SPIE10704E..2HW}, which is based on \texttt{IRAF}'s \citep{IRAF1993} \texttt{ccdproc} and follows the basic reduction steps of overscan, trim, bias, and flat field correction. 
Further analysis was done with the \texttt{AIJ} software using the fixed aperture of four pixels. 

\paragraph{MuSCAT2.}
We observed two transits of TOI-1468\,b simultaneously in $g$, $r$, $i$, and $z_\mathrm{s}$ bands with the MuSCAT2 multicolor imager  \citep{Narita2019} installed on the 1.52-meter Telescopio Carlos S\'anchez (TCS) at the Observatorio del Teide, Tenerife (Spain). 
The observations were carried out on the nights of 14 July 2021 and 30 August 2021, with exposure times optimized each night and per passband, and varied from 10\,s to 30\,s.
The airmass varied from a minimum of 1.1 to a maximum of 1.5 during the first night, and from a minimum of 1.01 to a maximum of 1.13 during the second night. 
The observing conditions were good through both nights, but the scatter in photometry was higher than expected. This high scatter is likely attributed to high levels of atmospheric dust.
The photometry was conducted using standard aperture photometry calibration and reduction steps with a dedicated MuSCAT2 photometry pipeline, as described in \citet{Parviainen2019}. The pipeline calculates aperture photometry for the target and a set of comparison stars and aperture sizes, and creates the final relative light curves via global optimization of a model that aims to find the optimal comparison stars and aperture size, while simultaneously modeling the transit and baseline variations modeled as a linear combination of a set of covariates.

\subsection{High-resolution spectroscopy }\label{subsec:rv}

The high-resolution spectroscopic data used for this paper were obtained with CARMENES\footnote{Calar Alto high-resolution search for M dwarfs with exoearths with near-infrared and optical \'echelle spectrographs\: \url{http://carmenes.caha.es}},
fiber-coupled to the Cassegrain focus of the 3.5-meter telescope at the Observatorio de Calar Alto in Almer\'ia (Spain), and MAROON-X, 
a new extreme precision RV spectrograph at the 8.1\,m Gemini North telescope in Maunakea, Hawai'i (USA).

\subsubsection{CARMENES}

CARMENES is a dual channel spectrograph operating in the optical wavelength band (VIS) between 0.52\,$\mu$m and 0.96\,$\mu$m, with a spectral resolving power of $\mathcal{R}$\,=\,94\,600, and in the near-infrared (NIR) between 0.96\,$\mu$m and 1.71\,$\mu$m at $\mathcal{R}$\,=\,80.400. 
With CARMENES, we obtained 65 spectra for TOI-1468 between 20 January 2020 and 09 October 2020. 
The exposure times were 1800\,s. 
The spectra followed the standard data flow \citep{Caballero2016} and were reduced with \texttt{caracal} \citep{2014A&A...561A..59Z}, while the RVs were produced with \texttt{serval} \citep{2018A&A...609A..12Z}. 
The reduction included the standard process of barycentric and instrumental drifts corrections. 
\texttt{serval} produces RVs were nightly-zero-point corrected as discussed by \citealt{Kaminski18}, \citealt{2019MNRAS.484L...8T},
and especially, \citealt{2020A&A...636A..74T}. 
Additional information, such as spectral activity indices, were also produced, as part of the science products from \texttt{serval}, such as the CRX chromatic index (VIS and NIR) and dLW, the differential line width. 
Following the method of \cite{Schoefer2019}, we additionally computed $\log{L_{{\rm H}\alpha} / L_{\rm bol}}$ and a number of atomic and molecular indices (H$\alpha$, He~{\sc i}~D$_3$, Na~{\sc i}~D$_1$ and D$_2$, Ca\,{\sc ii}\,IRT1, -2, and -3, He~{\sc i}~$\lambda$10833\,\AA, Pa$\beta$, CaH-2, CaH-3, TiO\,7050, TiO\,8430, TiO\,8860, VO\,7436, VO\,7942, and FeH Wing-Ford).  
The average signal-to-noise ratio (S/N) of the CARMENES spectra is 61 at 740\,nm, measured at the peak of the blaze function. 
The median error and the scatter of the time series of the VIS RVs are 2.0\,m\,s$^{-1}$ and $\sigma$ = 4.7\,m\,s$^{-1}$, respectively, while those of the NIR RVs are 8.0\,m\,s$^{-1}$ and $\sigma$ = 9.8\,m\,s$^{-1}$, respectively. The median errors on NIR RV data were larger than the predicted RV semi-amplitude, $K \sim$ 3--4\,m\,s$^{-1}$), and therefore we only used VIS RVs for all of our further analyses. 
The CARMENES RVs, along with their uncertainties and their respective BJD time stamps, are listed in Table\,\ref{tab:RV_table}.

\subsubsection{MAROON-X}

MAROON-X is a stabilized, fiber-fed \'echelle spectrograph, with a resolving power of $\mathcal{R}\,=\,85\,000$ and a wavelength range of 0.50--0.92\,$\mu$m covered by two arms. 
MAROON-X demonstrated an RV stability of at least 30\,cm\,s$^{-1}$ over the span of a few weeks during its first year of operations \citep{Seifahrt2020} and was used to determine the precise mass of the nearby transiting rocky planet Gl\,486\,b \citep{Trifonov2021}.
We obtained 16 spectra of TOI-1468 in two observing runs in August and October-November 2021.
The exposure time was typically 600\,s. 
The RVs from both runs were treated as independent data sets with their own RV offset. 
The spectra were reduced with a custom package and the RVs were produced with a Python 3 implementation of \texttt{serval} \citep{2018A&A...609A..12Z}. 
One-dimensional spectra and RVs were computed separately for the blue and red arms of MAROON-X. 
Barycentric corrections were calculated for the flux-weighted midpoint of each observation. Wavelength solutions and instrumental drift corrections were based on the MAROON-X etalon calibrator. 
In August 2021, an additional ad hoc drift correction of 0.19\,m\,s$^{-1}$\,day$^{-1}$ was applied, based on consistent systematics found in the observations of multiple RV standard stars. 
As for CARMENES, additional information, such as spectral activity indicators (CRX and dLW), as well as line indices for H$\alpha$, Na~{\sc i}~D, and Ca\,{\sc ii}\,IRT1, -2, and -3, were computed. Average S/N (at the peak of the blaze) for the spectra of TOI-1468 are $\sim$50 at 640\,nm in the blue arm and $\sim$125 at 800\,nm in the red arm. 
These large rations resulted in average RV uncertainties of 1.8\,m\,s$^{-1}$ for the blue arm and 0.95\,m\,s$^{-1}$ for the red arm of MAROON-X. 
The RVs, along with their uncertainties and activity indicators, are listed in Table~\ref{tab:RV_table_MAROONX}

\subsection{High-resolution imaging} \label{subsec:imaging}

For TOI-1468, the {\em Gaia} EDR3 renormalized unit weight error ({\tt RUWE}) value is 1.62, which is slightly above the critical value of 1.40.
This value might hint that the source could be non-single or problematic for the photometric solution \citep{2018A&A...616A..17A,2018A&A...616A...2L}. 
Due to the {\tt RUWE} value and the large pixel size of {\em TESS}, we obtained Gemini high-resolution speckle imaging in the visible, and Palomar adaptive optics imaging in the near-infrared, to detect and measure the contribution of any contaminating sources near TOI-1468.

\subsubsection{Gemini} \label{gemini} 

\begin{figure}
    \centering
\includegraphics[width=0.45\textwidth]{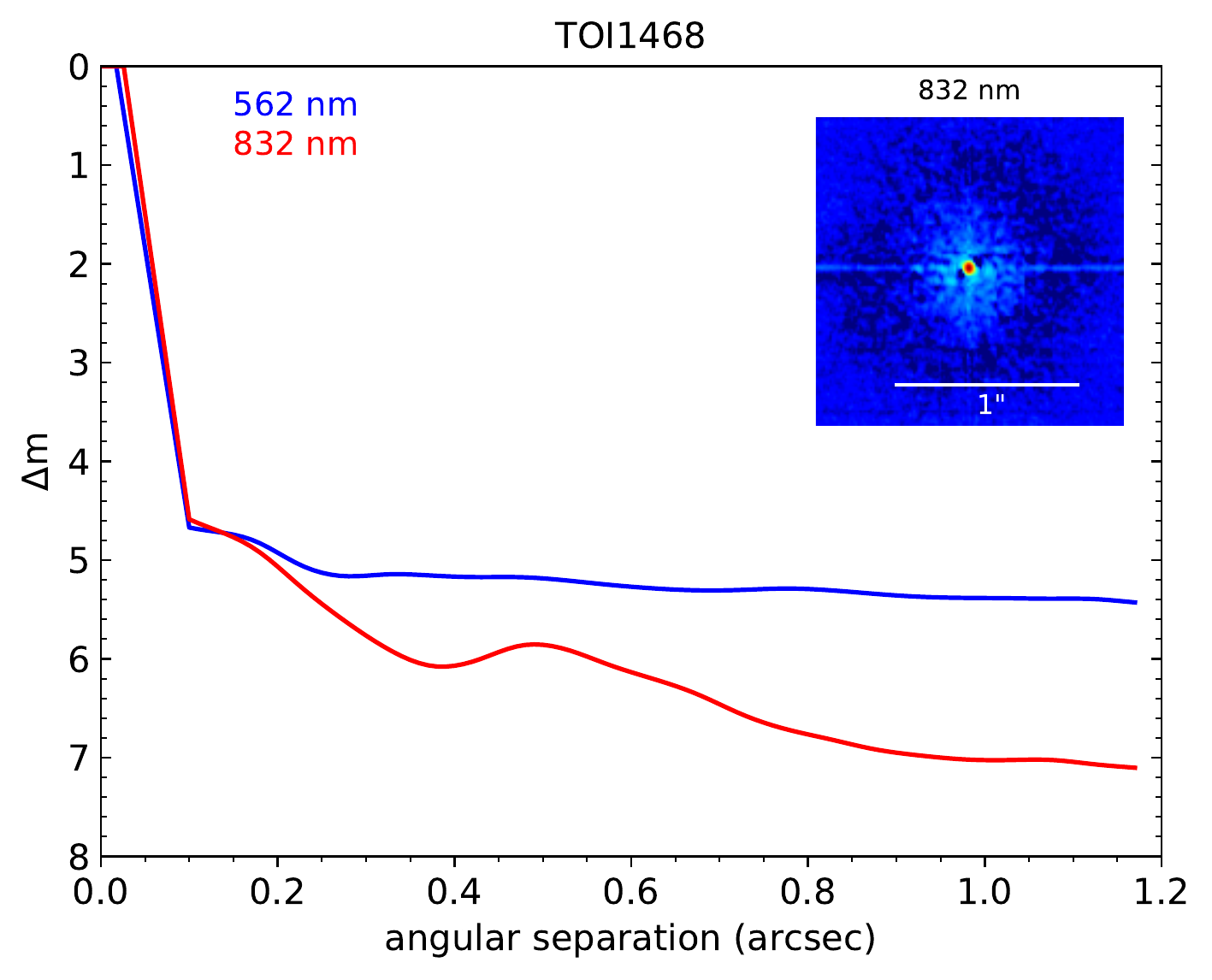}
      \caption{`Alopeke 832\,nm reconstructed image of TOI-1468 and 5$\sigma$ contrast curves for the simultaneous observations at 562\,nm (blue) and 832\,nm (red).}
        \label{fig:speckle}
\end{figure}

TOI-1468 was observed on 04 August 2020 with the `Alopeke speckle imager mounted on the 8.1-meter Gemini North telescope. 
The data were simultaneously acquired in two bands centered at 562\,nm and 832\,nm, with filter bandwidths of 54\,nm and 40\,nm, respectively, on which eight sets of 1000\,$\times$\,0.06\,s exposures were obtained. 
The images were reduced, as discussed by \cite{2011AJ....142...19H}.
The inner working angle (which is equal to the diffraction limit) is 0.02\,arcsec at 562\,nm and 0.03\,arcsec at 832\,nm. The inner spatial resolution is $\sim$0.5--0.7\,au at the TOI-1468 distance. 
Between 0.1\,arcsec and 1.2\,arcsec, we excluded nearby stars fainter than $\sim$5 mag at 562\,nm and $\sim$5--7\,mag at 832\,nm, as shown in Fig.~\ref{fig:speckle}.

\subsubsection{Palomar} \label{palomar}

\begin{figure}
    \centering
    \includegraphics[width=0.49\textwidth]{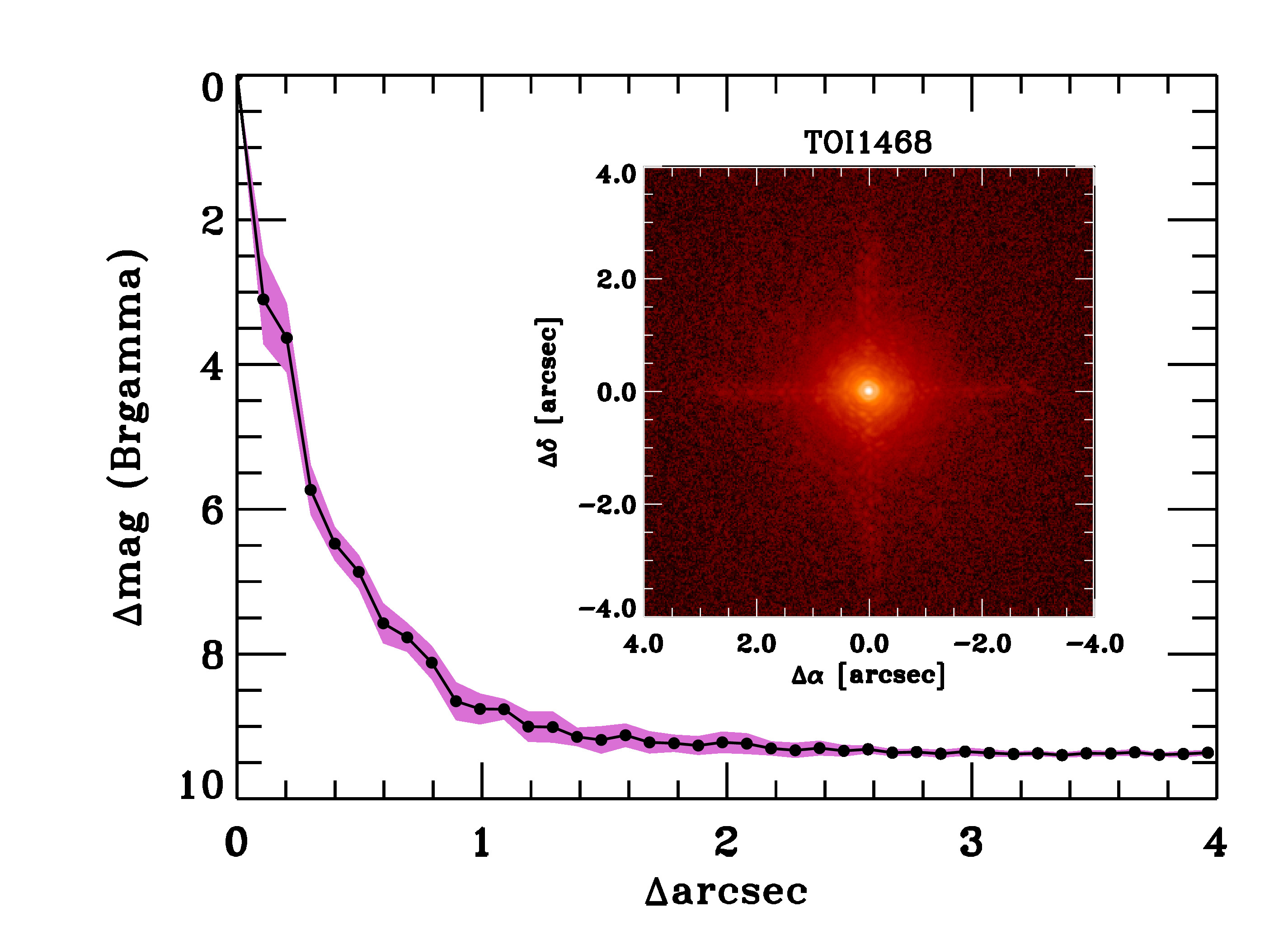}\label{Fig:AO_Palomar}
    \caption{Palomar PHARO Br$\gamma$ image of TOI-1468 and contrast curve ($5 \sigma$ limits in black dots, rms dispersion in magenta).}
    \label{fig:imaging}
\end{figure}    
    
Deeper high-resolution imaging observations of TOI-1468 were made at the 200-inch Hale telescope of the Palomar Observatory.
On 08~Aug~2021, we used the PHARO instrument \citep{hayward2001} behind the natural guide star adaptive optics system P3K \citep{dekany2013} in a standard five-point quincunx dither pattern with steps of 5\,arcsec in the narrow-band Br$\gamma$ filter 
($\lambda_0$ = 2168.6\,nm, $\Delta\lambda$ = 32.6\,nm).  
Each dither position was observed three times, offset in position from each other by 0.5\,arcsec for a total of 15 frames, with an integration time of 9.9\,s per frame, for a total on-source time of 148\,s. 
The images were taken in good seeing conditions.
PHARO has a pixel scale of 0.025\,arcsec per pixel for a total field of view of $\sim$25\,arcsec.
    
The science frames were flat-fielded and sky-subtracted. The flat fields were generated from a median average of dark subtracted flats taken on-sky. The flats were normalized such that the median value of the flats was unity. The sky frames were generated from the median average of the 15 dithered science frames; each science image was then sky-subtracted and flat-fielded. The reduced science frames were combined into a single combined image using an intra-pixel interpolation that conserves flux, shifts the individual dithered frames by the appropriate fractional pixels, and median-co-adds the frames.  
The final resolutions of the combined dithers were determined from the full width at half maximum (FWHM) of the point spread function (PSF), namely 0.099\,arcsec.  
The sensitivities of the final combined adaptive optics image were determined by injecting simulated sources azimuthally around the primary target every 20\,deg at separations of integer multiples of the central source's FWHM \citep{furlan2017}. 
The brightness of each injected source was scaled until standard aperture photometry detected it with $5\sigma$ significance.
The resulting brightness of the injected sources relative to TOI~1468 set the contrast limits at that injection location. 
The final $5\sigma$ limit at each separation was determined from the average of all of the determined limits at that separation, and the uncertainty on the limit was set by the root-mean-square (rms) dispersion of the azimuthal slices at a given radial separation. 
The final sensitivity curve for the Palomar data is shown in Fig.~\ref{Fig:AO_Palomar}.
        
While the Gemini speckle observations provide high spatial resolution, the Palomar adaptive optics data provide greater sensitivity in the region of 0.5--1.0\,arcsec.  
No additional stellar companions were detected to a depth of $\Delta m \approx$ 7\,mag at 0.5\,arcsec and $\Delta m \approx$ 9\,mag at 1.0\,arcsec, indicating that no companions down to the approximately mid-T dwarf were detected \citep{kirkpatrick2019}.
    
\section{Host star properties} \label{sec:star}

\begin{table}
\centering
\tiny
\caption{Stellar parameters of TOI-1468.} \label{tab:stellar-param}
\begin{tabular}{lcr}
\hline\hline
\noalign{\smallskip}
Parameter   & Value             & Reference \\ 
\hline
\noalign{\smallskip}
\multicolumn{3}{c}{\em Name and identifiers}\\
\noalign{\smallskip}
Name    & LSPM J0106+1913     & Lep05 \\
Karmn   & J01066+192            & AF15 \\
TOI     & 1468                  & ExoFOP-TESS \\  
TIC     & 243185500             & Sta18 \\  
\noalign{\smallskip}
\multicolumn{3}{c}{\em Coordinates and basic photometry}\\
\noalign{\smallskip}
$\alpha$ (J2016.0) & 01:06:36.93  & {\em Gaia} EDR3  \\
$\delta$ (J2016.0) & +19:13:29.6  & {\em Gaia} EDR3 \\
$G$ [mag]        & $12.1047\pm0.0007$ & {\em Gaia} EDR3 \\
$T$ [mag]       & $10.886\pm0.008$ & Sta19\\
$J$ [mag]        & $9.343\pm0.021$ & Skr06 \\
\noalign{\smallskip}
\multicolumn{3}{c}{\em Parallax and kinematics}\\
\noalign{\smallskip}
$\pi$ [mas]        & $40.45\pm0.04$ & Gaia EDR3 \\
$d$ [pc]        & $24.72\pm0.02$ & Gaia EDR3 \\
$\mu_\alpha \cos \delta$ [mas a$^{-1}$] & $-42.07\pm0.05$ & Gaia EDR3 \\
$\mu_\delta$ [mas a$^{-1}$] & $-222.79\pm0.03$ & Gaia EDR3 \\
$\gamma$ [km\,s$^{-1}$] & $+11.58\pm0.07$ & Mar21 \\
$U$ [km\,s$^{-1}$] & $+8.21\pm0.03$ & This work \\
$V$ [km\,s$^{-1}$] & $-6.01\pm0.04$ & This work \\
$W$ [km\,s$^{-1}$] & $-27.14\pm0.05$ & This work \\
Galactic population & Young disk & Mar21\\
\noalign{\smallskip}
\multicolumn{3}{c}{\em Photospheric parameters and spectral type}\\
\noalign{\smallskip}
Sp. type         & M3.0\,V  & This work \\
$T_{\mathrm{eff}}$ [K]  & $ 3496\pm25$ & Mar21   \\
$\log g$                & $5.00\pm0.11$ & Mar21   \\
{[Fe/H]}                & $-0.04\pm0.07$ & Mar21   \\
$v \sin i$ [$\mathrm{km\,s^{-1}}$]    & $<2.0$ & This work \\
\noalign{\smallskip}
\multicolumn{3}{c}{\em Stellar properties}\\
\noalign{\smallskip}
$L_\star$ [$10^{-4}\,L_\odot$] & $159.5\pm0.9$      & This work \\
$M_\star$ [$M_{\odot}$]       & $0.339\pm0.011$ & This work \\
$R_\star$ [$R_{\odot}$]       & $0.344\pm0.005$ & This work \\
$P_{\rm rot;GP}$ [d] & 41--44 & This work$^a$ \\
pEW'(H$\alpha$) [\AA] & $-0.11\pm0.03$ & This work\\
$\log{ L_{\rm X} / L_{\rm bol}}$ & $< -3.60$ & This work \\
\noalign{\smallskip}
\hline
\end{tabular}
\tablebib{
    AF15: \citet{AlonsoFlorian2015};
    Cif20: \citet{Cifuentes2020};
   Gaia EDR3: \citet{GaiaEDR3};
    Lep05: \citet{Lepine2005};
    Mar21: \citet{Marfil2021};
    Skr06: \citet{2MASS};
    Sta18: \citet{Stassun2018};
    Sta19: \citet{Stassun2019}. 
}
   \tablefoot{
   \tablefoottext{$^a$}{See Sect.~\ref{subsec:Prot} for a $P_{\rm rot}$ determination.}
   }
\end{table}

Situated at a distance of about 24.7\,pc {\citep{GaiaEDR3}}, TOI-1468 is a relatively bright ($J$ = 9.34\,mag) M1.0\,V-type star \citep{Lepine2011} that has been poorly investigated in the literature.  
It was discovered in a proper-motion survey by \citet{Lepine2005}, who tabulated it as \object{LSPM~J0106+1913}.
Afterward, it appeared (with the LSPM designation) only in the catalogs of bright M dwarfs of \citet{Lepine2011}, \citet{Frith2013}, and \citet{Cifuentes2020}.

Table\,\ref{tab:stellar-param} summarizes the stellar parameters of TOI-1468 with their corresponding uncertainties and references.
We took the photospheric parameters $T_{\rm eff}$, $\log g$, and [Fe/H] from \citet{Marfil2021}, who employed a Bayesian spectral synthesis implementation particularly designed to infer the stellar atmospheric parameters of late-type stars with a high S/N, high spectral resolution, co-added CARMENES VIS and NIR spectra of TOI-1468.
The bolometric luminosity was computed from the integration of the spectral energy distribution from the blue optical to the mid-infrared as in \citet{Cifuentes2020}, but with the latest {\em Gaia} EDR3 values of parallax and magnitudes.
A compilation of multiband photometry of TOI-1468 from $u'$ to $W4$ was also provided by \citet{Cifuentes2020}.  
After we obtained $T_{\rm eff}$ and $L_{\star}$, we derived the stellar  radius $R_{\star}$ by means of the Stefan-Boltzmann law, and finally  determined the stellar mass $M_{\star}$ using the mass-radius relation from \citet{Schweitzer2019}. 

\citet{Lepine2011} estimated an M1\,V spectral type from the $V-J$ color.
However, on the one hand, they used $V$ magnitudes estimated from photographic $B_J$, $R_F$, and $I_N$ magnitudes.
On the other hand, the $V$ band has some disadvantages in the M-dwarf domain according to \citet{Cifuentes2020}.
As a result, we estimated our own spectral type from the color- and absolute-magnitude spectral type relations of the latter authors.
Our spectral type, M3.0\,V, with about half a subtype uncertainty, better matches the measured $T_{\rm eff}$ and, especially, the $L_{\star}$ of TOI-1468 than the estimation by \citet{Lepine2011}.

TOI-1468 was not detected in the \textit{ROSAT} All-Sky Survey (RASS) and we estimated an upper limit of $L_{\rm X}\approx 1.5 \cdot 10^{28}$\,erg\,s$^{-1}$ using the characteristic
limiting RASS X-ray flux of $2 \cdot 10^{-13}$\,erg\,cm$^{-2}$\,s$^{-1}$ \citep{Schmitt1995},
resulting in an upper limit of $L_{\rm X}/L_{\rm bol}\approx 2.5 \cdot 10^{-4}$.
TOI-1468 was not detected in $FUV$ or in $NUV$ by {\em GALEX} \citep[cf.][]{Cifuentes2020}.
This lack of ultraviolet and X-ray emission, in spite of its closeness, is consistent with very weak activity.
In fact, all of the individual CARMENES spectra, with the exception of one, show a normalized H$\alpha$ pseudo-continuum, pEW'(H$\alpha$), as defined by \citet{Schoefer2019}, greater than --0.3\,{\AA} (negative values are in emission).
The outlier spectrum has a pEW'(H$\alpha$) just slightly above the activity boundary.

We looked for wide companions with {\em Gaia} EDR3 at projected physical separations up to 100\,000\,au and did not find any object with similar parallaxes and proper motions with the criteria of \citet{Montes2018}.
TOI-1468 appears single not only with adaptive optics, but also at larger separations. 
Based on the kinematic space velocities, the star belongs to the young disk population \citep{Marfil2021}, but this is at odds with its weak stellar activity.
As a result, the age of TOI-1468 is rather unconstrained (i.e., 1--10\,Ga).
Finally, the rotation period is determined to be 41--44\,d (Sect.~\ref{subsec:Prot}).

\section{Analysis and results} \label{sec:results}

\subsection{Rotation period of the host star}\label{subsec:Prot}

\subsubsection{Radial velocity data analysis}\label{subsubsec:periodogram}

\begin{figure*}
    \centering
    \includegraphics[width=\textwidth]{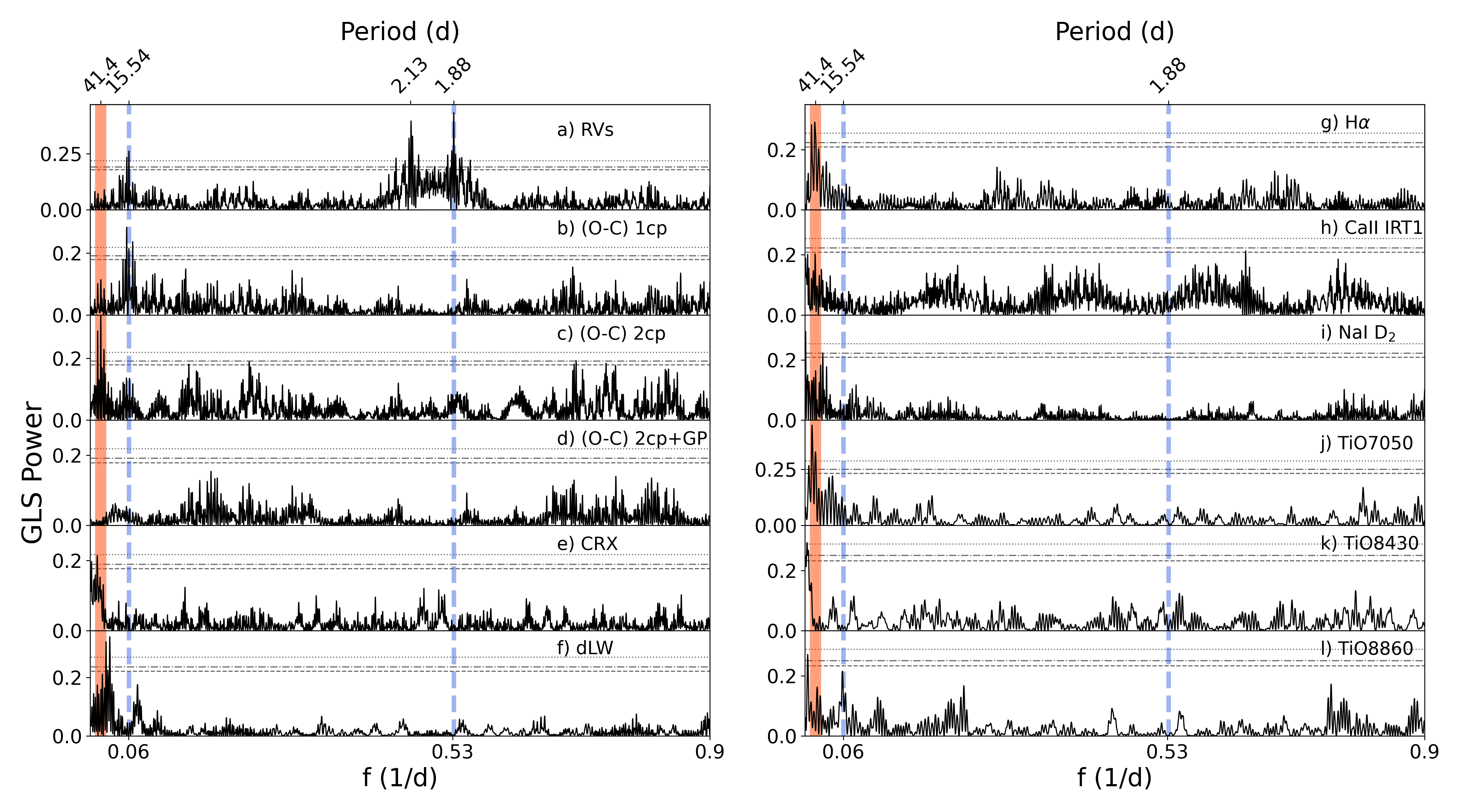}
    \caption{GLS periodograms of:
    ($a$) RV measurements from CARMENES VIS and MAROON-X data;
    ($b$) RV residuals after subtracting the inner planet signal at 1.88\,d;
    ($c$) RV residuals after subtracting the two planet signals at 1.88 and 15.53\,d;    ($d$) RV residuals after subtracting the two planet signals plus a GP at $\sim$41\,d;    ($e$-$g$) CRX, dLW, H$\alpha$, Ca~{\sc ii}~IRT$_1$, and Na~{\sc i}~D$_2$ activity indices from the combined CARMENES VIS and MAROON-X data; 
    ($j$-$l$) TiO7050, TiO8430, and TiO8860 activity indices from CARMENES VIS only.
    The ``cp'' in the residual models (panels $a$--$d$) corresponds to planets with circular orbits (see Sect.\,\ref{appsubsec:rv-juliet} for a detailed explanation).
    In all panels, the vertical dashed blue lines correspond to the periods of the inner and outer planets at 1.88\,d and 15.53\,d, respectively. 
    The orange shaded region corresponds to the stellar rotational period seen between 41\,d and 44\,d.
    The horizontal gray lines mark the theoretical FAP levels of 1\,\% (dotted), 5\,\% (dash-dotted), and 10\,\% (dashed).
    }    
    \label{fig:spectral_indices}
\end{figure*}

We performed a generalized Lomb-Scargle (GLS) periodogram \citep{Zechmeister2009} analysis on the CARMENES and MAROON-X data. 
The data sets included RV measurements from both instruments, and CARMENES photospheric and chromospheric activity indicators provided by \texttt{serval}, namely the CRX, dLW, H$\alpha$, Ca~{\sc ii}~IRT, Na~{\sc i}~D, and TiO7050, TiO8430, and TiO8430 indices.
As a first step, we searched for periodic signals in the RV data. 
The analysis was done in a sequential pre-whitening procedure where we computed the periodogram, removed the dominant signal, and searched for periodic signals in the residuals. 
This process is illustrated by panels b--d in Fig.\,\ref{fig:spectral_indices}. 
The first two signals seen in the RVs (panel a) correspond to the two transiting planets at 1.88\,d and 15.53\,d (an alias of the 1.88\,d is also visible at 2.13\,d). 
After subtracting these two signals from the data (see Sect.\,\ref{appsubsec:rv-juliet} for details), a signal at $\sim$41\,d showed up (panel c).

Stellar activity can induce RV variations that can influence the RV amplitude of planets, or even mimic a planetary signal \citep[see, e.g.,][, and references therein]{ Oshagh2017,Cale2021,Kossakowski2022}. 
We investigated the impact of stellar activity by performing two different analyses. 
The first of them was computing if there are statistically significant  correlations of the activity indicators with RV, and the second was by performing a periodogram analysis of activity indicators that may reveal periodic signals due to activity. 
For the first analysis, we used the Pearson $r$ coefficient on which we defined a strong correlation (or anticorrelation) if $r>0.7$ (or $r<0.7)$ \citep[][]{Jeffers2020}. 
For this analysis, we did not find any strong or moderate correlation between the RVs and any of the activity indicators.
For our second analysis, the investigation of periodicities in the activity indices in Fig.\,\ref{fig:spectral_indices} (panels e--l) revealed that some of them, such as CRX, H$\alpha$, and TiO7050, have a forest of significant signals around the 41--44\,d period, while others, such as TiO8860, have some peaks around 21\,d (related to the first harmonic of the 41--44\,d signal; \citealt{Schoefer2022}).
The activity indices and their uncertainties are listed in Table\,\ref{tab:RV_table}.

\begin{figure*}[!htp]
\centering
\includegraphics[width=0.33\textwidth]{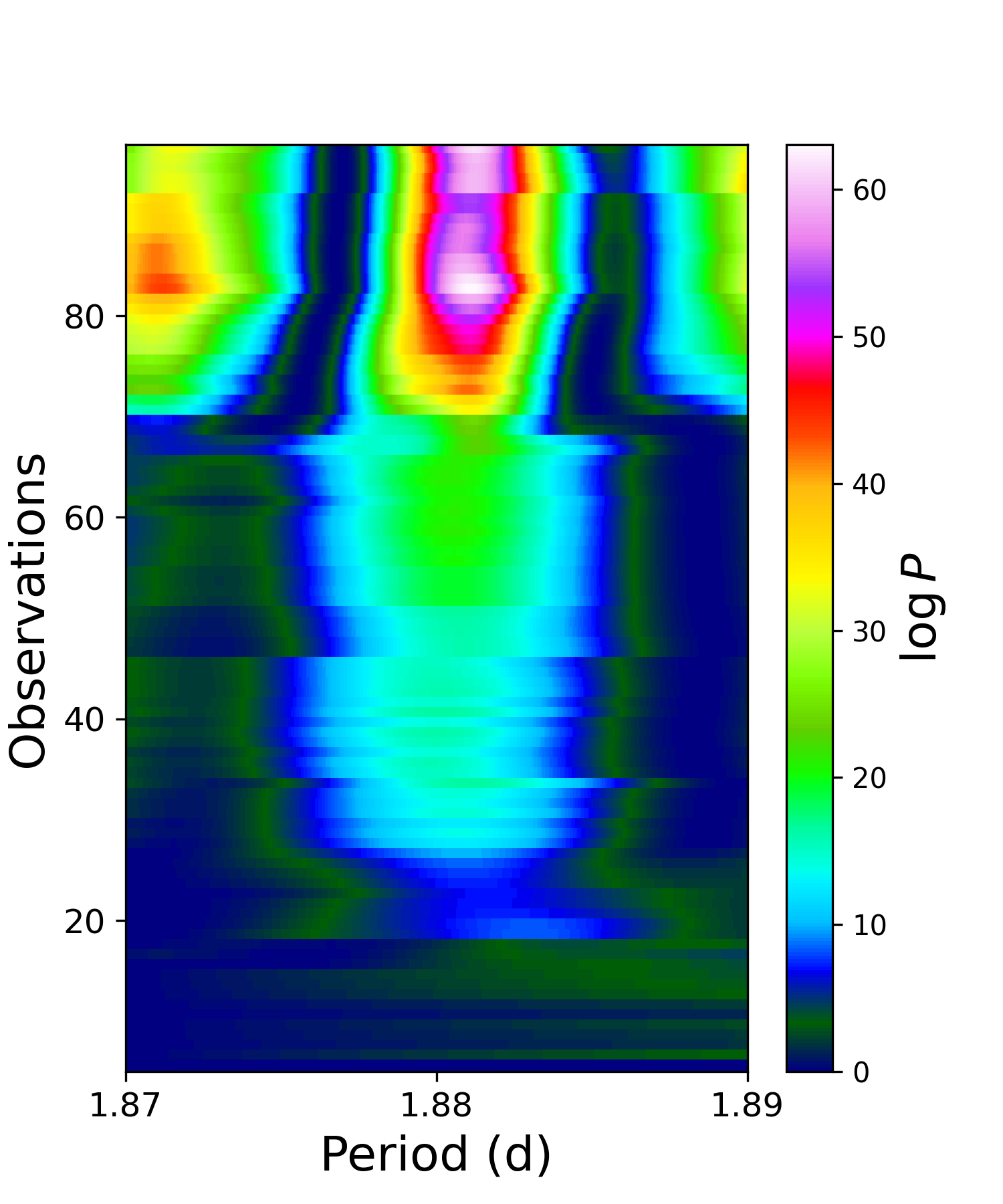}
\includegraphics[width=0.33\textwidth]{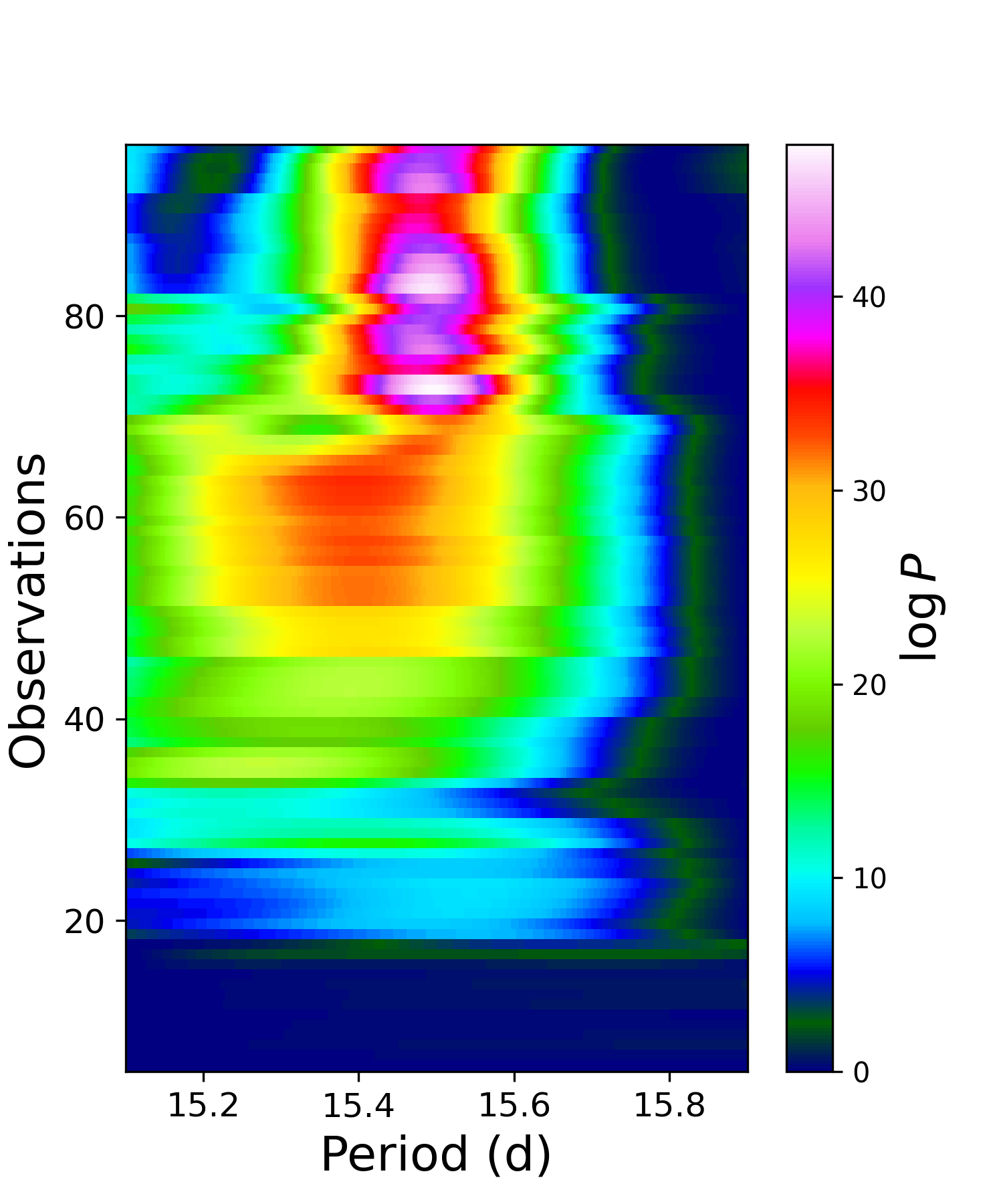}
\includegraphics[width=0.33\textwidth]{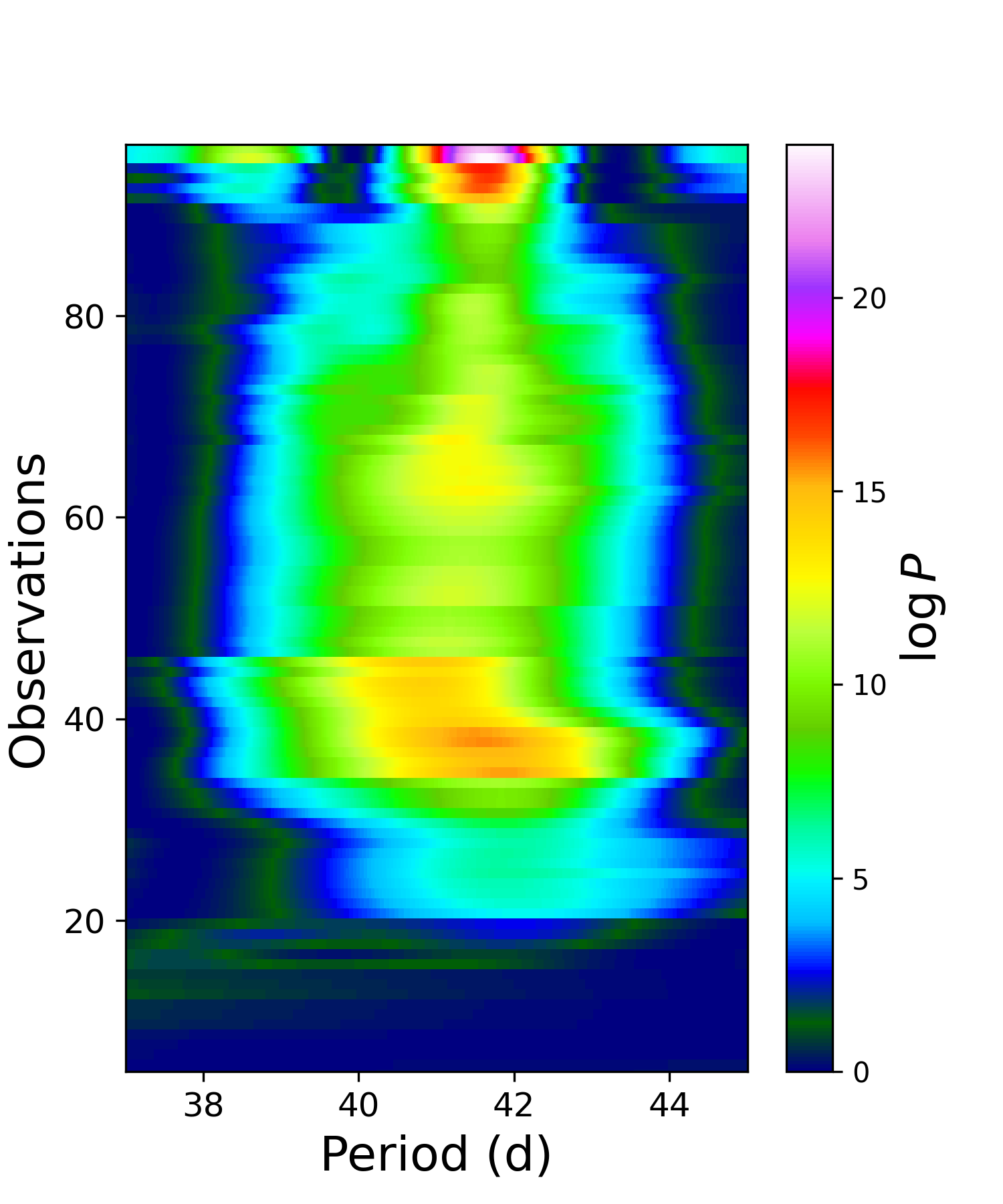}
\caption{Evolution of the s-BGLS periodogram for the CARMENES VIS plus MAROON-X RV data of TOI-1468 around the inner planet signal of 1.88\,d ({\em left}), the outer planet signal of 15.53\,d ({\em middle}), and the stellar rotation signal of $\sim$41\,d after removal of both planetary signals ({\em right}).} 
\label{fig:sBGLS_TOI1468}
\end{figure*}

Based on the upper limit of the projected rotational velocity and the radius of the star, we estimated a lower limit for the rotation period of roughly 9\,d, assuming null stellar obliquity.
To determine the actual rotational period of the star, we investigated the evolution of the 1.88\,d and 15.53\,d signals in the combined RV data set from CARMENES and MAROON-X. We plot the stacked Bayesian generalized Lomb-Scargle periodogram \citep[s-BGLS;][]{BGLS2015} with the normalization of \cite{sBGLS2017} in Fig.\,\ref{fig:sBGLS_TOI1468}. 
In this diagram, the RVs are plotted against their frequency axes centered around the inner planet signal of 1.88\,d (left) and the outer planet signal of 15.53\,d (middle). 
The planetary signals are subsequently removed from the RVs and the residuals are plotted centered around the third prominent signal seen in the RVs, (i.e., around 41\,d, right). 
From Fig.\,\ref{fig:sBGLS_TOI1468},
the s-BGLS of RVs for the 1.88\,d and 15.53\,d signals monotonically increase, which indicates the stability of the signal and provides further evidence of the planetary nature. 
However, the $\sim$41\,d signal does not show a monotonic behavior with time. First, the power of this signal tends to increase up to 46 observations, then the power decreases until 91 observations, and then drastically increases again.
This incoherence is characteristic for a non-planetary origin of the signal, and is supported by the evidence from several of the CARMENES activity indicators. Therefore, we attributed this signal to the rotation period of the star.

\begin{table*}
    \centering
    \small
    \caption{
    Data from public ground-based surveys used in this work.} \label{tab:phot-archive}
    \begin{tabular}{llllcccccc}
        \hline\hline
        \noalign{\smallskip}
Survey & Band & Start date & End date & $N$ & $\Delta t$ & $\overline{m}$ & $\sigma_m$ & $\overline{\delta {m}}$ \\
 & & & & & (d) & (mag) & (mag) & (mag) \\ 
        \noalign{\smallskip}
        \hline
        \noalign{\smallskip}

    NSVS & Clear & 11 July 1999 & 03 February 2000 & 225 & 206 & 12.015 & 0.028 & 0.018 \\
    SuperWASP & $V$ & 20 June 2004 & 04 January 2014 & 34\,109 & 3485 & ... & 0.026 & 0.024 \\
    OSN & $V,R$ & 01 September 2020 & 12 January 2021 & 2062 & 134 & ... & 0.003 & 0.003 \\
    TJO & $R$ & 21 August 2020 & 17 January 2021 & 327 & 150 & ... & 0.009 & 0.001 \\
    \noalign{\smallskip}         
        \hline
    \end{tabular}
   \tablefoot{ $\Delta t$ is the time span of the observations, $\overline{m}$ is the average magnitude, $\sigma_m$ is the standard deviation of the observed magnitudes, and $\overline{\delta {m}}$ is the average error bar associated with each observation.
   If not indicated, $\overline{m} = 0$ by construction.
   }
\end{table*}

\begin{figure}[] 
    \centering
    \includegraphics[width=\hsize]{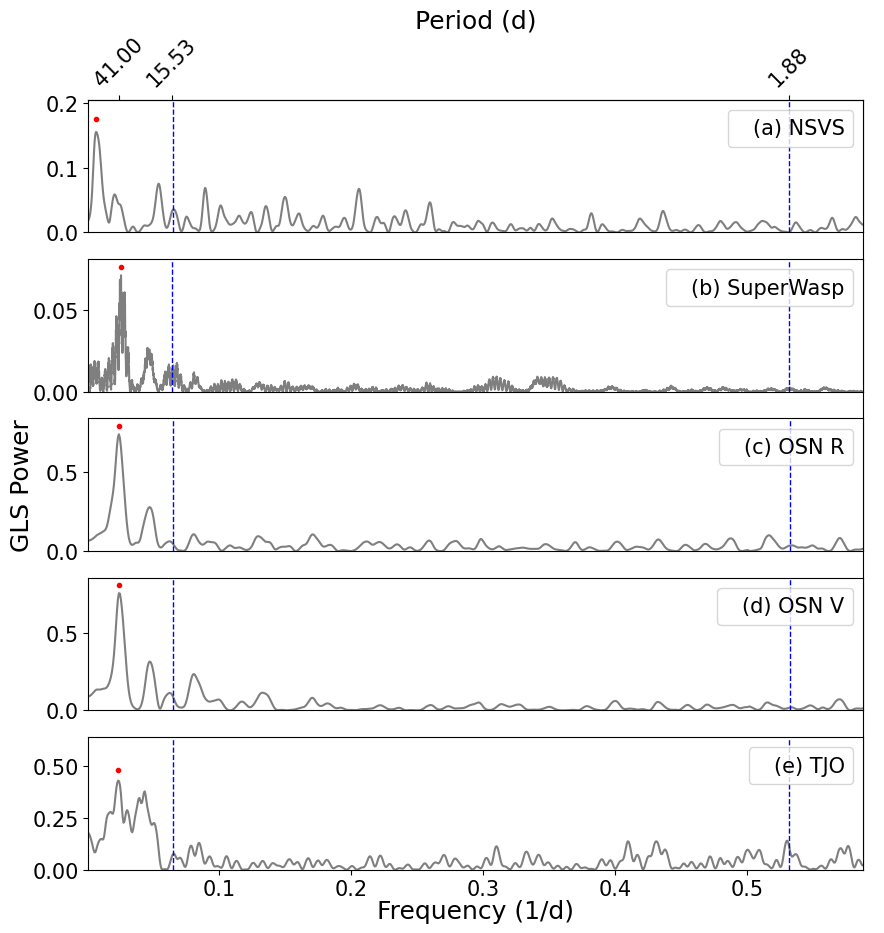}
    \caption{GLS periodograms of the long-term photometry data from NSVS, SuperWASP, OSN\,$R$, OSN\,$V$, and TJO ($a$--$e$, from top to bottom).
    Red dots indicate the most significant periods seen in each data set, and vertical dashed blue lines indicate the planet orbital periods of 1.88\,d and 15.53\,d.}
    \label{fig:gls_phot}
\end{figure}

\subsubsection{Long-term photometry}

To detect periodically modulated signals attributed to rotating surface manifestations of stellar magnetic activity such as dark spots and bright faculae, we examined archival  time-series photometry from the All-Sky Automated Survey for Supernovae \citep[ASAS-SN;][]{Shapee2014}, Northern Sky Variability Survey \citep[NSVS;][]{Wozniak2004}, Catalina Sky Survey \citep{Drake2009}, and Wide Angle Search for Planets \citep[WASP;][]{Butters2010}, in a similar fashion as \citet{DiezAlonso2019}.
In addition, we carried out follow-up photometry with the T150 telescope located at the Observatorio de Sierra Nevada (OSN) in Granada, and the
Telescopi Joan Or\'o (TJO) at the Observatori Astron\`omic del Montsec in Lleida, both in Spain. 
Since the data quality for the ASAS-SN and the Catalina survey was poor, we did not make use of these observations in our analyses. 
The instrumental setups, as well as the compiled data sets, are described below. 
We present the observation log in Table\,\ref{tab:phot-archive}. 

NSVS monitored approximately 14 million objects primarily in the northern hemisphere with $V$ magnitudes ranging from 8.0\,mag to 15.5\,mag. 
The main objective was a prompt response to gamma ray burst triggers from satellites to measure the early light curves of their optical counterparts. The robotic telescope array located in Los Alamos, NM, USA, consisted of four unfiltered telephoto lenses and covered a total field of view of $16 \times 16$\,deg$^2$. 
The photometric data of TOI-1468 were collected between July 1999 and February 2000, and encompass 267 measurements. 
Details on the basic characteristics and the reduction of the data set were provided by \citet{Wozniak2004}. 
    
SuperWASP-North is located in La Palma, Spain, and continuously monitors the sky for planetary transit events \citep{Butters2010}. 
It consists of eight lenses with a $2048\times2048$ CCD with pixel sizes of $13.5\,\mu$m, resulting in a field of view of $7.8 \times 7.8$\,deg$^2$ per camera. 
The observations were conducted with a broadband filter with a passband from  400\,nm to 700\,nm. 
The data set for TOI-1468 used in this work were provided by the SuperWASP consortium via the NASA Exoplanet Archive\footnote{\url{https://exoplanetarchive.ipac.caltech.edu/docs/SuperWASPMission.html}} and consists of 34\,109 measurements with a baseline of ten years. 

T150 is a 150-centimeter Ritchey-Chr\'etien telescope equipped with a 2k\,$\times$\,2k Andor Ikon-L DZ936N-BEX2-DD CCD camera with a field of view of 7.9$\,\times\,$7.9\,arcmin$^2$ \citep{Quirrenbach2022}.
The photometric observations were carried out in Johnson $V$ and $R$ filters, covering 52 epochs between September 2020 and January 2021, with typical exposure times of 70\,s in $V$ and 40\,s in $R$.
All CCD measurements were obtained by the method of synthetic aperture photometry using a 2\,$\times$\,2 binning. 
Each CCD frame was corrected in a standard way for bias and flat fielding. 
Different aperture sizes were tested to find the optimal one for our observations. 
After removing $3 \sigma$ outliers due to bad weather conditions, the rms on each night was about 3.0\,mmag and 2.5\,mmag in $V$ and $R$ bands, respectively. 

TJO is a 0.8-meter robotic telescope equipped with the 4k\,$\times$\,4k back-illuminated CCD camera LAIA, which has a pixel scale of 0.4\,arcsec and a squared field of view of 30\,arcmin$^2$. 
Several blocks of five images were collected between August 2020 and January 2021 over the course of 150 nights using the Johnson $R$ filter. 
The images were calibrated with bias, darks, and flat fields with the {\tt ICAT} pipeline \citep{Colome2006}. 
Differential photometry was extracted with \texttt{AIJ} \citep{Collins:2017}, with the aperture size and the set of comparison stars that minimized the rms of the photometry.

Figure\,\ref{fig:gls_phot} shows the most significant signal of the GLS periodograms of the long-term photometry. 
Almost all data sets show pronounced peaks between 38\,d and 41\,d, as well as a strong secondary signal at half this range, at $\sim$21\,d,  which would correspond to the first harmonic. 
However, other secondary peaks at $\sim20$\,d and $\sim45$\,d also are also present. 
These periods may be associated with the rotation period of the star, since for old M dwarfs these values are typically in the range of 10 to 150 d \citep{Jeffers18}.
The only exception is the NSVS light curve, which also shows a dominant peak at 147\,d (not shown in the diagram), longer than half its time baseline. 
However, these data are much noisier and shorter than the others, and since the rest of the photometry data and spectroscopic activity indicators share a common periodicity of about $\sim$41\,d, we question the reliability of this peak.

Next, we applied a more sophisticated approach and modeled the light curves with a Gaussian process (GP). 
We used the fitting tool \texttt{juliet} \citep{juliet}, which incorporates the Python library \texttt{george} \citep{Ambikasaran2015} for the in-built kernels. 
For our purpose, we selected the quasi-periodic (QP) kernel, which is an exponential-sine-squared kernel multiplied by a squared-exponential kernel, which allows complex periodic signals to be modeled.
This kernel is suitable for accounting for the effects of active regions present on the surface of stars, which often mimic
a sinusoidal-like signal \citep{Angus2018}.
It has the form:

\begin{equation}\label{eq:QP-GP}
k(\tau)\,=\,\sigma^{2}_{\rm GP}\,\exp\left(-\alpha_{\rm GP}\,\tau^2 -\Gamma \sin^2\left[\frac{\pi \tau}{P_{\rm rot;GP}}\right] \right)~,~
\end{equation}

\noindent where $\sigma_{\rm GP}$ is the GP amplitude (in parts per million, ppm), $\Gamma$ is the dimensionless amplitude of the GP sine-squared component, $\alpha$ is the inverse length scale of the GP exponential component (d$^{-2}$), $\tau$ is the time lag (d), and $P_{\rm rot;GP}$ is the rotational period of the star (d). 

\begin{figure}[]
    \centering
  \includegraphics[width=0.45\textwidth]{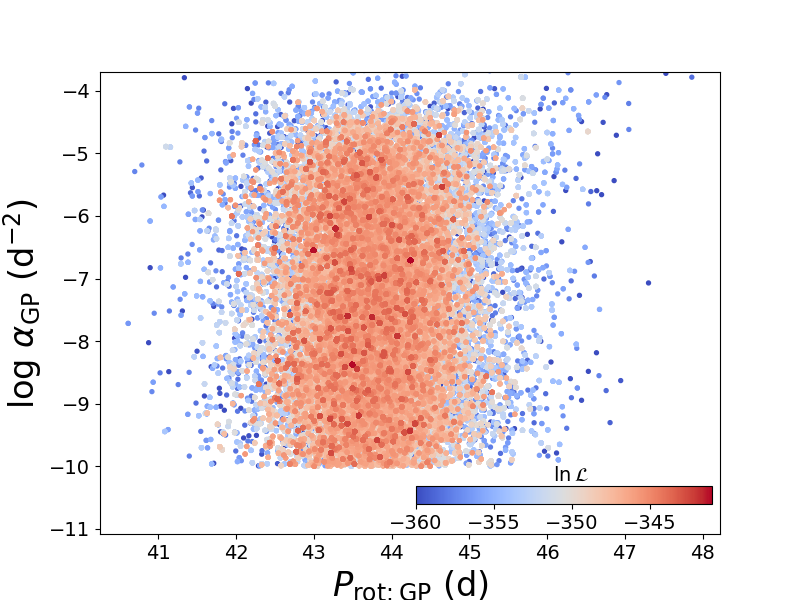}
    \caption{Posterior distribution in the $\alpha_{\rm{GP}}$ vs. $P_{\rm rot;GP}$  plane for the joint GP modeling of NSVS, SuperWASP, OSN\,$R$, OSN\,$V$, and TJO photometry data. The normalized log-likelihood increases from blue to red.}
    \label{fig:GP-Prot}
\end{figure} 

All the data sets displayed in Fig.\,\ref{fig:gls_phot} were used for the GP modeling. 
Considering that the photometry data were collected by different instruments and with different filters, we let the values of $\sigma_{\rm{GP}}$ and $\Gamma$ be variable for each data set and kept the $\alpha$ and $P_{\rm rot;GP,Phot}$ as common parameters. 
As justified by \cite{Stock2020}, wide uninformative priors were used for all parameters: $\sigma_{\rm{GP}}$ (Jeffreys distribution between $10^{-2}$\,ppm and $10^2$\,ppm), $\Gamma$ (Jeffreys distribution between $10^{-6}$ and $10$), instrumental jitter (Jeffreys distribution between 0.01\,ppm and $100$\,ppm), $\alpha$ (uniform between $10^{-10}$\,d$^{-2}$ and $10^{-2}$\,d$^{-2}$), and $P_{\rm rot;GP}$ (uniform between 30\,d and 50\,d). 
The relative offset between fluxes of different instruments was chosen to have a normal distribution between 0 and $1000$.

We determined the rotational period from the GP analysis as $P_{\rm rot;GP}$\,=\,$43.8\pm0.7$\,d. 
We plot the $\alpha_{\rm GP}$ vs $P_{\rm rot;GP}$ diagram in Fig.\,\ref{fig:GP-Prot}, similar to previously discussed by \cite{Stock2020}, \cite{Bluhm2021}, and \cite{Kossakowski2021}. 
This diagram gives an idea if a strong correlated noise (small $\alpha$) would favor any periodicity. 
As seen in Fig.\,\ref{fig:GP-Prot}, a peak is centered around 44\,d with log\,$\alpha$ values spread between $-4$ to $-10$, which are indicative of the fact that the GP is modeling a periodic signal in the entire $\alpha$ range.

Based on our analysis of the $P_{\rm rot;GP}$ derived by the GP, the photometric GLS periodogram, and spectroscopic activity indicators, we conclude that the rotational period of the star should be around 41--44\,d, indicating that TOI-1468 is a slow rotator. Such a long rotation period is consistent with the object being older than the Praesepe open cluster \citep{Curtis2019}, whose age ranges from 590\,Ma to 900\,Ma \citep{Delorme2011, Lodieu2019}. 

\subsection{Orbital fits of the TOI-1468 planets}
\label{subsec:orbit-model}

To determine the orbital elements of the TOI-1468 system, we used the Python-based library \texttt{juliet} and modeled the transit data, the RV data, as well as both data sets in a joint manner. The \texttt{juliet} library makes use of other Python packages: \texttt{radvel} \citep{Fulton2018} for RV modeling, and \texttt{batman} \citep{Kreidberg2015} for transit modeling.  Based on the initial supplied prior inputs, \texttt{juliet} uses dynamic nested sampling from \texttt{dynesty} \citep{Speagle2020} to compute the Bayesian model log evidence, $\ln \mathcal{Z}$, along with posterior samples. There is a provision to model GPs that are implemented through \texttt{george} \citep{Ambikasaran2015} and \texttt{celerite} \citep{Foreman-Mackey2017}.

    \begin{table}
    \centering
    \tiny
    \caption{Posterior parameters of the 
    joint fit for TOI-1468\,b and\,c.}
    \label{tab:posteriors_planet}
    \begin{tabular}{lcc} 
        \hline
        \hline
        \noalign{\smallskip}
        Parameter$^{a}$ & TOI-1468\,b & TOI-1468\,c\\
        \noalign{\smallskip}
        \hline
        \noalign{\smallskip}
        $P$ (d) & $ 1.8805136^{+0.0000024}_{-0.0000026}$ & $15.532482^{+0.000034}_{-0.000033}$\\[0.1 cm]
        $t_0$ (BJD)&$2458765.68079^{+0.00070}_{-0.00069}$ & $2458766.9269^{+0.0012}_{-0.0012}$\\[0.1 cm]
        $a/R_\star$ &$13.14^{+0.21}_{-0.24}$ &  $53.69^{+0.84}_{-0.97}$ \\[0.1 cm]
        $b\,=\,(a/R_\star)\cos i_{\rm p}$  & $0.350^{+0.062}_{-0.075}$ & $0.623^{+0.023}_{-0.024}$\\[0.1 cm]
        $i_{\rm p}$ (deg)   & $88.47^{+0.34}_{-0.29}$  & $89.335^{+0.032}_{-0.035}$\\[0.1 cm]
        $r_1$                                &$0.567^{+0.041}_{-0.050}$ & $0.749^{+0.015}_{-0.016}$ \\[0.1 cm]
        $r_2$                                & $0.0341^{+0.0009}_{-0.0009}$  & $0.055^{+0.0008}_{-0.0009}$\\[0.1 cm]
        $K$ ($\mathrm{m\,s^{-1}}$)           & $3.40^{+0.25}_{-0.24}$  & $3.48^{+0.34}_{-0.35}$\\[0.1 cm] 
        \noalign{\smallskip}
        \multicolumn{3}{c}{\it Derived physical parameters} \\
        \noalign{\smallskip}
        $M\,(M\oplus)$       &$3.21^{+0.24}_{-0.24}$  &$6.64^{+0.67}_{-0.68}$ \\[0.1 cm]  
        $R\,(R\oplus)$       &$1.280^{+0.038}_{-0.039}$  &$2.064^{+0.044}_{-0.044}$ \\[0.1 cm]
        $g\,(\rm{m\,s^{-2}})$      &$19.12^{+1.93}_{-1.76}$  &$15.26^{+1.68}_{-1.63}$ \\[0.1 cm]
        $S\,({S\oplus})$     &$36.0^{+1.6}_{-1.4}$  &$2.15^{+0.09}_{-0.09}$ \\[0.1 cm]
       $T_{\rm eq}\,\rm(K)^b$    &$682.2^{+7.4}_{-6.9}$  &$337.5^{+3.7}_{-3.4}$ \\[0.1 cm]
    \noalign{\smallskip}
        \hline
    \end{tabular}
    \tablefoot{
      \tablefoottext{a}{Parameters obtained with the posterior values from Table\,\ref{tab:posteriors}.
      Error bars denote the 68\,\% posterior credibility intervals.}
      \tablefoottext{b}{The equilibrium temperature was calculated assuming zero Bond albedo.}
      }    
\end{table}

    \begin{table}
    \centering
    \tiny
    \caption{Posterior distributions of the \texttt{juliet} joint fit for the instrumental and GP fit parameters obtained for the TOI-1468 system.}
    \label{tab:posteriors}
    \begin{tabular}{lc} 
        \hline
        \hline
        \noalign{\smallskip}
        Parameter$^{a}$ & TOI-1468\\
        \noalign{\smallskip}
        \hline
        \noalign{\smallskip}    
        \multicolumn{2}{c}{\it Stellar parameters} \\[0.1cm]
        \noalign{\smallskip}
        $\rho_\star$ ($\mathrm{g\,cm\,^{-3}}$)& $12.13^{+0.58}_{-0.65}$\\[0.1 cm]
        \noalign{\smallskip}
        \multicolumn{2}{c}{\it Photometry parameters} \\[0.1cm]
        \noalign{\smallskip}
        $M_{\mathrm{TESS\,17}}$ ($10^{-6}$)   & $-72^{+25}_{-27}$\\[0.1 cm]
        $\sigma_{\mathrm{TESS\,17}}$ (ppm)     & $0.29^{+6.42}_{-0.25}$\\[0.1 cm]
        $q_{1,\mathrm{TESS\,17}}$               & $0.70^{+0.19}_{-0.23}$\\[0.1 cm]
        $q_{2,\mathrm{TESS\,17}}$               & $0.57^{+0.26}_{-0.29}$\\[0.1 cm]
        $M_{\mathrm{TESS\,42}}$ ($10^{-6}$)   & $-64^{+15}_{-16}$\\[0.1 cm]
        $\sigma_{\mathrm{TESS\,42}}$ (ppm)     & $0.39^{+13.73}_{-0.38}$\\[0.1 cm]
        $q_{1,\mathrm{TESS\,42}}$               & $0.086^{+0.126}_{-0.060}$\\[0.1 cm]
        $q_{2,\mathrm{TESS\,42}}$               & $0.39^{+0.35}_{-0.25}$\\[0.1 cm]
        $M_{\mathrm{TESS\,43}}$ ($10^{-6}$)   & $-48^{+14}_{-15}$\\[0.1 cm]
        $\sigma_{\mathrm{TESS\,43}}$ (ppm)     & $0.27^{+10.63}_{-0.26}$\\[0.1 cm]
        $q_{1,\mathrm{TESS\,43}}$               & $0.31^{+0.26}_{-0.18}$\\[0.1 cm]
        $q_{2,\mathrm{TESS\,43}}$               & $0.29^{+0.27}_{-0.19}$\\[0.1 cm]   
        $M_{\mathrm{LCO-SAAO}}$ ($10^{-6}$)      & $-74^{+67}_{-67}$\\[0.1 cm]
        $\sigma_{\mathrm{LCO-SAAO}}$ (ppm)        & $1195^{+ 63}_{- 60}$\\[0.1 cm] 
        $q_{1,\mathrm{LCO-SAAO}}$                 & $0.68^{+0.20}_{-0.30}$\\[0.1 cm]
        $M_{\mathrm{LCO-SSO}}$ ($10^{-6}$)      & $-7^{+76}_{-80}$\\[0.1 cm]
        $\sigma_{\mathrm{LCO-SSO}}$ (ppm)        & $1873^{+64}_{-65}$\\[0.1 cm]
        $q_{1,\mathrm{LCO-SSO}}$                 & $0.72^{+0.16}_{-0.22}$\\[0.1 cm]
        $M_{\mathrm{LCO-MCD}}$ ($10^{-6}$)      & $-42^{+50}_{-51}$\\[0.1 cm]
        $\sigma_{\mathrm{LCO-MCD}}$ (ppm)        & $842^{+47}_{-46}$\\[0.1 cm]
        $q_{1,\mathrm{LCO-MCD}}$                 & $0.58^{+0.17}_{-0.19}$\\[0.1 cm]
        $M_{\mathrm{MUSCAT2}}$ ($10^{-6}$)   & $-610^{+ 120}_{-120}$\\[0.1 cm]
        $\sigma_{\mathrm{MUSCAT2}}$ (ppm)     & $0.57^{+19.58}_{-0.56}$\\[0.1 cm]
        $q_{1,\mathrm{MUSCAT2}}$               & $0.40^{+0.28}_{-0.25}$\\[0.1 cm]
        $M_{\mathrm{KUIPER}}$ ($10^{-6}$)   & $-670^{+190}_{-190}$\\[0.1 cm]
        $\sigma_{\mathrm{KUIPER}}$ (ppm)     & $1.9^{+187.0}_{-1.9}$\\[0.1 cm]
        $q_{1,\mathrm{KUIPER}}$               & $0.74^{+0.17}_{-0.27}$\\[0.1 cm]
        \noalign{\smallskip}
        \multicolumn{2}{c}{\it RV parameters}\\[0.1cm]
        \noalign{\smallskip}
        $\mu_{\mathrm{CARMENES}}$ ($\mathrm{m\,s^{-1}}$)       & $-0.14^{+4.14}_{-4.31}$\\[0.1 cm]
        $\sigma_{\mathrm{CARMENES}}$ ($\mathrm{m\,s^{-1}}$)    & $1.553^{+0.384}_{-0.399}$\\[0.1cm]
        $\mu_{\mathrm{MAROON-X,Blue,1}}$ ($\mathrm{m\,s^{-1}}$)       & $-0.54^{+4.14}_{-4.39}$\\[0.1 cm]
        $\sigma_{\mathrm{MAROON-X,Blue,1}}$ ($\mathrm{m\,s^{-1}}$)    & $0.425^{+1.102}_{-0.379}$\\[0.1cm]
        $\mu_{\mathrm{MAROON-X,Blue,2}}$ ($\mathrm{m\,s^{-1}}$)       & $1.93^{+ 4.12}_{-4.46}$\\[0.1 cm]
        $\sigma_{\mathrm{MAROON-X,Blue,2}}$ ($\mathrm{m\,s^{-1}}$)    & $0.210^{+1.156}_{-0.178}$\\[0.1cm]
        $\mu_{\mathrm{MAROON-X,Red,1}}$ ($\mathrm{m\,s^{-1}}$)       & $-0.88^{+4.21}_{-4.37}$\\[0.1 cm]
        $\sigma_{\mathrm{MAROON-X,Red,1}}$ ($\mathrm{m\,s^{-1}}$)    & $0.143^{+0.470}_{-0.115}$\\[0.1cm]
        $\mu_{\mathrm{MAROON-X,Red,2}}$ ($\mathrm{m\,s^{-1}}$)       & $1.01^{+4.22}_{-4.55}$\\[0.1 cm]
        $\sigma_{\mathrm{MAROON-X,Red,2}}$ ($\mathrm{m\,s^{-1}}$)    & $1.731^{+1.104}_{-0.850}$\\[0.1cm]
        \noalign{\smallskip}
        \multicolumn{2}{c}{\it GP hyperparameters} \\
        \noalign{\smallskip}
        $\sigma_\mathrm{GP,RV}$ ($\mathrm{m\,s^{-1}}$)              & $6.602^{+5.56}_{-2.89}$\\[0.1 cm]
        $\alpha_\mathrm{GP,RV}$ ($10^{-6}\,\mathrm{d^{-2}}$)        & $0.17^{+ 252}_{-0.16}$\\[0.1 cm]
        $\Gamma_\mathrm{GP,RV}$                           & $0.33^{+0.76}_{-0.17}$\\[0.1 cm]
        $P_\mathrm{rot;GP,RV}$ (d)                              & $41.48^{+ 0.16}_{-0.17}$\\[0.1 cm]
        \noalign{\smallskip}
        \hline
    \end{tabular}    \tablefoot{
      \tablefoottext{a}{Priors and descriptions for each parameter are in Table\,\ref{tab:priors}. Error bars denote the 68\,\% posterior credibility intervals.}}
    \end{table}

\begin{figure*}
    \centering
    \includegraphics[width=\textwidth]{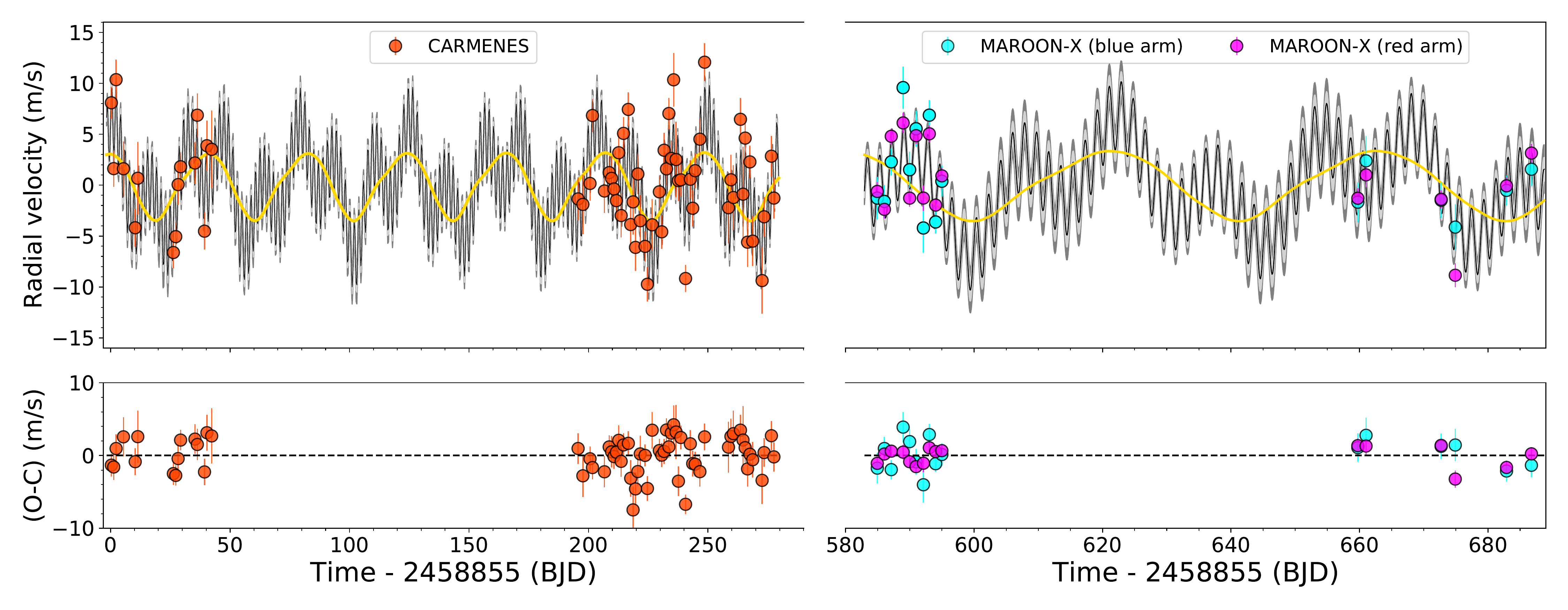}
    \includegraphics[width=\textwidth]{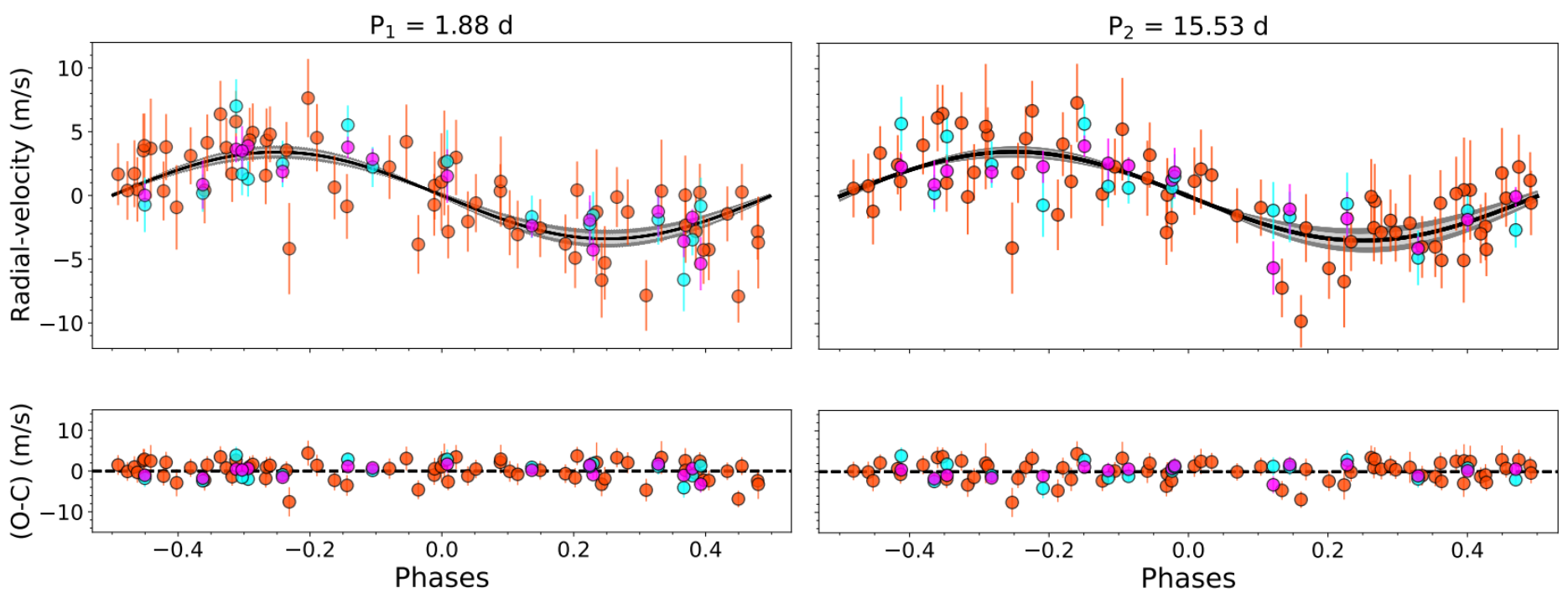}   
    \caption{Joint modeling of the RV data from CARMENES (VIS: orange), MAROON-X (Blue arm: cyan; Red arm: magenta) for TOI-1468, along with their residuals.
    In both panels, the solid curve is the median best-fit \texttt{juliet} model. 
    The light and dark gray shaded regions represent $68\,\%$ and $95\,\%$ credibility bands, respectively.
    {\em Top panel}:
    RV time-series with the GP component (solid yellow curve). 
    {\em Bottom panel}:
    Phase-folded RVs for TOI-1468 for the inner 1.88\,d-period planet and the outer 15.53\,d-period planet, along with their residuals folded at their respective periods.}
    \label{fig:RV-joint-fit}
\end{figure*}

Mutually independent parameters were constrained through transit-only and RV-only fits with \texttt{juliet} (see Appendix\,\ref{appA:data_modelling} for details). More precise values of $P$, $T_{c}$, $\omega$, and $e$ were obtained by performing a simultaneous fit to all parameters.
For the purpose of joint fitting, we used the RV points from CARMENES and MAROON-X, the light curves from {\em TESS}, and ground-based photometry. The best-fit results from the transit-only and the RV-only analyses were used as priors. The 2cp+GP model was chosen for modeling the RV points, as discussed in \ref{appsubsec:rv-juliet}. The complete list of priors used for the joint fit are described in Table\,\ref{tab:priors}. The RV semi-amplitude, $K$ value for the inner planet is $3.403^{+0.246}_{-0.244}$\,m\,$\rm s^{-1}$, and the $K$ value for the outer planet is $3.485^{+0.344}_{-0.351}$\,m\,$\rm s^{-1}$. The posterior planet parameters for the joint orbital fit are presented in Table\,\ref{tab:posteriors_planet} and in Table\,\ref{tab:posteriors}. The covariance plot for the fitted parameters is presented in Fig.\,\ref{Fig:corner_plot-2pl}.
However, uncertainties in planet mass and radius depend on the input uncertainties in stellar mass, radius, and equilibrium temperature, which in this case may be underestimated.
As a result, the actual planet densities of TOI-1468\,b and\,c may differ by more than $1 \sigma$ from the values derived by us, and a better characterization of the planet-host star would be desirable.
See \cite{Caballero2022} for an exhaustive analysis on sources of error and propagation of uncertainty of parameters of transiting planets with RV follow-up.

As described by the posterior parameters of our joint fit, and the resulting RV model presented in Fig.\,\ref{fig:RV-joint-fit}, the maximum posteriori of the GP periodic component, $P_\mathrm{rot,GP:RV}$, is about 41\,d, which is in agreement with the signal observed in the GLS periodogram of the RVs (Fig.\,\ref{fig:spectral_indices}) and  corresponds to the stellar rotation period. The best-fit results obtained from joint modeling are displayed in Figs.\,\ref{fig:phot-ground-1} and \ref{fig:phot-ground-2} for transits, and Fig.\,\ref{fig:RV-joint-fit} for RVs.

\section{Discussion} \label{sec:discussion}
\subsection{The radius valley}
\label{sec:mass_radius_discussion}

\begin{figure*}
\includegraphics[width=\textwidth]{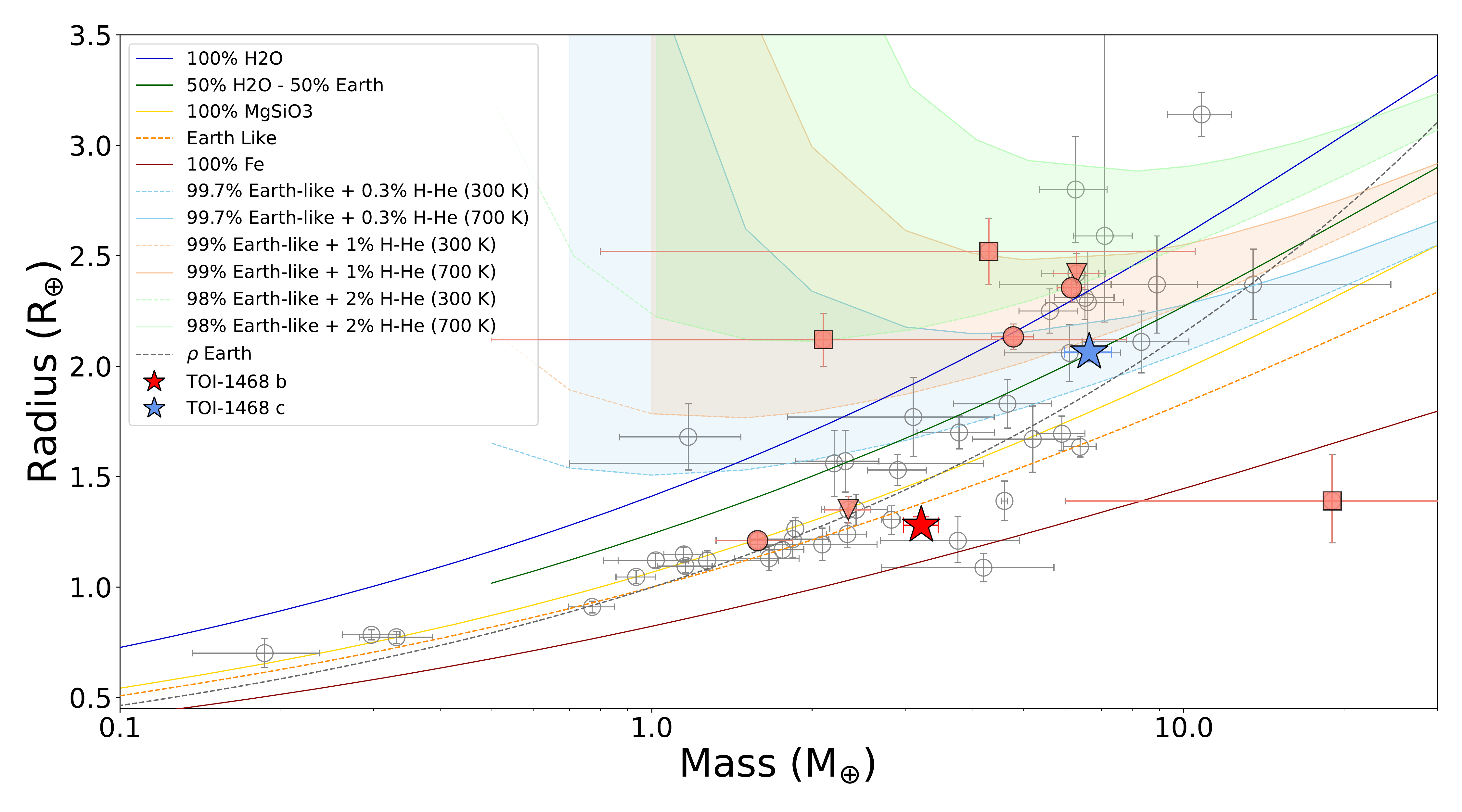}
\caption{Mass-radius diagram of well-characterized planets with radii $R < 3\,R_{\oplus}$ and masses $M < 10\,M_{\oplus}$. 
All the planets plotted in gray are planets transiting M dwarfs taken from \cite{Trifonov2021}, last updated on 08 April 2021, with $\Delta M < 20\,\%$ and $\Delta R < 20\,\%$. 
The ``$\rho$\,Earth'' is the constant Earth-density line. 
TOI-1468\,b and~c are marked with red and blue stars, respectively, and LTT3780\,b;c, L231-32\,b;c;d, and TOI1749\,b;c;d are marked with salmon filled circles, inverted triangles and squares, respectively. 
Theoretical mass-radius relations are taken from \cite{Zeng2019}.}
\label{fig:mr_diagram}
\end{figure*}

We plot all the planets transiting M dwarfs determined with a precision of better than $20\%$ for masses and radii in Fig.\,\ref{fig:mr_diagram}. We used the transiting M dwarfs as listed by \cite{Trifonov2021}, updated on 08 April 2021. The compositional models from \citep{Zeng2019} are also shown. TOI-1468\,b and c are marked with red and blue stars, respectively. The inner planet has a bulk density consistent with a composition ranging from $50\%$ silicates and $50\%$ iron, to $100\%$ silicates. The planet appears to have been irradiated, which is indicative of atmospheric losses. The mass and radius for TOI-1468\,c indicates that it must have a low-density envelope. As the losses depend on the amount of XUV radiation that the planet receives, the evolution of the two planets orbiting  the same star at different distances can be different. For example, the inner one may lose the H/He envelope, while the outer one keeps it. When XUV erosion is considered as a possible explanation for the radius valley, the valley must depend on the orbital separation of the host star and also its spectral type (FGK or M), with the same planetary composition. There have been a handful of discoveries for systems with multiple planets straddling the radius valley around different spectral type stars: K2-36 b c \citep{2019A&A...624A..38D}, K2-106\,b, c \citep{Guenther2017}, HD3167\,b, c \citep{2017AJ....154..123G}, GJ9827\,b, c, d \citep{2017AJ....154..266N}, or Kepler10\,b, c \citep{2014ApJ...789..154D}, to name a few. There are only a few such examples for planets on the opposite sides of the radius valley for transiting M dwarfs, such as LTT3780\,b, c {\citep{2020A&A...642A.173N,Cloutier2020},} L231-32\,b, c, d \citep{2021MNRAS.507.2154V}, and TOI1749\,b, c, d \citep{2021AJ....162..167F}. A common observation governing all the discoveries is the fact that, in most of these systems, the inner planet has a rocky Earth-like composition, and the outer planet or planets have solid cores with an outer envelope composed of lighter gases such as H and He. 

It was also demonstrated by \cite{vaneylen2018} that this radius valley narrows for smaller orbital periods. Both these observations are consistent with the photoevaporation model, although it cannot be excluded that it is due to formation. Core-powered mass loss \citep{Ginzburg+2018, gupta2020} has been suggested as an alternate hypothesis for the origin of the radius valley. There are also suggestions of different formation mechanisms for the planets on both sides of the radius valley, where one side of the valley consists of water-worlds, and the other consists of rocky and terrestrial planets \citep{Zeng2019}. It is imperative to find out if the radius valley is lower for M stars than for FGK stars, this system being an important contribution. This would imply a strong argument for the atmospheric erosion via XUV radiation. A core-powered model would be able to explain this, if it is assumed that the ratio of rocky to icy planets is different for M stars than for FGK stars. It is interesting to focus on multi-planet systems on two sides of the radius valley, which could be the key to answering similar questions. For example, the evolution of a planet orbiting a young active star should be different from a planet orbiting a young but inactive star. Measurements of the isotope ratios $\rm ^{36}Ar/^{38}Ar$, $\rm ^{20}Ne/^{22}Ne$, and $\rm ^{36}Ar/^{22}Ne$ on Earth and Venus, and the abundances of sodium and potassium of the lunar regolith both indicate that our Sun was only weakly active in the first 100\,Ma \citep{Lammer2019}. Thus, the evolution of the planets in our Solar-System could quite be different from those orbiting M stars that were very active when they were young and also stayed in this high-activity phase for a long time. For this reason, it is important to study the properties of low-mass planets orbiting particularly low-mass stars. 

\subsection{System architecture}

The scaled orbital separation ($a/R_{\star}$) for the inner planet, TOI-1468\,b, is $13.139^{+ 0.205}_{-0.238}$. The light curves from {\em TESS} and the several ground-based transit measurements taken for the inner planet results in
$R_{\rm b}$ as $1.280^{+0.038}_{-0.039}\,R_{\oplus}$. The mass for the planet, as derived by the RV measurements from CARMENES, is $M_{\rm b}$\,=\,$3.21\pm0.24\,M_\oplus$. This gives us the bulk planet density as $\rho_{\rm b}$\,=\,$8.39^{+ 1.05}_{- 0.92}$\,g\,cm$^{-3}$. The total amount of insolation received by the inner planet is 36 times that of Earth. Assuming a zero albedo and a uniform dayside temperature, the equilibrium temperature of TOI-1468\,b is $\sim$682\,K. 

Similarly, the scaled orbital separation for the outer planet, TOI-1468\,c, is $53.687^{+0.839}_{-0.975}$. The radius and mass for the planet are $2.06\pm 0.04$\,$R_\oplus$ and \,$6.64^{+0.67}_{-0.68}\,M_\oplus$, respectively. This results in a bulk planet density of $\rho_{\rm c}$\,=\,$2.00^{+ 0.21}_{- 0.19}$\,g cm$^{-3}$. The stellar insolation for the outer planet TOI-1468\,c is twice that of the Earth and, with a zero albedo, has an equilibrium temperature of $\sim$337\,K. Apparently TOI-1468\,c could be located close to the inner edge of of the habitable zone \citep{Kasting1993,Kasting1998,Kasting2010,Kasting2013,Kopparapu2014,Kasting2021}, and probably the actual temperature should be much higher than the equilibrium temperature, due to atmospheric heating effects. However, since the planet is most likely tidally locked, this does not exclude the possibility of surface liquid water \citep[e.g.,][]{Wandel2018,Martinez-Rodriguez2019}.

Systems similar to the TOI-1468 system are interesting from the point of view of planet formation: two planets that orbit the same host star on close-in orbits but have different densities. It could be that both these planets formed in different environments. It is possible that TOI-1468\,b formed at its current location, whereas TOI-1468\,c could have formed further out and eventually migrated in \citep{Ida2010}. The other explanation is that both planets could have formed in similar environments, but the photoevaporation due to the XUV radiation could have stripped off a substantial portion of the inner planet's gaseous envelope due to hydrodynamic losses \citep{Lopez2013}. The mass loss history of planets depends on the amount of incident radiation they receive from the host star and the mass of the planet core. The critical mass loss timescale \citep{Lopez2013} for TOI-1468\,b is $\sim$\,2.5 Ga, which is on the order of the age of the star, and the critical mass loss timescale for the outer planet is twenty times larger, which suggests the survival of the outer atmosphere. This theory is also supported by the fact that there are no low-density exoplanets found in close-in orbits to their host stars where they would face extreme irradiation \citep{Lopez2012}. Moreover, in many multi-planet systems it is commonly observed that inner planets are smaller than the outer planet, which can be better explained by a photoevaporation model \citep{Ciardi2013}.

As discussed by \citet{Cubillos2017} and \citet{Guenther2017}, another useful parameter to look at is the thermal escape for a hydrodynamic atmosphere subjected to the gravitational perturbation from the host star in terms of the restricted Jeans escape rate \citep{Fossati2017},
\begin{equation}
\Lambda\,=\,\frac{G M_{p} m_{\rm H} }{ k_{\rm b} T_{\rm eq} R_{p} }~,~  
\end{equation}
where $\Lambda$ is the Jeans escape parameter for a hydrogen atom with mass ($m_{\rm H}$) evaluated at the planet with its mass ($M_{\rm p}$), radius ($R_{\rm p}$), and equilibrium temperature ($T_{\rm eq}$). $G$ and $k_{\rm b}$ are the gravitational and Boltzmann constants, respectively. The value of $\Lambda$ gives an understanding on the stability of the planetary atmosphere against evaporation. 
In the case of TOI-1468, $\Lambda$ is $\sim29$ for the inner planet and $\sim80$ for the outer one.  
This result puts the inner planet in the $\Lambda$ regime of 20--40, which is typical for the boil-off regime planets \citep{Owen2016} where the atmosphere escape is driven by the thermal energy and low planetary gravity. 
Systems such as TOI-1468 are excellent test beds to study planets that straddle the radius valley, offering further insights into their formation mechanisms.

\subsection{Additional planet candidates}\label{additional-signals}

In a study by \cite{Dietrich_Apai2020}, a model was created with population statistics to predict previously undetected planets in the existing multi-planetary {\em TESS} systems. Their model predicted TOI-1468 to have an additional planet at an orbital period of $3.82^{+0.93}_{-0.75}$\,d with a planet radius of $1.63^{+0.57}_{-0.42}$\,$R_\oplus$, whereas the clustered periods model predicted an orbital period of $2.68^{+0.15}_{-0.01}$\,d with a planet radius of $1.63^{+0.57}_{-0.42}\,R_\oplus$ for the additional planet. We decided to apply the box least-square \citep[BLS;][]{box_least} algorithm 
to the PDCSAP {\em TESS} light curves to search for additional transits. After removing the two transiting planets, there was no indication of any significant signal in the data corresponding to a planet with an upper limit of $\sim1.0$\,$R_{\oplus}$ in the similar period range.

\begin{figure}
   \centering
   \includegraphics[width=\hsize]{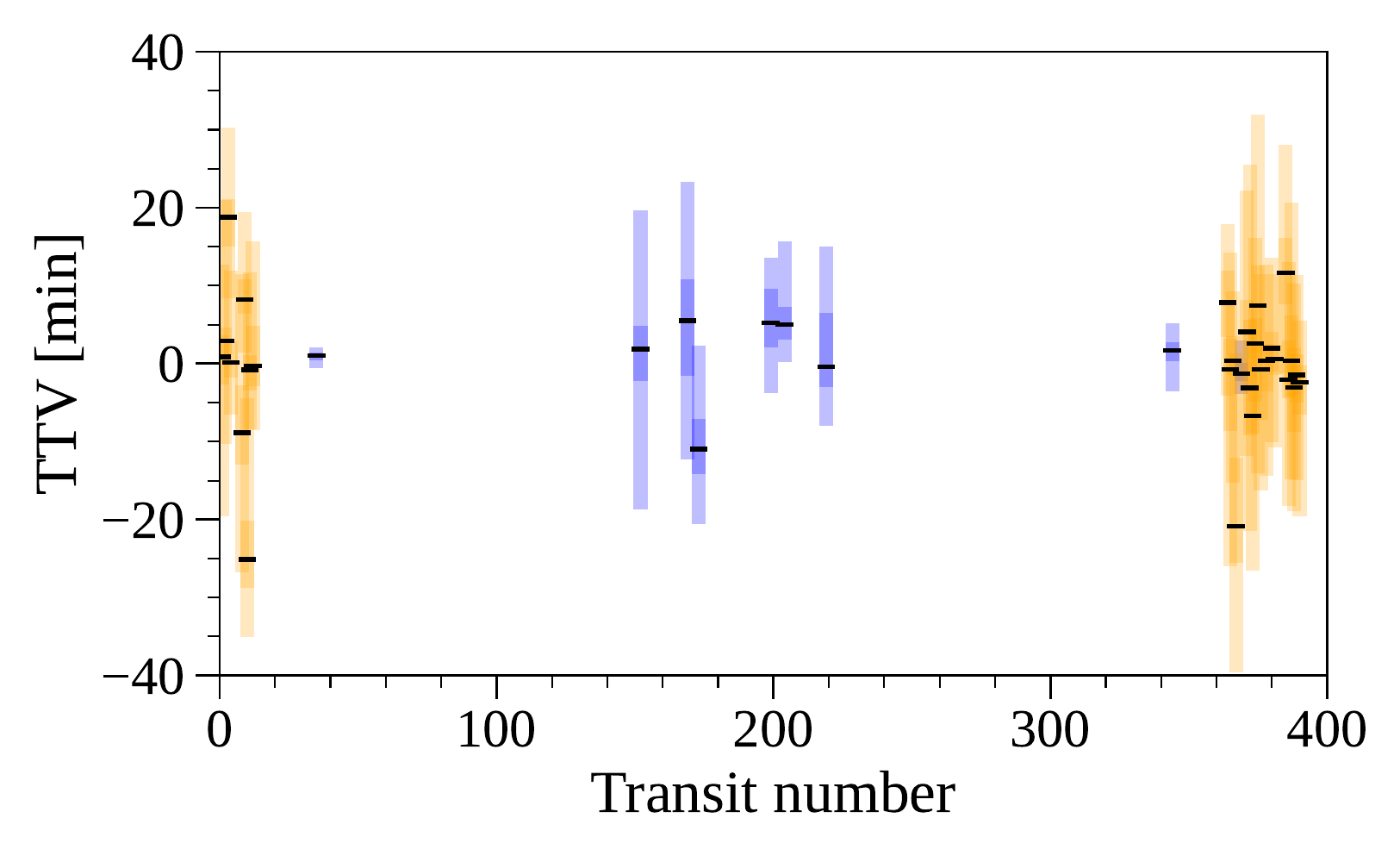}
   \caption{TOI-1468\,b transit-timing variations from {\em TESS} data (orange) and follow-up observations (blue). The different shades of color of the error bars represent the 1$\sigma$ (dark) and 3$\sigma$ (light) levels of uncertainty of the measurements. No significant TTVs are detected in the $\sim 750$\,d of baseline covered by the observations.}
    \label{Fig:TOI1468b_TTV}
\end{figure}

Since we did not find any further statistically significant signals, except the known transit signals, in the data set,  this hints that the hypothetical planet either does not exist, or would be likely non-transiting. Since the predicted planet should have an orbital period that covers the 2:1 period commensurability with the known inner planet, we searched the {\em TESS} and ground-based light curves for transit timing variations (TTVs). We did not detect any significant hints for TTVs (Fig.\,\ref{Fig:TOI1468b_TTV}). We note that, depending on the period of the hypothetical planet, the TTV period would be longer than the baseline covered by the observations ($\sim$\,750\,d) and that the S/N was so far not sufficient to detect variations in the minute range. We also did not find any evidence of this planet in our RV data. 

Planet formation models of core accretion predict an enhanced giant planet occurrence in systems with high-density rocky planets \citep{Schlecker2021}. The different bulk densities of TOI-1468 b and c do not allow a clear prediction in this regard. However, the possible high abundance of volatiles in TOI-1468 c allows us to make the assumption that no gas giant is present in the system, which would have prevented the transport of volatile-rich material into the inner system. No such planet is expected from simulated systems with host stars with masses similar to that of TOI-1468 \citep{Burn2021}, and we do not observe evidence for an outer giant planet companion.

\subsection{Atmospheric characterization}

Multi-planet systems provide additional opportunities for atmospheric characterization. Satellite missions such as the {\em James Webb Space Telescope} ({\em JWST})\footnote{\url{https://jwst.nasa.gov/science.html}} or the upcoming {\em Atmospheric Remote-sensing Infrared Exoplanet Large-survey} ({\em ARIEL})\footnote{\url{https://arielmission.space}} \citep{2016SPIE.9904E..1XT} offer excellent space-based laboratories for such studies.
To qualitatively assess the suitability of both planets for atmospheric investigations, we calculated the transmission spectroscopy metrics  (TSMs) and emission spectroscopy metrics (ESMs), as defined by \cite{Kempton2018}. We generated 10$^5$ random extractions of the planetary, orbital, and stellar parameters according to their error bars, thus obtaining the probability density function for each TSM and ESM factor. For the inner planet, we obtained $\mathrm{TSM}_{\mathrm{b}} = 9 \pm 1$ and $\mathrm{ESM}_{\mathrm{b}} = 6.3_{-0.6}^{+0.8}$. Both values are close to the recommended thresholds of ten and 7.5, respectively, defining the top-ranked atmospheric targets in the terrestrial sample. The outer planet is a small sub-Neptune with $\mathrm{TSM}_{\mathrm{c}} = 59_{-10}^{+12}$, $90$ being the threshold for its category.
It is worth noticing that these metrics rank the planets based solely on the predicted strength of an atmospheric detection. Having TSM and/or ESM values slightly below the threshold does not indicate that detailed atmospheric studies are impossible or challenging with current facilities. In other words, these metrics are not the unique criteria for determining the best targets for atmospheric studies. Scientific interest can also inspire observing proposals, for example the opportunity to explore a system with small temperate planets straddling the radius valley around an M dwarf.

To quantitatively assess the potential for atmospheric characterization of both planets, we generated synthetic {\em JWST} spectra for a range of atmospheric scenarios. Our simulations made use of the photo-chemical model \texttt{ChemKM} \citep{molaverdikhani2019a, molaverdikhani2019b, molaverdikhani2020}, the radiative transfer code \texttt{petitRADTRANS} \citep{molliere2019}, and \texttt{ExoTETHyS} \citep{morello2021} to incorporate the overall response of the {\em JWST} system, including realistic noise and error bar estimates.
For each planet, we considered a benchmark model with H/He gaseous envelope and solar abundances, and other two models showing the effect of haze or enhanced metallicity (100$\times$ solar abundances).

\begin{figure}
   \centering
   \includegraphics[width=\hsize]{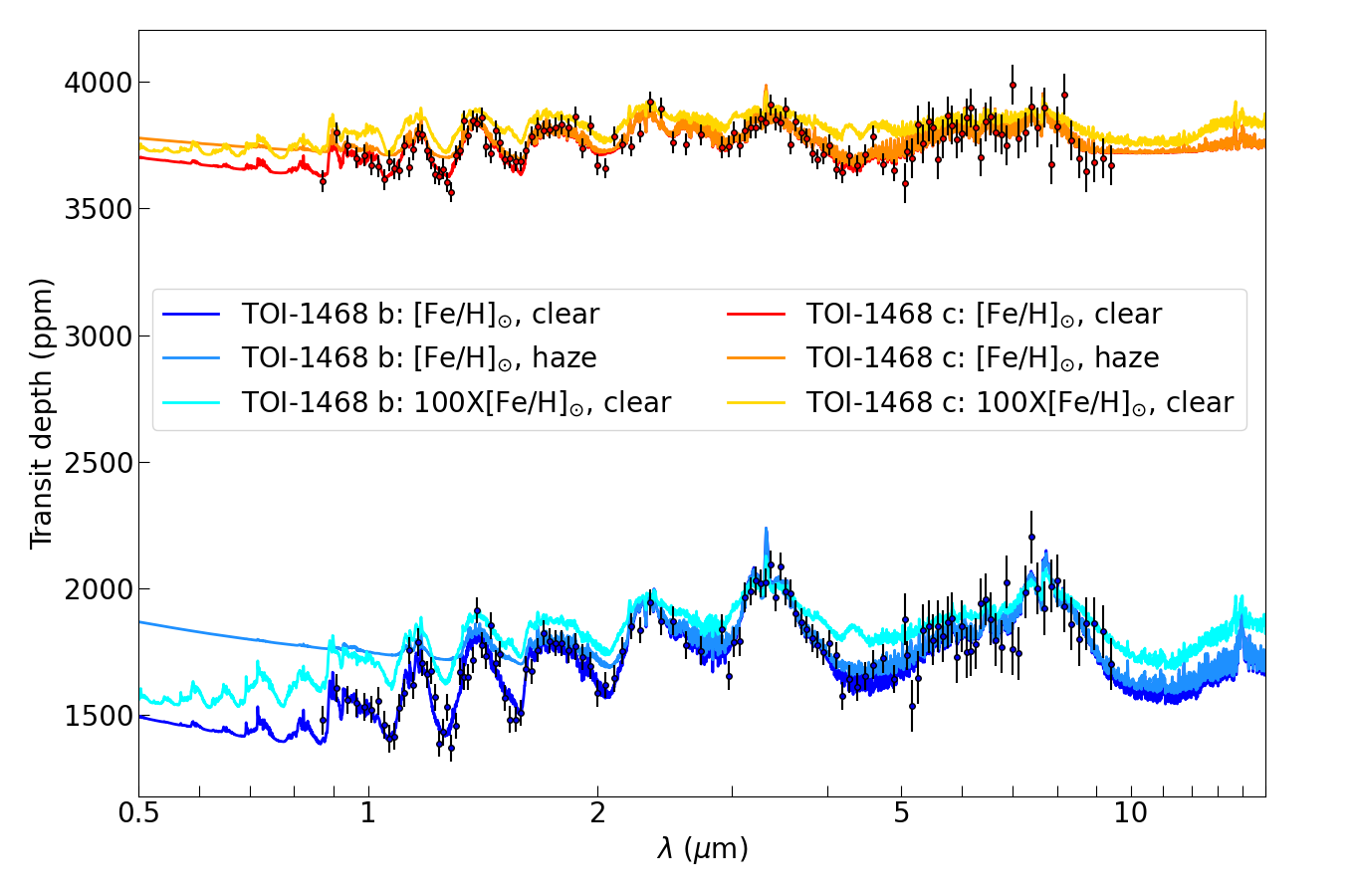}
   \caption{Synthetic {\em JWST} transmission atmospheric spectra of TOI-1468\,b and~c. 
   Fiducial models with solar abundances and no haze (solid blue and red lines), with haze (dodger blue and orange), and enhanced metallicity by a factor of 100 (cyan and gold). Estimated uncertainties are shown for the simulated observation of one transit with {\em JWST} NIRISS-SOSS, NIRSpec-G395M, and MIRI-LRS configurations.}
    \label{Fig:TOI1468bc_JWSTspectra}
\end{figure}

The transmission spectra for the H/He-dominated atmospheres show strong absorption features due to H$_2$O and CH$_4$ over the wavelength range 0.5--12\,$\mu$m (see Fig.\,\ref{Fig:TOI1468bc_JWSTspectra}). The spectroscopic modulations are on the order of 400--600\,ppm and 100--200\,ppm for TOI-1468 b and c, respectively, with a relatively modest dampening effect due to haze or metallicity, particularly at shorter wavelengths. Similar trends with enhanced metallicity or haze were also observed in simulations made for other planets (e.g., \citealp{2020A&A...642A.173N,Trifonov2021, Espinoza2022}), but the features are essentially muted in the cases with 100$\times$ solar metallicity and haze (not shown here).

We simulated {\em JWST} spectra for the NIRISS/SOSS (0.6--2.8\,$\mu$m), NIRSpec/G395M (2.88--5.20\,$\mu$m), and MIRI-LRS (5--12\,$\mu$m) instrumental modes. The wavelength bins were specifically determined, through \texttt{ExoTETHyS}, to have similar counts, leading to nearly uniform error bars per spectral point. We also used \texttt{PandExo} \citep{batalha2017} to check the best setups for each instrumental mode and the corresponding observing efficiencies (i.e., the fraction of effective integration time per given observing interval). Finally, we inflated the error bars by a factor of 1.2 to account for correlated noise. In particular, the spectral error bars estimated for just one transit observation per instrument configuration are 40-60 ppm at wavelengths $<$5 $\mu$m, with a median resolving power $R\sim 50$ and 75-100 ppm at wavelengths $>$5\,$\mu$m with bin sizes of $\sim$0.1--0.2\,$\mu$m. The lower error bars are estimated for the outer planet owing to its longer transit duration. Based on these numbers and the visualization of the simulated spectra in Fig. \ref{Fig:TOI1468bc_JWSTspectra}, we conclude that a single transit observation in any of these {\em JWST} modes would be sufficient for robust detection of the molecular features in the H/He-dominated scenarios, and the larger wavelength coverage provided by the three modes can help distinguish between the effects of metallicity and haze. However, the possible lack of a H/He envelope around the inner planet would represent a challenge for detecting its atmosphere, if any, even with {\em JWST}, unless many observations are stacked together.

Even if TOI-1468 b may have lost its primordial atmosphere, resupply of H can occur under favorable circumstances. A possible mechanism consists in the dissolution of H/He in the magma ocean of young planets and subsequent outgassing that can recreate a substantial atmosphere \citep{chachan2018, kite2019, kite2020}. Recently, this scenario has been proposed to explain the tentative detection of the HCN absorption feature on the terrestrial planet GJ 1132 b \citep{swain2021}, although the authenticity of the spectral feature has been debated \citep{mugnai2021}. Tentative evidence of H$_2$O vapor in a H/He envelope has been reported for the habitable-zone super-Earth LHS 1140 b \citep{edwards2021}, which, similar to GJ 1132 b and TOI-1468 b, belongs to the left side (at the very edge) of the radius valley.

\section{Summary} \label{sec:summary}

The TOI-1468 system consists of an early-to-mid-type M dwarf (LSPM J0106+1913) with two transiting planets in circular orbits. 
The host star has a surface temperature of $T_{\rm eff} = 3496\pm25$\,K, surface gravity of $\log{g} = 5.00\pm0.11$\,dex, and a metallicity of [Fe/H]\,=\,$-0.04\pm0.07$\,dex. 
We thereby determine a stellar mass of $0.339\pm0.011\,M_{\odot}$ and a stellar radius of $0.344\pm0.005$\,$R_{\odot}$. 
The relatively bright star ($G$\,=\,12.10\,mag, $J$\,=\,9.34\,mag) is located at a distance of $24.72\pm0.02$\,pc and has a high proper motion of $332$\,mas\,a$^{-1}$.
We also determine that the star is inactive with a relatively long rotational period of around 41--44\,d.

This multi-planet system consists of an inner super-Earth having a mass of $M_{\rm b}$\,=\,$3.21\pm0.24$\,$M_\oplus$ and a radius of $R_{\rm b} =1.280^{+0.038}_{-0.039}\,R_{\oplus}$, with an orbital period of 1.88\,d, and an outer planet with a mass of $M_{\rm c}$\,=\,$6.64 ^{+ 0.67}_{- 0.68}$\,$M_\oplus$ and a radius of $R_{\rm c}$\,=\,$2.06\pm0.04$\,$R_\oplus$, with an orbital period of 15.53\,d, and is therefore close to the inner edge of the habitable zone. 
The bulk densities of the inner and outer planets are $8.39^{+ 1.05}_{- 0.92}$\,g cm$^{-3}$ and $2.00^{+ 0.21}_{- 0.19}$\,g cm$^{-3}$, respectively. 
Multi-planet systems with planets lying on opposite sides of the radius valley are interesting laboratories to probe planet formation models through atmospheric studies. 
For example, according to the photoevaporation theory, the atmosphere of the outer planet is likely to be primordial metal enriched, while the inner one may host a secondary atmosphere, or none. 
Thus, accurate measurements of planetary masses and radii, such as those presented in this work, are required in order to estimate their density and determine the extent to which their atmosphere has been retained or removed. 
Finally, spectroscopic observations of just a few transits and eclipses of TOI-1468\,b and\,c with the {\em JWST} would provide an excellent opportunity to test photoevaporation, as well as other formation and evolution scenarios. 

\begin{acknowledgements}

CARMENES is an instrument at the Centro Astron\'omico Hispano en Andaluc\'ia (CAHA) at Calar Alto (Almer\'{\i}a, Spain), operated jointly by the Junta de Andaluc\'ia and the Instituto de Astrof\'isica de Andaluc\'ia (CSIC). CARMENES was funded by the Max-Planck-Gesellschaft (MPG), the Consejo Superior de Investigaciones Cient\'{\i}ficas (CSIC),
  the Ministerio de Econom\'ia y Competitividad (MINECO) and the European Regional Development Fund (ERDF) through projects FICTS-2011-02, ICTS-2017-07-CAHA-4, and CAHA16-CE-3978, 
  and the members of the CARMENES Consortium (Max-Planck-Institut f\"ur Astronomie, Instituto de Astrof\'{\i}sica de Andaluc\'{\i}a,
  Landessternwarte K\"onigstuhl,
  Institut de Ci\`encies de l'Espai,
  Institut f\"ur Astrophysik G\"ottingen,
  Universidad Complutense de Madrid,
  Th\"uringer Landessternwarte Tautenburg,
  Instituto de Astrof\'{\i}sica de Canarias,
  Hamburger Sternwarte,
  Centro de Astrobiolog\'{\i}a and
  Centro Astron\'omico Hispano-Alem\'an), 
  with additional contributions by the MINECO, 
  the Deutsche Forschungsgemeinschaft (DFG) through the Major Research Instrumentation Programme and Research Unit FOR2544 ``Blue Planets around Red Stars'', 
  the Klaus Tschira Stiftung, 
  the states of Baden-W\"urttemberg and Niedersachsen, 
  and by the Junta de Andaluc\'{\i}a. This work was based on data from the CARMENES data archive at CAB (CSIC-INTA).
 
Funding for the {\em TESS} mission is provided by NASA's Science Mission Directorate. We acknowledge the use of public {\em TESS} data from pipelines at the {\em TESS} Science Office and at the {\em TESS} Science Processing Operations Center. This research has made use of the Exoplanet Follow-up Observation Program website, which is operated by the California Institute of Technology, under contract with the National Aeronautics and Space Administration under the Exoplanet Exploration Program. 
Resources supporting this work were provided by the NASA High-End Computing (HEC) Program through the NASA Advanced Supercomputing (NAS) Division at Ames Research Center for the production of the SPOC data products. This paper includes data collected by the {\em TESS} mission that are publicly available from the Mikulski Archive for Space Telescopes (MAST).  

The development of the MAROON-X spectrograph was funded by the David and Lucile Packard Foundation, the Heising-Simons Foundation, the Gemini Observatory, and the University of Chicago. The MAROON-X team acknowledges support for this work from the NSF (award number 2108465) and NASA (through the \textit{TESS} Cycle 4 GI program, grant number 80NSSC22K0117). This work was enabled by observations made from the Gemini North telescope, located within the Maunakea Science Reserve and adjacent to the summit of Maunakea. We are grateful for the privilege of observing the Universe from a place that is unique in both its astronomical quality and its cultural significance.

Data were partly collected with the 150-cm telescope at Observatorio de Sierra Nevada (OSN), operated by the Instituto de Astrof\'\i fica de 
Andaluc\'\i a (IAA, CSIC), 
with the MuSCAT2 instrument, developed by ABC, at Telescopio Carlos S\'anchez operated on the island of Tenerife by the IAC in the Spanish Observatorio del Teide,
with the Telescopi Joan Or\'o (TJO) of the Observatori Astron\'omic del Montsed (OdM), which is owned by the Generalitat de Catalunya and operated by the Institute for Space Studies of Catalonia (IEEC),
and with the LCOFT network (part of the LCOGT telescope time was granted by NOIRLab through the Mid-Scale Innovations Program (MSIP), which is funded by the National Science Foundation).

Some of the Observations in the paper made use of the High-Resolution Imaging instrument. ‘Alopeke. ‘Alopeke was funded by the NASA Exoplanet Exploration Program and built at the NASA Ames Research Center by Steve B. Howell, Nic Scott, Elliott P. Horch, and Emmett Quigley. Data were reduced using a software pipeline originally written by Elliott Horch and Mark Everett. ‘Alopeke was mounted on the Gemini North telescope of the international Gemini Observatory, a program of NSF’s OIR Lab, which is managed by the Association of Universities for Research in Astronomy (AURA) under a cooperative agreement with the National Science Foundation. on behalf of the Gemini partnership: the National Science Foundation (United States), National Research Council (Canada), Agencia Nacional de Investigación y Desarrollo (Chile),
Ministerio de Ciencia, Tecnolog\'ia e Innovaci\'on (Argentina), Minist\'erio da Ci\^encia, Tecnologia, Inova\c{c}\~oes e Comunica\c{c}\~oes (Brazil), and Korea Astronomy and Space Science Institute (Republic of Korea). 

We acknowledge financial support from:
the Th\"uringer Ministerium f\"ur Wirtschaft, Wissenschaft und Digitale Gesellschaft; 
the Spanish Agencia Estatal de Investigaci\'on of the Ministerio de Ciencia e Innovaci\'on and the ERDF ``A way of making Europe'' through projects 
  PID2019-109522GB-C5[1:4],
  PID2019-107061GB-C64, 
  PID2019-110689RB-100,
  PGC2018-098153-B-C31, 
and the Centre of Excellence ``Severo Ochoa'' and ``Mar\'ia de Maeztu'' awards to the Instituto de Astrof\'isica de Canarias (CEX2019-000920-S), Instituto de Astrof\'isica de Andaluc\'ia (SEV-2017-0709), and Centro de Astrobiolog\'ia (MDM-2017-0737); 
the Generalitat de Catalunya/CERCA programme;
the European Union's Horizon 2020 research and innovation programme under the Marie Sk\l{}odowska-Curie grant agreement No. 895525; 
the DFG through grant CH~2636/1-1, 
the Excellence Cluster ORIGINS under Germany's Excellence Strategy (EXC-2094 - 390783311),
and priority programme SPP~1992 ``Exploring the Diversity of Extrasolar Planets'' (JE 701/5-1); 
the Swedish National Space Agency (SNSA; DNR 2020-00104); 
the JSPS KAKENHI grants JP17H04574, JP18H05439, JP21K13975, Grant-in-Aid for JSPS fellows grant JP20J21872, JST CREST Grant Number JPMJCR1761, and the Astrobiology Center of National Institutes of Natural Sciences (NINS) through grants AB031010 and AB031014; and the program ``Alien Earths'' (supported by the 
National Aeronautics and Space Administration under agreement No. 80NSSC21K0593) for NASA’s Nexus for Exoplanet System Science (NExSS) research coordination network sponsored by NASA’s Science Mission Directorate.

\end{acknowledgements}

\bibliographystyle{aa} 
\bibliography{bibliography}

\begin{thebibliography}{163}
\expandafter\ifx\csname natexlab\endcsname\relax\def\natexlab#1{#1}\fi

\bibitem[{{Aller} {et~al.}(2020){Aller}, {Lillo-Box}, {Jones}, {Miranda}, \&
  {Barcel{\'o} Forteza}}]{2020A&A...635A.128A}
{Aller}, A., {Lillo-Box}, J., {Jones}, D., {Miranda}, L.~F., \& {Barcel{\'o}
  Forteza}, S. 2020, \aap, 635, A128

\bibitem[{{Alonso-Floriano} {et~al.}(2015){Alonso-Floriano}, {Morales},
  {Caballero}, {Montes}, {Klutsch}, {Mundt}, {Cort{\'e}s-Contreras}, {Ribas},
  {Reiners}, {Amado}, {Quirrenbach}, \& {Jeffers}}]{AlonsoFlorian2015}
{Alonso-Floriano}, F.~J., {Morales}, J.~C., {Caballero}, J.~A., {et~al.} 2015,
  \aap, 577, A128

\bibitem[{{Ambikasaran} {et~al.}(2015){Ambikasaran}, {Foreman-Mackey},
  {Greengard}, {Hogg}, \& {O'Neil}}]{Ambikasaran2015}
{Ambikasaran}, S., {Foreman-Mackey}, D., {Greengard}, L., {Hogg}, D.~W., \&
  {O'Neil}, M. 2015, IEEE Transactions on Pattern Analysis and Machine
  Intelligence, 38, 252

\bibitem[{{Angus} {et~al.}(2018){Angus}, {Morton}, {Aigrain}, {Foreman-Mackey},
  \& {Rajpaul}}]{Angus2018}
{Angus}, R., {Morton}, T., {Aigrain}, S., {Foreman-Mackey}, D., \& {Rajpaul},
  V. 2018, \mnras, 474, 2094

\bibitem[{{Arenou} {et~al.}(2018){Arenou}, {Luri}, {Babusiaux}, {Fabricius},
  {Helmi}, {Muraveva}, {Robin}, {Spoto}, {Vallenari}, {Antoja},
  {Cantat-Gaudin}, {Jordi}, {Leclerc}, {Reyl{\'e}}, {Romero-G{\'o}mez}, {Shih},
  {Soria}, {Barache}, {Bossini}, {Bragaglia}, {Breddels}, {Fabrizio},
  {Lambert}, {Marrese}, {Massari}, {Moitinho}, {Robichon}, {Ruiz-Dern},
  {Sordo}, {Veljanoski}, {Eyer}, {Jasniewicz}, {Pancino}, {Soubiran}, {Spagna},
  {Tanga}, {Turon}, \& {Zurbach}}]{2018A&A...616A..17A}
{Arenou}, F., {Luri}, X., {Babusiaux}, C., {et~al.} 2018, \aap, 616, A17

\bibitem[{{Baglin} {et~al.}(2006){Baglin}, {Auvergne}, {Boisnard}, {Lam-Trong},
  {Barge}, {Catala}, {Deleuil}, {Michel}, \& {Weiss}}]{2006cosp...36.3749B}
{Baglin}, A., {Auvergne}, M., {Boisnard}, L., {et~al.} 2006, in 36th COSPAR
  Scientific Assembly, Vol.~36, 3749

\bibitem[{{Batalha} {et~al.}(2017){Batalha}, {Mandell}, {Pontoppidan},
  {Stevenson}, {Lewis}, {Kalirai}, {Earl}, {Greene}, {Albert}, \&
  {Nielsen}}]{batalha2017}
{Batalha}, N.~E., {Mandell}, A., {Pontoppidan}, K., {et~al.} 2017, \pasp, 129,
  064501

\bibitem[{{Batalha} {et~al.}(2013){Batalha}, {Rowe}, {Bryson}, {Barclay},
  {Burke}, {Caldwell}, {Christiansen}, {Mullally}, {Thompson}, {Brown},
  {Dupree}, {Fabrycky}, {Ford}, {Fortney}, {Gilliland}, {Isaacson}, {Latham},
  {Marcy}, {Quinn}, {Ragozzine}, {Shporer}, {Borucki}, {Ciardi}, {Gautier},
  {Haas}, {Jenkins}, {Koch}, {Lissauer}, {Rapin}, {Basri}, {Boss}, {Buchhave},
  {Carter}, {Charbonneau}, {Christensen-Dalsgaard}, {Clarke}, {Cochran},
  {Demory}, {Desert}, {Devore}, {Doyle}, {Esquerdo}, {Everett}, {Fressin},
  {Geary}, {Girouard}, {Gould}, {Hall}, {Holman}, {Howard}, {Howell},
  {Ibrahim}, {Kinemuchi}, {Kjeldsen}, {Klaus}, {Li}, {Lucas}, {Meibom},
  {Morris}, {Pr{\v{s}}a}, {Quintana}, {Sanderfer}, {Sasselov}, {Seader},
  {Smith}, {Steffen}, {Still}, {Stumpe}, {Tarter}, {Tenenbaum}, {Torres},
  {Twicken}, {Uddin}, {Van Cleve}, {Walkowicz}, \&
  {Welsh}}]{2013ApJS..204...24B}
{Batalha}, N.~M., {Rowe}, J.~F., {Bryson}, S.~T., {et~al.} 2013, \apjs, 204, 24

\bibitem[{{Berger} {et~al.}(2018){Berger}, {Huber}, {Gaidos}, \& {van
  Saders}}]{Berger2018}
{Berger}, T.~A., {Huber}, D., {Gaidos}, E., \& {van Saders}, J.~L. 2018, \apj,
  866, 99

\bibitem[{{Bluhm} {et~al.}(2021){Bluhm}, {Pall{\'e}}, {Molaverdikhani},
  {Kemmer}, {Hatzes}, {Kossakowski}, {Stock}, {Caballero}, {Lillo-Box},
  {B{\'e}jar}, {Soto}, {Amado}, {Brown}, {Cadieux}, {Cloutier}, {Collins},
  {Collins}, {Cort{\'e}s-Contreras}, {Doyon}, {Dreizler}, {Espinoza}, {Fukui},
  {Gonz{\'a}lez-{\'A}lvarez}, {Henning}, {Horne}, {Jeffers}, {Jenkins},
  {Jensen}, {Kaminski}, {Kielkopf}, {Kusakabe}, {K{\"u}rster},
  {Lafreni{\`e}re}, {Luque}, {Murgas}, {Montes}, {Morales}, {Narita},
  {Passegger}, {Quirrenbach}, {Sch{\"o}fer}, {Reffert}, {Reiners}, {Ribas},
  {Ricker}, {Seager}, {Schweitzer}, {Schwarz}, {Tamura}, {Trifonov},
  {Vanderspek}, {Winn}, {Zechmeister}, \& {Zapatero Osorio}}]{Bluhm2021}
{Bluhm}, P., {Pall{\'e}}, E., {Molaverdikhani}, K., {et~al.} 2021, \aap, 650,
  A78

\bibitem[{Bonfils {et~al.}(2013)Bonfils, Delfosse, Udry, Forveille, Mayor,
  Perrier, Bouchy, Gillon, Lovis, Pepe, {et~al.}}]{Bonfils2013}
Bonfils, X., Delfosse, X., Udry, S., {et~al.} 2013, \aap, 549, A109

\bibitem[{Borucki {et~al.}(2010)Borucki, Koch, Basri, Batalha, Brown, Caldwell,
  Caldwell, Christensen-Dalsgaard, Cochran, DeVore, {et~al.}}]{Borucki2010}
Borucki, W.~J., Koch, D., Basri, G., {et~al.} 2010, Science, 327, 977

\bibitem[{{Brown} {et~al.}(2013){Brown}, {Baliber}, {Bianco}, {Bowman},
  {Burleson}, {Conway}, {Crellin}, {Depagne}, {De Vera}, {Dilday}, {Dragomir},
  {Dubberley}, {Eastman}, {Elphick}, {Falarski}, {Foale}, {Ford}, {Fulton},
  {Garza}, {Gomez}, {Graham}, {Greene}, {Haldeman}, {Hawkins}, {Haworth},
  {Haynes}, {Hidas}, {Hjelstrom}, {Howell}, {Hygelund}, {Lister}, {Lobdill},
  {Martinez}, {Mullins}, {Norbury}, {Parrent}, {Paulson}, {Petry}, {Pickles},
  {Posner}, {Rosing}, {Ross}, {Sand}, {Saunders}, {Shobbrook}, {Shporer},
  {Street}, {Thomas}, {Tsapras}, {Tufts}, {Valenti}, {Vander Horst}, {Walker},
  {White}, \& {Willis}}]{Brown2013}
{Brown}, T.~M., {Baliber}, N., {Bianco}, F.~B., {et~al.} 2013, \pasp, 125, 1031

\bibitem[{{Burn} {et~al.}(2021){Burn}, {Schlecker}, {Mordasini}, {Emsenhuber},
  {Alibert}, {Henning}, {Klahr}, \& {Benz}}]{Burn2021}
{Burn}, R., {Schlecker}, M., {Mordasini}, C., {et~al.} 2021, \aap, 656, A72

\bibitem[{{Butters} {et~al.}(2010){Butters}, {West}, {Anderson}, {Collier
  Cameron}, {Clarkson}, {Enoch}, {Haswell}, {Hellier}, {Horne}, {Joshi},
  {Kane}, {Lister}, {Maxted}, {Parley}, {Pollacco}, {Smalley}, {Street},
  {Todd}, {Wheatley}, \& {Wilson}}]{Butters2010}
{Butters}, O.~W., {West}, R.~G., {Anderson}, D.~R., {et~al.} 2010, \aap, 520,
  L10

\bibitem[{{Caballero} {et~al.}(2016){Caballero}, {Cort{\'e}s-Contreras},
  {Alonso-Floriano}, {Montes}, {Quirrenbach}, {Amado}, {Ribas}, {Reiners},
  {Abellan}, {B{\'e}jar}, {Brinkm{\"o}ller}, {Czesla}, {Dorda}, {Gallardo},
  {Gonz{\'a}lez-{\'A}lvarez}, {Hidalgo}, {Holgado}, {Jeffers}, {Kim},
  {Klutsch}, {Lamert}, {Llamas}, {L{\'o}pez-Santiago},
  {Mart{\'{\i}}nez-Rodr{\'{\i}}guez}, {Morales}, {Mundt}, {Passegger},
  {Sch{\"o}fer}, {Seifert}, \& {Zechmeister}}]{Caballero2016}
{Caballero}, J.~A., {Cort{\'e}s-Contreras}, M., {Alonso-Floriano}, F.~J.,
  {et~al.} 2016, in 19th Cambridge Workshop on Cool Stars, Stellar Systems, and
  the Sun (CS19), 148

\bibitem[{{Caballero} {et~al.}(2022){Caballero}, {Gonzalez-Alvarez}, {Brady},
  {Trifonov}, {Ellis}, {Dorn}, {Cifuentes}, {Molaverdikhani}, {Bean},
  {Boyajian}, {Rodriguez}, {Sanz-Forcada}, {Zapatero Osorio}, {Abia}, {Amado},
  {Anugu}, {Bejar}, {Davies}, {Dreizler}, {Dubois}, {Ennis}, {Espinoza},
  {Farrington}, {Garcia Lopez}, {Gardner}, {Hatzes}, {Henning}, {Herrero},
  {Herrero-Cisneros}, {Kaminski}, {Kasper}, {Klement}, {Kraus}, {Labdon},
  {Lanthermann}, {Le Bouquin}, {Lopez Gonzalez}, {Luque}, {Mann}, {Marfil},
  {Monnier}, {Montes}, {Morales}, {Palle}, {Pedraz}, {Quirrenbach}, {Reffert},
  {Reiners}, {Ribas}, {Rodriguez-Lopez}, {Schaefer}, {Schweitzer}, {Seifahrt},
  {Setterholm}, {Shan}, {Shulyak}, {Solano}, {Sreenivas}, {Stefansson},
  {Stuermer}, {Tabernero}, {Tal-Or}, {ten Brummelaar}, {Vanaverbeke}, {von
  Braun}, {Youngblood}, \& {Zechmeister}}]{Caballero2022}
{Caballero}, J.~A., {Gonzalez-Alvarez}, E., {Brady}, M., {et~al.} 2022, arXiv
  e-prints, arXiv:2206.09990

\bibitem[{{Cale} {et~al.}(2021){Cale}, {Reefe}, {Plavchan}, {Tanner}, {Gaidos},
  {Gagn{\'e}}, {Gao}, {Kane}, {B{\'e}jar}, {Lodieu}, {Anglada-Escud{\'e}},
  {Ribas}, {Pall{\'e}}, {Quirrenbach}, {Amado}, {Reiners}, {Caballero}, {Rosa
  Zapatero Osorio}, {Dreizler}, {Howard}, {Fulton}, {Xuesong Wang}, {Collins},
  {El Mufti}, {Wittrock}, {Gilbert}, {Barclay}, {Klein}, {Martioli},
  {Wittenmyer}, {Wright}, {Addison}, {Hirano}, {Tamura}, {Kotani}, {Narita},
  {Vermilion}, {Lee}, {Geneser}, {Teske}, {Quinn}, {Latham}, {Esquerdo},
  {Calkins}, {Berlind}, {Zohrabi}, {Stibbards}, {Kotnana}, {Jenkins},
  {Twicken}, {Henze}, {Kidwell}, {Burke}, {Villase{\~n}or}, \&
  {Boyd}}]{Cale2021}
{Cale}, B.~L., {Reefe}, M., {Plavchan}, P., {et~al.} 2021, \aj, 162, 295

\bibitem[{{Chabrier}(2003)}]{Chabrier2003}
{Chabrier}, G. 2003, \pasp, 115, 763

\bibitem[{{Chachan} \& {Stevenson}(2018)}]{chachan2018}
{Chachan}, Y. \& {Stevenson}, D.~J. 2018, \apj, 854, 21

\bibitem[{{Charbonneau} {et~al.}(2008){Charbonneau}, {Irwin}, {Nutzman}, \&
  {Falco}}]{Charbonneau2008}
{Charbonneau}, D., {Irwin}, J., {Nutzman}, P., \& {Falco}, E.~E. 2008, in AAS
  Meeting Abstract, Vol. 212, AAS Meeting, 44.02

\bibitem[{{Ciardi} {et~al.}(2013){Ciardi}, {Fabrycky}, {Ford}, {Gautier},
  {Howell}, {Lissauer}, {Ragozzine}, \& {Rowe}}]{Ciardi2013}
{Ciardi}, D.~R., {Fabrycky}, D.~C., {Ford}, E.~B., {et~al.} 2013, \apj, 763, 41

\bibitem[{{Cifuentes} {et~al.}(2020){Cifuentes}, {Caballero},
  {Cort{\'e}s-Contreras}, {Montes}, {Abell{\'a}n}, {Dorda}, {Holgado},
  {Zapatero Osorio}, {Morales}, {Amado}, {Passegger}, {Quirrenbach}, {Reiners},
  {Ribas}, {Sanz-Forcada}, {Schweitzer}, {Seifert}, \&
  {Solano}}]{Cifuentes2020}
{Cifuentes}, C., {Caballero}, J.~A., {Cort{\'e}s-Contreras}, M., {et~al.} 2020,
  \aap, 642, A115

\bibitem[{{Cloutier} {et~al.}(2018){Cloutier}, {Artigau}, {Delfosse}, {Malo},
  {Moutou}, {Doyon}, {Donati}, {Cumming}, {Dumusque}, {H{\'e}brard}, \&
  {Menou}}]{Cloutier2018}
{Cloutier}, R., {Artigau}, {\'E}., {Delfosse}, X., {et~al.} 2018, \aj, 155, 93

\bibitem[{{Cloutier} {et~al.}(2020){Cloutier}, {Eastman}, {Rodriguez},
  {Astudillo-Defru}, {Bonfils}, {Mortier}, {Watson}, {Stalport}, {Pinamonti},
  {Lienhard}, {Harutyunyan}, {Damasso}, {Latham}, {Collins}, {Massey}, {Irwin},
  {Winters}, {Charbonneau}, {Ziegler}, {Matthews}, {Crossfield}, {Kreidberg},
  {Quinn}, {Ricker}, {Vanderspek}, {Seager}, {Winn}, {Jenkins}, {Vezie},
  {Udry}, {Twicken}, {Tenenbaum}, {Sozzetti}, {S{\'e}gransan}, {Schlieder},
  {Sasselov}, {Santos}, {Rice}, {Rackham}, {Poretti}, {Piotto}, {Phillips},
  {Pepe}, {Molinari}, {Mignon}, {Micela}, {Melo}, {de Medeiros}, {Mayor},
  {Matson}, {Martinez Fiorenzano}, {Mann}, {Magazz{\'u}}, {Lovis},
  {L{\'o}pez-Morales}, {Lopez}, {Lissauer}, {L{\'e}pine}, {Law}, {Kielkopf},
  {Johnson}, {Jensen}, {Howell}, {Gonzales}, {Ghedina}, {Forveille},
  {Figueira}, {Dumusque}, {Dressing}, {Doyon}, {D{\'\i}az}, {Fabrizio},
  {Delfosse}, {Cosentino}, {Conti}, {Collins}, {Cameron}, {Ciardi}, {Caldwell},
  {Burke}, {Buchhave}, {Brice{\~n}o}, {Boyd}, {Bouchy}, {Beichman}, {Artigau},
  \& {Almenara}}]{Cloutier2020}
{Cloutier}, R., {Eastman}, J.~D., {Rodriguez}, J.~E., {et~al.} 2020, \aj, 160,
  3

\bibitem[{{Collins} {et~al.}(2017){Collins}, {Kielkopf}, {Stassun}, \&
  {Hessman}}]{Collins:2017}
{Collins}, K.~A., {Kielkopf}, J.~F., {Stassun}, K.~G., \& {Hessman}, F.~V.
  2017, \aj, 153, 77

\bibitem[{{Colome} \& {Ribas}(2006)}]{Colome2006}
{Colome}, J. \& {Ribas}, I. 2006, IAU Special Session, 6, 11

\bibitem[{{Cubillos} {et~al.}(2017){Cubillos}, {Erkaev}, {Juvan}, {Fossati},
  {Johnstone}, {Lammer}, {Lendl}, {Odert}, \& {Kislyakova}}]{Cubillos2017}
{Cubillos}, P., {Erkaev}, N.~V., {Juvan}, I., {et~al.} 2017, \mnras, 466, 1868

\bibitem[{{Curtis} {et~al.}(2019){Curtis}, {Ag{\"u}eros}, {Mamajek}, {Wright},
  \& {Cummings}}]{Curtis2019}
{Curtis}, J.~L., {Ag{\"u}eros}, M.~A., {Mamajek}, E.~E., {Wright}, J.~T., \&
  {Cummings}, J.~D. 2019, \aj, 158, 77

\bibitem[{{Damasso} {et~al.}(2018){Damasso}, {Bonomo}, {Astudillo-Defru},
  {Bonfils}, {Malavolta}, {Sozzetti}, {Lopez}, {Zeng}, {Haywood}, {Irwin},
  {Mortier}, {Vanderburg}, {Maldonado}, {Lanza}, {Affer}, {Almenara},
  {Benatti}, {Biazzo}, {Bignamini}, {Borsa}, {Bouchy}, {Buchhave}, {Cameron},
  {Carleo}, {Charbonneau}, {Claudi}, {Cosentino}, {Covino}, {Delfosse},
  {Desidera}, {Di Fabrizio}, {Dressing}, {Esposito}, {Fares}, {Figueira},
  {Fiorenzano}, {Forveille}, {Giacobbe}, {Gonz{\'a}lez-{\'A}lvarez}, {Gratton},
  {Harutyunyan}, {Johnson}, {Latham}, {Leto}, {Lopez-Morales}, {Lovis},
  {Maggio}, {Mancini}, {Masiero}, {Mayor}, {Micela}, {Molinari}, {Motalebi},
  {Murgas}, {Nascimbeni}, {Pagano}, {Pepe}, {Phillips}, {Piotto}, {Poretti},
  {Rainer}, {Rice}, {Santos}, {Sasselov}, {Scandariato}, {S{\'e}gransan},
  {Smareglia}, {Udry}, {Watson}, \& {W{\"u}nsche}}]{Damasso2018}
{Damasso}, M., {Bonomo}, A.~S., {Astudillo-Defru}, N., {et~al.} 2018, \aap,
  615, A69

\bibitem[{{Damasso} {et~al.}(2019){Damasso}, {Zeng}, {Malavolta}, {Mayo},
  {Sozzetti}, {Mortier}, {Buchhave}, {Vanderburg}, {Lopez-Morales}, {Bonomo},
  {Cameron}, {Coffinet}, {Figueira}, {Latham}, {Mayor}, {Molinari}, {Pepe},
  {Phillips}, {Poretti}, {Rice}, {Udry}, \& {Watson}}]{2019A&A...624A..38D}
{Damasso}, M., {Zeng}, L., {Malavolta}, L., {et~al.} 2019, \aap, 624, A38

\bibitem[{{Dekany} {et~al.}(2013){Dekany}, {Roberts}, {Burruss}, {Bouchez},
  {Truong}, {Baranec}, {Guiwits}, {Hale}, {Angione}, {Trinh}, {Zolkower},
  {Shelton}, {Palmer}, {Henning}, {Croner}, {Troy}, {McKenna}, {Tesch},
  {Hildebrandt}, \& {Milburn}}]{dekany2013}
{Dekany}, R., {Roberts}, J., {Burruss}, R., {et~al.} 2013, \apj, 776, 130

\bibitem[{{Delorme} {et~al.}(2011){Delorme}, {Cameron}, {Hebb}, {Rostron},
  {Lister}, {Norton}, {Pollacco}, \& {West}}]{Delorme2011}
{Delorme}, P., {Cameron}, A.~C., {Hebb}, L., {et~al.} 2011, in Astronomical
  Society of the Pacific Conference Series, Vol. 448, 16th Cambridge Workshop
  on Cool Stars, Stellar Systems, and the Sun, ed. C.~{Johns-Krull}, M.~K.
  {Browning}, \& A.~A. {West}, 841

\bibitem[{{Dietrich} \& {Apai}(2020)}]{Dietrich_Apai2020}
{Dietrich}, J. \& {Apai}, D. 2020, \aj, 160, 107

\bibitem[{{D{\'\i}ez Alonso} {et~al.}(2019){D{\'\i}ez Alonso}, {Caballero},
  {Montes}, {de Cos Juez}, {Dreizler}, {Dubois}, {Jeffers}, {Lalitha}, {Naves},
  {Reiners}, {Ribas}, {Vanaverbeke}, {Amado}, {B{\'e}jar},
  {Cort{\'e}s-Contreras}, {Herrero}, {Hidalgo}, {K{\"u}rster}, {Logie},
  {Quirrenbach}, {Rau}, {Seifert}, {Sch{\"o}fer}, \& {Tal-Or}}]{DiezAlonso2019}
{D{\'\i}ez Alonso}, E., {Caballero}, J.~A., {Montes}, D., {et~al.} 2019, \aap,
  621, A126

\bibitem[{{Drake} {et~al.}(2009){Drake}, {Djorgovski}, {Mahabal}, {Beshore},
  {Larson}, {Graham}, {Williams}, {Christensen}, {Catelan}, {Boattini},
  {Gibbs}, {Hill}, \& {Kowalski}}]{Drake2009}
{Drake}, A.~J., {Djorgovski}, S.~G., {Mahabal}, A., {et~al.} 2009, \apj, 696,
  870

\bibitem[{{Dressing} \& {Charbonneau}(2013)}]{Dressing2013}
{Dressing}, C.~D. \& {Charbonneau}, D. 2013, \apj, 767, 95

\bibitem[{{Dumusque} {et~al.}(2014){Dumusque}, {Bonomo}, {Haywood},
  {Malavolta}, {S{\'e}gransan}, {Buchhave}, {Collier Cameron}, {Latham},
  {Molinari}, {Pepe}, {Udry}, {Charbonneau}, {Cosentino}, {Dressing},
  {Figueira}, {Fiorenzano}, {Gettel}, {Harutyunyan}, {Horne}, {Lopez-Morales},
  {Lovis}, {Mayor}, {Micela}, {Motalebi}, {Nascimbeni}, {Phillips}, {Piotto},
  {Pollacco}, {Queloz}, {Rice}, {Sasselov}, {Sozzetti}, {Szentgyorgyi}, \&
  {Watson}}]{2014ApJ...789..154D}
{Dumusque}, X., {Bonomo}, A.~S., {Haywood}, R.~D., {et~al.} 2014, \apj, 789,
  154

\bibitem[{{Dumusque} {et~al.}(2011){Dumusque}, {Udry}, {Lovis}, {Santos}, \&
  {Monteiro}}]{2011A&A...525A.140D}
{Dumusque}, X., {Udry}, S., {Lovis}, C., {Santos}, N.~C., \& {Monteiro},
  M.~J.~P.~F.~G. 2011, \aap, 525, A140

\bibitem[{{Edwards} {et~al.}(2021){Edwards}, {Changeat}, {Mori}, {Anisman},
  {Morvan}, {Yip}, {Tsiaras}, {Al-Refaie}, {Waldmann}, \&
  {Tinetti}}]{edwards2021}
{Edwards}, B., {Changeat}, Q., {Mori}, M., {et~al.} 2021, \aj, 161, 44

\bibitem[{{Espinoza}(2018)}]{Espinoza2018}
{Espinoza}, N. 2018, \rnaas, 2, 209

\bibitem[{{Espinoza} {et~al.}(2019){Espinoza}, {Kossakowski}, \&
  {Brahm}}]{juliet}
{Espinoza}, N., {Kossakowski}, D., \& {Brahm}, R. 2019, \mnras, 490, 2262

\bibitem[{{Espinoza} {et~al.}(2022){Espinoza}, {Pall{\'e}}, {Kemmer}, {Luque},
  {Caballero}, {Cifuentes}, {Herrero}, {S{\'a}nchez B{\'e}jar}, {Stock},
  {Molaverdikhani}, {Morello}, {Kossakowski}, {Schlecker}, {Amado}, {Bluhm},
  {Cort{\'e}s-Contreras}, {Henning}, {Kreidberg}, {K{\"u}rster}, {Lafarga},
  {Lodieu}, {Morales}, {Oshagh}, {Passegger}, {Pavlov}, {Quirrenbach},
  {Reffert}, {Reiners}, {Ribas}, {Rodr{\'\i}guez}, {L{\'o}pez}, {Schweitzer},
  {Trifonov}, {Chaturvedi}, {Dreizler}, {Jeffers}, {Kaminski},
  {L{\'o}pez-Gonz{\'a}lez}, {Lillo-Box}, {Montes}, {Nowak}, {Pedraz},
  {Vanaverbeke}, {Zapatero Osorio}, {Zechmeister}, {Collins}, {Girardin},
  {Guerra}, {Naves}, {Crossfield}, {Matthews}, {Howell}, {Ciardi}, {Gonzales},
  {Matson}, {Beichman}, {Schlieder}, {Barclay}, {Vezie}, {Villase{\~n}or},
  {Daylan}, {Mireies}, {Dragomir}, {Twicken}, {Jenkins}, {Winn}, {Latham},
  {Ricker}, \& {Seager}}]{Espinoza2022}
{Espinoza}, N., {Pall{\'e}}, E., {Kemmer}, J., {et~al.} 2022, \aj, 163, 133

\bibitem[{{Fischer} {et~al.}(2005){Fischer}, {Laughlin}, {Butler}, {Marcy},
  {Johnson}, {Henry}, {Valenti}, {Vogt}, {Ammons}, {Robinson}, {Spear},
  {Strader}, {Driscoll}, {Fuller}, {Johnson}, {Manrao}, {McCarthy},
  {Mu{\~n}oz}, {Tah}, {Wright}, {Ida}, {Sato}, {Toyota}, \&
  {Minniti}}]{Fischer2005}
{Fischer}, D.~A., {Laughlin}, G., {Butler}, P., {et~al.} 2005, \apj, 620, 481

\bibitem[{{Foreman-Mackey} {et~al.}(2017){Foreman-Mackey}, {Agol},
  {Ambikasaran}, \& {Angus}}]{Foreman-Mackey2017}
{Foreman-Mackey}, D., {Agol}, E., {Ambikasaran}, S., \& {Angus}, R. 2017,
  {celerite: Scalable 1D Gaussian Processes in C++, Python, and Julia}

\bibitem[{{Fossati} {et~al.}(2017){Fossati}, {Erkaev}, {Lammer}, {Cubillos},
  {Odert}, {Juvan}, {Kislyakova}, {Lendl}, {Kubyshkina}, \&
  {Bauer}}]{Fossati2017}
{Fossati}, L., {Erkaev}, N.~V., {Lammer}, H., {et~al.} 2017, \aap, 598, A90

\bibitem[{{Frith} {et~al.}(2013){Frith}, {Pinfield}, {Jones}, {Barnes},
  {Pavlenko}, {Martin}, {Brown}, {Kuznetsov}, {Marocco}, {Tata}, \&
  {Cappetta}}]{Frith2013}
{Frith}, J., {Pinfield}, D.~J., {Jones}, H.~R.~A., {et~al.} 2013, \mnras, 435,
  2161

\bibitem[{{Fukui} {et~al.}(2021){Fukui}, {Korth}, {Livingston}, {Twicken},
  {Osorio}, {Jenkins}, {Mori}, {Murgas}, {Ogihara}, {Narita}, {Pall{\'e}},
  {Stassun}, {Nowak}, {Ciardi}, {Alvarez-Hernandez}, {B{\'e}jar},
  {Casasayas-Barris}, {Crouzet}, {de Leon}, {Esparza-Borges}, {Soto}, {Isogai},
  {Kawauchi}, {Klagyivik}, {Kodama}, {Kurita}, {Kusakabe}, {Luque},
  {Madrigal-Aguado}, {Rodriguez}, {Morello}, {Nishiumi}, {Orell-Miquel},
  {Oshagh}, {Parviainen}, {S{\'a}nchez-Benavente}, {Stangret}, {Terada},
  {Watanabe}, {Chen}, {Tamura}, {Bosch-Cabot}, {Bowen}, {Eastridge}, {Freour},
  {Gonzales}, {Guerra}, {Jundiyeh}, {Kim}, {Kroer}, {Levine}, {Morgan},
  {Reefe}, {Tronsgaard}, {Wedderkopp}, {Wittrock}, {Collins}, {Hesse},
  {Latham}, {Ricker}, {Seager}, {Vanderspek}, {Winn}, {Bachelet}, {Bowman},
  {McCully}, {Daily}, {Harbeck}, \& {Volgenau}}]{2021AJ....162..167F}
{Fukui}, A., {Korth}, J., {Livingston}, J.~H., {et~al.} 2021, \aj, 162, 167

\bibitem[{{Fulton} {et~al.}(2018){Fulton}, {Petigura}, {Blunt}, \&
  {Sinukoff}}]{Fulton2018}
{Fulton}, B.~J., {Petigura}, E.~A., {Blunt}, S., \& {Sinukoff}, E. 2018, \pasp,
  130, 044504

\bibitem[{{Fulton} {et~al.}(2017){Fulton}, {Petigura}, {Howard}, {Isaacson},
  {Marcy}, {Cargile}, {Hebb}, {Weiss}, {Johnson}, {Morton}, {Sinukoff},
  {Crossfield}, \& {Hirsch}}]{Fulton17}
{Fulton}, B.~J., {Petigura}, E.~A., {Howard}, A.~W., {et~al.} 2017, \aj, 154,
  109

\bibitem[{{Furlan} {et~al.}(2017){Furlan}, {Ciardi}, {Everett}, {Saylors},
  {Teske}, {Horch}, {Howell}, {van Belle}, {Hirsch}, {Gautier}, {Adams},
  {Barrado}, {Cartier}, {Dressing}, {Dupree}, {Gilliland}, {Lillo-Box},
  {Lucas}, \& {Wang}}]{furlan2017}
{Furlan}, E., {Ciardi}, D.~R., {Everett}, M.~E., {et~al.} 2017, \aj, 153, 71

\bibitem[{{Gaia Collaboration} {et~al.}(2021){Gaia Collaboration}, {Brown},
  {Vallenari}, {Prusti}, {de Bruijne}, {Babusiaux}, {Biermann}, {Creevey},
  {Evans}, {Eyer}, {Hutton}, {Jansen}, {Jordi}, {Klioner}, {Lammers},
  {Lindegren}, {Luri}, {Mignard}, {Panem}, {Pourbaix}, {Randich}, {Sartoretti},
  {Soubiran}, {Walton}, {Arenou}, {Bailer-Jones}, {Bastian}, {Cropper},
  {Drimmel}, {Katz}, {Lattanzi}, {van Leeuwen}, {Bakker}, {Cacciari},
  {Casta{\~n}eda}, {De Angeli}, {Ducourant}, {Fabricius}, {Fouesneau},
  {Fr{\'e}mat}, {Guerra}, {Guerrier}, {Guiraud}, {Jean-Antoine Piccolo},
  {Masana}, {Messineo}, {Mowlavi}, {Nicolas}, {Nienartowicz}, {Pailler},
  {Panuzzo}, {Riclet}, {Roux}, {Seabroke}, {Sordo}, {Tanga}, {Th{\'e}venin},
  {Gracia-Abril}, {Portell}, {Teyssier}, {Altmann}, {Andrae}, {Bellas-Velidis},
  {Benson}, {Berthier}, {Blomme}, {Brugaletta}, {Burgess}, {Busso}, {Carry},
  {Cellino}, {Cheek}, {Clementini}, {Damerdji}, {Davidson}, {Delchambre},
  {Dell'Oro}, {Fern{\'a}ndez-Hern{\'a}ndez}, {Galluccio}, {Garc{\'\i}a-Lario},
  {Garcia-Reinaldos}, {Gonz{\'a}lez-N{\'u}{\~n}ez}, {Gosset}, {Haigron},
  {Halbwachs}, {Hambly}, {Harrison}, {Hatzidimitriou}, {Heiter},
  {Hern{\'a}ndez}, {Hestroffer}, {Hodgkin}, {Holl}, {Jan{\ss}en}, {Jevardat de
  Fombelle}, {Jordan}, {Krone-Martins}, {Lanzafame}, {L{\"o}ffler}, {Lorca},
  {Manteiga}, {Marchal}, {Marrese}, {Moitinho}, {Mora}, {Muinonen}, {Osborne},
  {Pancino}, {Pauwels}, {Petit}, {Recio-Blanco}, {Richards}, {Riello},
  {Rimoldini}, {Robin}, {Roegiers}, {Rybizki}, {Sarro}, {Siopis}, {Smith},
  {Sozzetti}, {Ulla}, {Utrilla}, {van Leeuwen}, {van Reeven}, {Abbas}, {Abreu
  Aramburu}, {Accart}, {Aerts}, {Aguado}, {Ajaj}, {Altavilla}, {{\'A}lvarez},
  {{\'A}lvarez Cid-Fuentes}, {Alves}, {Anderson}, {Anglada Varela}, {Antoja},
  {Audard}, {Baines}, {Baker}, {Balaguer-N{\'u}{\~n}ez}, {Balbinot}, {Balog},
  {Barache}, {Barbato}, {Barros}, {Barstow}, {Bartolom{\'e}}, {Bassilana},
  {Bauchet}, {Baudesson-Stella}, {Becciani}, {Bellazzini}, {Bernet}, {Bertone},
  {Bianchi}, {Blanco-Cuaresma}, {Boch}, {Bombrun}, {Bossini}, {Bouquillon},
  {Bragaglia}, {Bramante}, {Breedt}, {Bressan}, {Brouillet}, {Bucciarelli},
  {Burlacu}, {Busonero}, {Butkevich}, {Buzzi}, {Caffau}, {Cancelliere},
  {C{\'a}novas}, {Cantat-Gaudin}, {Carballo}, {Carlucci}, {Carnerero},
  {Carrasco}, {Casamiquela}, {Castellani}, {Castro-Ginard}, {Castro Sampol},
  {Chaoul}, {Charlot}, {Chemin}, {Chiavassa}, {Cioni}, {Comoretto}, {Cooper},
  {Cornez}, {Cowell}, {Crifo}, {Crosta}, {Crowley}, {Dafonte}, {Dapergolas},
  {David}, {David}, {de Laverny}, {De Luise}, {De March}, {De Ridder}, {de
  Souza}, {de Teodoro}, {de Torres}, {del Peloso}, {del Pozo}, {Delbo},
  {Delgado}, {Delgado}, {Delisle}, {Di Matteo}, {Diakite}, {Diener},
  {Distefano}, {Dolding}, {Eappachen}, {Edvardsson}, {Enke}, {Esquej}, {Fabre},
  {Fabrizio}, {Faigler}, {Fedorets}, {Fernique}, {Fienga}, {Figueras},
  {Fouron}, {Fragkoudi}, {Fraile}, {Franke}, {Gai}, {Garabato},
  {Garcia-Gutierrez}, {Garc{\'\i}a-Torres}, {Garofalo}, {Gavras}, {Gerlach},
  {Geyer}, {Giacobbe}, {Gilmore}, {Girona}, {Giuffrida}, {Gomel}, {Gomez},
  {Gonzalez-Santamaria}, {Gonz{\'a}lez-Vidal}, {Granvik},
  {Guti{\'e}rrez-S{\'a}nchez}, {Guy}, {Hauser}, {Haywood}, {Helmi}, {Hidalgo},
  {Hilger}, {H{\l}adczuk}, {Hobbs}, {Holland}, {Huckle}, {Jasniewicz},
  {Jonker}, {Juaristi Campillo}, {Julbe}, {Karbevska}, {Kervella}, {Khanna},
  {Kochoska}, {Kontizas}, {Kordopatis}, {Korn}, {Kostrzewa-Rutkowska},
  {Kruszy{\'n}ska}, {Lambert}, {Lanza}, {Lasne}, {Le Campion}, {Le Fustec},
  {Lebreton}, {Lebzelter}, {Leccia}, {Leclerc}, {Lecoeur-Taibi}, {Liao},
  {Licata}, {Lindstr{\o}m}, {Lister}, {Livanou}, {Lobel}, {Madrero Pardo},
  {Managau}, {Mann}, {Marchant}, {Marconi}, {Marcos Santos}, {Marinoni},
  {Marocco}, {Marshall}, {Martin Polo}, {Mart{\'\i}n-Fleitas}, {Masip},
  {Massari}, {Mastrobuono-Battisti}, {Mazeh}, {McMillan}, {Messina},
  {Michalik}, {Millar}, {Mints}, {Molina}, {Molinaro}, {Moln{\'a}r},
  {Montegriffo}, {Mor}, {Morbidelli}, {Morel}, {Morris}, {Mulone}, {Munoz},
  {Muraveva}, {Murphy}, {Musella}, {Noval}, {Ord{\'e}novic}, {Orr{\`u}},
  {Osinde}, {Pagani}, {Pagano}, {Palaversa}, {Palicio}, {Panahi}, {Pawlak},
  {Pe{\~n}alosa Esteller}, {Penttil{\"a}}, {Piersimoni}, {Pineau}, {Plachy},
  {Plum}, {Poggio}, {Poretti}, {Poujoulet}, {Pr{\v{s}}a}, {Pulone}, {Racero},
  {Ragaini}, {Rainer}, {Raiteri}, {Rambaux}, {Ramos}, {Ramos-Lerate}, {Re
  Fiorentin}, {Regibo}, {Reyl{\'e}}, {Ripepi}, {Riva}, {Rixon}, {Robichon},
  {Robin}, {Roelens}, {Rohrbasser}, {Romero-G{\'o}mez}, {Rowell}, {Royer},
  {Rybicki}, {Sadowski}, {Sagrist{\`a} Sell{\'e}s}, {Sahlmann}, {Salgado},
  {Salguero}, {Samaras}, {Sanchez Gimenez}, {Sanna}, {Santove{\~n}a},
  {Sarasso}, {Schultheis}, {Sciacca}, {Segol}, {Segovia}, {S{\'e}gransan},
  {Semeux}, {Shahaf}, {Siddiqui}, {Siebert}, {Siltala}, {Slezak}, {Smart},
  {Solano}, {Solitro}, {Souami}, {Souchay}, {Spagna}, {Spoto}, {Steele},
  {Steidelm{\"u}ller}, {Stephenson}, {S{\"u}veges}, {Szabados}, {Szegedi-Elek},
  {Taris}, {Tauran}, {Taylor}, {Teixeira}, {Thuillot}, {Tonello}, {Torra},
  {Torra}, {Turon}, {Unger}, {Vaillant}, {van Dillen}, {Vanel}, {Vecchiato},
  {Viala}, {Vicente}, {Voutsinas}, {Weiler}, {Wevers}, {Wyrzykowski}, {Yoldas},
  {Yvard}, {Zhao}, {Zorec}, {Zucker}, {Zurbach}, \& {Zwitter}}]{GaiaEDR3}
{Gaia Collaboration}, {Brown}, A.~G.~A., {Vallenari}, A., {et~al.} 2021, \aap,
  649, A1

\bibitem[{{Gandolfi} {et~al.}(2017){Gandolfi}, {Barrag{\'a}n}, {Hatzes},
  {Fridlund}, {Fossati}, {Donati}, {Johnson}, {Nowak}, {Prieto-Arranz},
  {Albrecht}, {Dai}, {Deeg}, {Endl}, {Grziwa}, {Hjorth}, {Korth}, {Nespral},
  {Saario}, {Smith}, {Antoniciello}, {Alarcon}, {Bedell}, {Blay}, {Brems},
  {Cabrera}, {Csizmadia}, {Cusano}, {Cochran}, {Eigm{\"u}ller}, {Erikson},
  {Gonz{\'a}lez Hern{\'a}ndez}, {Guenther}, {Hirano}, {Su{\'a}rez
  Mascare{\~n}o}, {Narita}, {Palle}, {Parviainen}, {P{\"a}tzold}, {Persson},
  {Rauer}, {Saviane}, {Schmidtobreick}, {Van Eylen}, {Winn}, \&
  {Zakhozhay}}]{2017AJ....154..123G}
{Gandolfi}, D., {Barrag{\'a}n}, O., {Hatzes}, A.~P., {et~al.} 2017, \aj, 154,
  123

\bibitem[{{Ginzburg} {et~al.}(2018){Ginzburg}, {Schlichting}, \&
  {Sari}}]{Ginzburg+2018}
{Ginzburg}, S., {Schlichting}, H.~E., \& {Sari}, R. 2018, \mnras, 476, 759

\bibitem[{{Guenther} {et~al.}(2017){Guenther}, {Barrag{\'a}n}, {Dai},
  {Gandolfi}, {Hirano}, {Fridlund}, {Fossati}, {Chau}, {Helled}, {Korth},
  {Prieto-Arranz}, {Nespral}, {Antoniciello}, {Deeg}, {Hjorth}, {Grziwa},
  {Albrecht}, {Hatzes}, {Rauer}, {Csizmadia}, {Smith}, {Cabrera}, {Narita},
  {Arriagada}, {Burt}, {Butler}, {Cochran}, {Crane}, {Eigm{\"u}ller},
  {Erikson}, {Johnson}, {Kiilerich}, {Kubyshkina}, {Palle}, {Persson},
  {P{\"a}tzold}, {Sabotta}, {Sato}, {Shectman}, {Teske}, {Thompson}, {Van
  Eylen}, {Nowak}, {Vanderburg}, {Winn}, \& {Wittenmyer}}]{Guenther2017}
{Guenther}, E.~W., {Barrag{\'a}n}, O., {Dai}, F., {et~al.} 2017, \aap, 608, A93

\bibitem[{{Guerrero} {et~al.}(2021){Guerrero}, {Seager}, {Huang}, {Vanderburg},
  {Garcia Soto}, {Mireles}, {Hesse}, {Fong}, {Glidden}, {Shporer}, {Latham},
  {Collins}, {Quinn}, {Burt}, {Dragomir}, {Crossfield}, {Vanderspek},
  {Fausnaugh}, {Burke}, {Ricker}, {Daylan}, {Essack}, {G{\"u}nther}, {Osborn},
  {Pepper}, {Rowden}, {Sha}, {Villanueva}, {Yahalomi}, {Yu}, {Ballard},
  {Batalha}, {Berardo}, {Chontos}, {Dittmann}, {Esquerdo}, {Mikal-Evans},
  {Jayaraman}, {Krishnamurthy}, {Louie}, {Mehrle}, {Niraula}, {Rackham},
  {Rodriguez}, {Rowden}, {Sousa-Silva}, {Watanabe}, {Wong}, {Zhan},
  {Zivanovic}, {Christiansen}, {Ciardi}, {Swain}, {Lund}, {Mullally},
  {Fleming}, {Rodriguez}, {Boyd}, {Quintana}, {Barclay}, {Col{\'o}n},
  {Rinehart}, {Schlieder}, {Clampin}, {Jenkins}, {Twicken}, {Caldwell},
  {Coughlin}, {Henze}, {Lissauer}, {Morris}, {Rose}, {Smith}, {Tenenbaum},
  {Ting}, {Wohler}, {Bakos}, {Bean}, {Berta-Thompson}, {Bieryla}, {Bouma},
  {Buchhave}, {Butler}, {Charbonneau}, {Doty}, {Ge}, {Holman}, {Howard},
  {Kaltenegger}, {Kane}, {Kjeldsen}, {Kreidberg}, {Lin}, {Minsky}, {Narita},
  {Paegert}, {P{\'a}l}, {Palle}, {Sasselov}, {Spencer}, {Sozzetti}, {Stassun},
  {Torres}, {Udry}, \& {Winn}}]{Guerrero2021}
{Guerrero}, N.~M., {Seager}, S., {Huang}, C.~X., {et~al.} 2021, \apjs, 254, 39

\bibitem[{{Gupta} \& {Schlichting}(2019)}]{Gupta2019}
{Gupta}, A. \& {Schlichting}, H.~E. 2019, \mnras, 487, 24

\bibitem[{{Gupta} \& {Schlichting}(2020)}]{gupta2020}
{Gupta}, A. \& {Schlichting}, H.~E. 2020, \mnras, 493, 792

\bibitem[{{Hatzes}(2019)}]{2019dmde.book.....H}
{Hatzes}, A.~P. 2019, {The Doppler Method for the Detection of Exoplanets}
  (Institute of Physics Publishing), doi=10.1088/2514--3433/ab46a3

\bibitem[{{Hatzes} {et~al.}(2010){Hatzes}, {Dvorak}, {Wuchterl}, {Guterman},
  {Hartmann}, {Fridlund}, {Gandolfi}, {Guenther}, \&
  {P{\"a}tzold}}]{2010A&A...520A..93H}
{Hatzes}, A.~P., {Dvorak}, R., {Wuchterl}, G., {et~al.} 2010, \aap, 520, A93

\bibitem[{{Hayward} {et~al.}(2001){Hayward}, {Brandl}, {Pirger}, {Blacken},
  {Gull}, {Schoenwald}, \& {Houck}}]{hayward2001}
{Hayward}, T.~L., {Brandl}, B., {Pirger}, B., {et~al.} 2001, \pasp, 113, 105

\bibitem[{{Haywood} {et~al.}(2014){Haywood}, {Collier Cameron}, {Queloz},
  {Barros}, {Deleuil}, {Fares}, {Gillon}, {Lanza}, {Lovis}, {Moutou}, {Pepe},
  {Pollacco}, {Santerne}, {S{\'e}gransan}, \& {Unruh}}]{2014MNRAS.443.2517H}
{Haywood}, R.~D., {Collier Cameron}, A., {Queloz}, D., {et~al.} 2014, \mnras,
  443, 2517

\bibitem[{{Henry} {et~al.}(2018){Henry}, {Jao}, {Winters}, {Dieterich},
  {Finch}, {Ianna}, {Riedel}, {Silverstein}, {Subasavage}, \&
  {Vrijmoet}}]{Henry2018}
{Henry}, T.~J., {Jao}, W.-C., {Winters}, J.~G., {et~al.} 2018, \aj, 155, 265

\bibitem[{{Howard} {et~al.}(2010){Howard}, {Johnson}, {Marcy}, {Fischer},
  {Wright}, {Bernat}, {Henry}, {Peek}, {Isaacson}, {Apps}, {Endl}, {Cochran},
  {Valenti}, {Anderson}, \& {Piskunov}}]{Howard2010}
{Howard}, A.~W., {Johnson}, J.~A., {Marcy}, G.~W., {et~al.} 2010, \apj, 721,
  1467

\bibitem[{{Howard} {et~al.}(2012){Howard}, {Marcy}, {Bryson}, {Jenkins},
  {Rowe}, {Batalha}, {Borucki}, {Koch}, {Dunham}, {Gautier}, {Van Cleve},
  {Cochran}, {Latham}, {Lissauer}, {Torres}, {Brown}, {Gilliland}, {Buchhave},
  {Caldwell}, {Christensen-Dalsgaard}, {Ciardi}, {Fressin}, {Haas}, {Howell},
  {Kjeldsen}, {Seager}, {Rogers}, {Sasselov}, {Steffen}, {Basri},
  {Charbonneau}, {Christiansen}, {Clarke}, {Dupree}, {Fabrycky}, {Fischer},
  {Ford}, {Fortney}, {Tarter}, {Girouard}, {Holman}, {Johnson}, {Klaus},
  {Machalek}, {Moorhead}, {Morehead}, {Ragozzine}, {Tenenbaum}, {Twicken},
  {Quinn}, {Isaacson}, {Shporer}, {Lucas}, {Walkowicz}, {Welsh}, {Boss},
  {Devore}, {Gould}, {Smith}, {Morris}, {Prsa}, {Morton}, {Still}, {Thompson},
  {Mullally}, {Endl}, \& {MacQueen}}]{Howard2012}
{Howard}, A.~W., {Marcy}, G.~W., {Bryson}, S.~T., {et~al.} 2012, \apjs, 201, 15

\bibitem[{{Howell} {et~al.}(2011){Howell}, {Everett}, {Sherry}, {Horch}, \&
  {Ciardi}}]{2011AJ....142...19H}
{Howell}, S.~B., {Everett}, M.~E., {Sherry}, W., {Horch}, E., \& {Ciardi},
  D.~R. 2011, \aj, 142, 19

\bibitem[{{Hsu} {et~al.}(2020){Hsu}, {Ford}, \& {Terrien}}]{Hsu2020}
{Hsu}, D.~C., {Ford}, E.~B., \& {Terrien}, R. 2020, \mnras, 498, 2249

\bibitem[{{Ida} \& {Lin}(2010)}]{Ida2010}
{Ida}, S. \& {Lin}, D.~N.~C. 2010, \apj, 719, 810

\bibitem[{{Jeffers} {et~al.}(2020){Jeffers}, {Dreizler}, {Barnes}, {Haswell},
  {Nelson}, {Rodr{\'\i}guez}, {L{\'o}pez-Gonz‧lez}, {Morales}, {Luque},
  {Zechmeister}, {Vogt}, {Jenkins}, {Palle}, {Berdi {\~n}as}, {Coleman},
  {D{\'\i}az}, {Ribas}, {Jones}, {Butler}, {Tinney}, {Bailey}, {Carter},
  {O'Toole}, {Wittenmyer}, {Crane}, {Feng}, {Shectman}, {Teske}, {Reiners},
  {Amado}, \& {Anglada-Escud{\'e}}}]{Jeffers2020}
{Jeffers}, S.~V., {Dreizler}, S., {Barnes}, J.~R., {et~al.} 2020, Science, 368,
  1477

\bibitem[{{Jeffers} {et~al.}(2018){Jeffers}, {Sch{\"o}fer}, {Lamert},
  {Reiners}, {Montes}, {Caballero}, {Cort{\'e}s-Contreras}, {Marvin},
  {Passegger}, {Zechmeister}, {Quirrenbach}, {Alonso-Floriano}, {Amado},
  {Bauer}, {Casal}, {Alonso}, {Herrero}, {Morales}, {Mundt}, {Ribas}, \&
  {Sarmiento}}]{Jeffers18}
{Jeffers}, S.~V., {Sch{\"o}fer}, P., {Lamert}, A., {et~al.} 2018, \aap, 614,
  A76

\bibitem[{{Jenkins} {et~al.}(2016){Jenkins}, {Twicken}, {McCauliff},
  {Campbell}, {Sanderfer}, {Lung}, {Mansouri-Samani}, {Girouard}, {Tenenbaum},
  {Klaus}, {Smith}, {Caldwell}, {Chacon}, {Henze}, {Heiges}, {Latham},
  {Morgan}, {Swade}, {Rinehart}, \& {Vanderspek}}]{SPOC}
{Jenkins}, J.~M., {Twicken}, J.~D., {McCauliff}, S., {et~al.} 2016, in
  \procspie, Vol. 9913, Software and Cyberinfrastructure for Astronomy IV,
  99133E

\bibitem[{{Kaminski} {et~al.}(2018){Kaminski}, {Trifonov}, {Caballero},
  {Quirrenbach}, {Ribas}, {Reiners}, {Amado}, {Zechmeister}, {Dreizler},
  {Perger}, {Tal-Or}, {Bonfils}, {Mayor}, {Astudillo-Defru}, {Bauer},
  {B{\'e}jar}, {Cifuentes}, {Colom{\'e}}, {Cort{\'e}s-Contreras}, {Delfosse},
  {D{\'{\i}}ez-Alonso}, {Forveille}, {Guenther}, {Hatzes}, {Henning},
  {Jeffers}, {K{\"u}rster}, {Lafarga}, {Luque}, {Mandel}, {Montes}, {Morales},
  {Passegger}, {Pedraz}, {Reffert}, {Sadegi}, {Schweitzer}, {Seifert}, {Stahl},
  \& {Udry}}]{Kaminski18}
{Kaminski}, A., {Trifonov}, T., {Caballero}, J.~A., {et~al.} 2018, \aap, 618,
  A115

\bibitem[{{Kasting}(1998)}]{Kasting1998}
{Kasting}, J. 1998, in American Astronomical Society Meeting Abstracts, Vol.
  193, American Astronomical Society Meeting Abstracts, 50.03

\bibitem[{{Kasting}(2010)}]{Kasting2010}
{Kasting}, J.~F. 2010, in Astronomical Society of the Pacific Conference
  Series, Vol. 430, Pathways Towards Habitable Planets, ed. V.~{Coud{\'e} du
  Foresto}, D.~M. {Gelino}, \& I.~{Ribas}, 3

\bibitem[{{Kasting}(2021)}]{Kasting2021}
{Kasting}, J.~F. 2021, in ExoFrontiers; Big Questions in Exoplanetary Science,
  ed. N.~{Madhusudhan}, 22--1

\bibitem[{{Kasting} \& {Harman}(2013)}]{Kasting2013}
{Kasting}, J.~F. \& {Harman}, C.~E. 2013, \nat, 504, 221

\bibitem[{{Kasting} {et~al.}(1993){Kasting}, {Whitmire}, \&
  {Reynolds}}]{Kasting1993}
{Kasting}, J.~F., {Whitmire}, D.~P., \& {Reynolds}, R.~T. 1993, \icarus, 101,
  108

\bibitem[{{Kempton} {et~al.}(2018){Kempton}, {Bean}, {Louie}, {Deming}, {Koll},
  {Mansfield}, {Christiansen}, {L{\'o}pez-Morales}, {Swain}, {Zellem},
  {Ballard}, {Barclay}, {Barstow}, {Batalha}, {Beatty}, {Berta-Thompson},
  {Birkby}, {Buchhave}, {Charbonneau}, {Cowan}, {Crossfield}, {de Val-Borro},
  {Doyon}, {Dragomir}, {Gaidos}, {Heng}, {Hu}, {Kane}, {Kreidberg}, {Mallonn},
  {Morley}, {Narita}, {Nascimbeni}, {Pall{\'e}}, {Quintana}, {Rauscher},
  {Seager}, {Shkolnik}, {Sing}, {Sozzetti}, {Stassun}, {Valenti}, \& {von
  Essen}}]{Kempton2018}
{Kempton}, E. M.~R., {Bean}, J.~L., {Louie}, D.~R., {et~al.} 2018, \pasp, 130,
  114401

\bibitem[{{Kipping}(2013)}]{Kipping2013}
{Kipping}, D.~M. 2013, \mnras, 435, 2152

\bibitem[{{Kirkpatrick} {et~al.}(2019){Kirkpatrick}, {Martin}, {Smart},
  {Cayago}, {Beichman}, {Marocco}, {Gelino}, {Faherty}, {Cushing}, {Schneider},
  {Mace}, {Tinney}, {Wright}, {Lowrance}, {Ingalls}, {Vrba}, {Munn}, {Dahm}, \&
  {McLean}}]{kirkpatrick2019}
{Kirkpatrick}, J.~D., {Martin}, E.~C., {Smart}, R.~L., {et~al.} 2019, \apjs,
  240, 19

\bibitem[{{Kite} \& {Barnett}(2020)}]{kite2020}
{Kite}, E.~S. \& {Barnett}, M.~N. 2020, Proceedings of the National Academy of
  Science, 117, 18264

\bibitem[{{Kite} {et~al.}(2019){Kite}, {Fegley}, {Schaefer}, \&
  {Ford}}]{kite2019}
{Kite}, E.~S., {Fegley}, Bruce, J., {Schaefer}, L., \& {Ford}, E.~B. 2019,
  \apjl, 887, L33

\bibitem[{{Kopparapu} {et~al.}(2014){Kopparapu}, {Ramirez}, {SchottelKotte},
  {Kasting}, {Domagal-Goldman}, \& {Eymet}}]{Kopparapu2014}
{Kopparapu}, R.~K., {Ramirez}, R.~M., {SchottelKotte}, J., {et~al.} 2014,
  \apjl, 787, L29

\bibitem[{{Kossakowski} {et~al.}(2022){Kossakowski}, {Henning}, \&
  {Kuerster}}]{Kossakowski2022}
{Kossakowski}, D., {Henning}, T., \& {Kuerster}, M. 2022, \aap

\bibitem[{{Kossakowski} {et~al.}(2021){Kossakowski}, {Kemmer}, {Bluhm},
  {Stock}, {Caballero}, {B{\'e}jar}, {Guill{\'e}n}, {Lodieu}, {Collins},
  {Oshagh}, {Schlecker}, {Espinoza}, {Pall{\'e}}, {Henning}, {Kreidberg},
  {K{\"u}rster}, {Amado}, {Anderson}, {Morales}, {Cartwright}, {Charbonneau},
  {Chaturvedi}, {Cifuentes}, {Conti}, {Cort{\'e}s-Contreras}, {Dreizler},
  {Galad{\'\i}-Enr{\'\i}quez}, {Guerra}, {Hart}, {Hellier}, {Henze}, {Herrero},
  {Jeffers}, {Jenkins}, {Jensen}, {Kaminski}, {Kielkopf}, {Kunimoto},
  {Lafarga}, {Latham}, {Lillo-Box}, {Luque}, {Molaverdikhani}, {Montes},
  {Morello}, {Morgan}, {Nowak}, {Pavlov}, {Perger}, {Quintana}, {Quirrenbach},
  {Reffert}, {Reiners}, {Ricker}, {Ribas}, {L{\'o}pez}, {Osorio}, {Seager},
  {Sch{\"o}fer}, {Schweitzer}, {Trifonov}, {Vanaverbeke}, {Vanderspek}, {West},
  {Winn}, \& {Zechmeister}}]{Kossakowski2021}
{Kossakowski}, D., {Kemmer}, J., {Bluhm}, P., {et~al.} 2021, \aap, 656, A124

\bibitem[{{Kov{\'a}cs} {et~al.}(2002){Kov{\'a}cs}, {Zucker}, \&
  {Mazeh}}]{box_least}
{Kov{\'a}cs}, G., {Zucker}, S., \& {Mazeh}, T. 2002, \aap, 391, 369

\bibitem[{{Kreidberg}(2015)}]{Kreidberg2015}
{Kreidberg}, L. 2015, \pasp, 127, 1161

\bibitem[{{K{\"u}rster} {et~al.}(2003){K{\"u}rster}, {Endl}, {Rouesnel}, {Els},
  {Kaufer}, {Brillant}, {Hatzes}, {Saar}, \& {Cochran}}]{Kuerster2003}
{K{\"u}rster}, M., {Endl}, M., {Rouesnel}, F., {et~al.} 2003, in ESA Special
  Publication, Vol. 539, Earths: DARWIN/TPF and the Search for Extrasolar
  Terrestrial Planets, ed. M.~{Fridlund}, T.~{Henning}, \& H.~{Lacoste},
  485--489

\bibitem[{{Lammer} {et~al.}(2019){Lammer}, {Spro{\ss}}, {Grenfell}, {Scherf},
  {Fossati}, {Lendl}, \& {Cubillos}}]{Lammer2019}
{Lammer}, H., {Spro{\ss}}, L., {Grenfell}, J.~L., {et~al.} 2019, Astrobiology,
  19, 927

\bibitem[{{Lammer} {et~al.}(2014){Lammer}, {St{\"o}kl}, {Erkaev}, {Dorfi},
  {Odert}, {G{\"u}del}, {Kulikov}, {Kislyakova}, \& {Leitzinger}}]{Lammer2014}
{Lammer}, H., {St{\"o}kl}, A., {Erkaev}, N.~V., {et~al.} 2014, \mnras, 439,
  3225

\bibitem[{{L{\'e}pine} \& {Gaidos}(2011)}]{Lepine2011}
{L{\'e}pine}, S. \& {Gaidos}, E. 2011, \aj, 142, 138

\bibitem[{{L{\'e}pine} \& {Shara}(2005)}]{Lepine2005}
{L{\'e}pine}, S. \& {Shara}, M.~M. 2005, \aj, 129, 1483

\bibitem[{{Li} {et~al.}(2019){Li}, {Tenenbaum}, {Twicken}, {Burke}, {Jenkins},
  {Quintana}, {Rowe}, \& {Seader}}]{Li2019}
{Li}, J., {Tenenbaum}, P., {Twicken}, J.~D., {et~al.} 2019, \pasp, 131, 024506

\bibitem[{{Lindegren} {et~al.}(2018){Lindegren}, {Hern{\'a}ndez}, {Bombrun},
  {Klioner}, {Bastian}, {Ramos-Lerate}, {de Torres}, {Steidelm{\"u}ller},
  {Stephenson}, {Hobbs}, {Lammers}, {Biermann}, {Geyer}, {Hilger}, {Michalik},
  {Stampa}, {McMillan}, {Casta{\~n}eda}, {Clotet}, {Comoretto}, {Davidson},
  {Fabricius}, {Gracia}, {Hambly}, {Hutton}, {Mora}, {Portell}, {van Leeuwen},
  {Abbas}, {Abreu}, {Altmann}, {Andrei}, {Anglada}, {Balaguer-N{\'u}{\~n}ez},
  {Barache}, {Becciani}, {Bertone}, {Bianchi}, {Bouquillon}, {Bourda},
  {Br{\"u}semeister}, {Bucciarelli}, {Busonero}, {Buzzi}, {Cancelliere},
  {Carlucci}, {Charlot}, {Cheek}, {Crosta}, {Crowley}, {de Bruijne}, {de
  Felice}, {Drimmel}, {Esquej}, {Fienga}, {Fraile}, {Gai}, {Garralda},
  {Gonz{\'a}lez-Vidal}, {Guerra}, {Hauser}, {Hofmann}, {Holl}, {Jordan},
  {Lattanzi}, {Lenhardt}, {Liao}, {Licata}, {Lister}, {L{\"o}ffler},
  {Marchant}, {Martin-Fleitas}, {Messineo}, {Mignard}, {Morbidelli}, {Poggio},
  {Riva}, {Rowell}, {Salguero}, {Sarasso}, {Sciacca}, {Siddiqui}, {Smart},
  {Spagna}, {Steele}, {Taris}, {Torra}, {van Elteren}, {van Reeven}, \&
  {Vecchiato}}]{2018A&A...616A...2L}
{Lindegren}, L., {Hern{\'a}ndez}, J., {Bombrun}, A., {et~al.} 2018, \aap, 616,
  A2

\bibitem[{{Linsky} \& {G{\"u}del}(2015)}]{Linsky2015}
{Linsky}, J.~L. \& {G{\"u}del}, M. 2015, in Astrophysics and Space Science
  Library, Vol. 411, Characterizing Stellar and Exoplanetary Environments, ed.
  H.~{Lammer} \& M.~{Khodachenko}, 3

\bibitem[{{Lodieu} {et~al.}(2019){Lodieu}, {P{\'e}rez-Garrido}, {Smart}, \&
  {Silvotti}}]{Lodieu2019}
{Lodieu}, N., {P{\'e}rez-Garrido}, A., {Smart}, R.~L., \& {Silvotti}, R. 2019,
  \aap, 628, A66

\bibitem[{{L\'opez} \& {Fortney}(2013)}]{Lopez2013}
{L\'opez}, E.~D. \& {Fortney}, J.~J. 2013, \apj, 776, 2

\bibitem[{{Lopez} {et~al.}(2012){Lopez}, {Fortney}, \& {Miller}}]{Lopez2012}
{Lopez}, E.~D., {Fortney}, J.~J., \& {Miller}, N. 2012, \apj, 761, 59

\bibitem[{{Marfil} {et~al.}(2021){Marfil}, {Tabernero}, {Montes}, {Caballero},
  {L{\'a}zaro}, {Gonz{\'a}lez Hern{\'a}ndez}, {Nagel}, {Passegger},
  {Schweitzer}, {Ribas}, {Reiners}, {Quirrenbach}, {Amado}, {Cifuentes},
  {Cort{\'e}s-Contreras}, {Dreizler}, {Duque-Arribas},
  {Galad{\'\i}-Enr{\'\i}quez}, {Henning}, {Jeffers}, {Kaminski}, {K{\"u}rster},
  {Lafarga}, {L{\'o}pez-Gallifa}, {Morales}, {Shan}, \&
  {Zechmeister}}]{Marfil2021}
{Marfil}, E., {Tabernero}, H.~M., {Montes}, D., {et~al.} 2021, \aap, 656, A162

\bibitem[{{Mart{\'\i}nez-Rodr{\'\i}guez}
  {et~al.}(2019){Mart{\'\i}nez-Rodr{\'\i}guez}, {Caballero}, {Cifuentes},
  {Piro}, \& {Barnes}}]{Martinez-Rodriguez2019}
{Mart{\'\i}nez-Rodr{\'\i}guez}, H., {Caballero}, J.~A., {Cifuentes}, C.,
  {Piro}, A.~L., \& {Barnes}, R. 2019, \apj, 887, 261

\bibitem[{{McCully} {et~al.}(2018){McCully}, {Volgenau}, {Harbeck}, {Lister},
  {Saunders}, {Turner}, {Siiverd}, \& {Bowman}}]{McCully:2018}
{McCully}, C., {Volgenau}, N.~H., {Harbeck}, D.-R., {et~al.} 2018, in Society
  of Photo-Optical Instrumentation Engineers (SPIE) Conference Series, Vol.
  10707, \procspie, 107070K

\bibitem[{{Molaverdikhani} {et~al.}(2019{\natexlab{a}}){Molaverdikhani},
  {Henning}, \& {Molli{\`e}re}}]{molaverdikhani2019b}
{Molaverdikhani}, K., {Henning}, T., \& {Molli{\`e}re}, P. 2019{\natexlab{a}},
  \apj, 883, 194

\bibitem[{{Molaverdikhani} {et~al.}(2019{\natexlab{b}}){Molaverdikhani},
  {Henning}, \& {Molli{\`e}re}}]{molaverdikhani2019a}
{Molaverdikhani}, K., {Henning}, T., \& {Molli{\`e}re}, P. 2019{\natexlab{b}},
  \apj, 873, 32

\bibitem[{{Molaverdikhani} {et~al.}(2020){Molaverdikhani}, {Henning}, \&
  {Molli{\`e}re}}]{molaverdikhani2020}
{Molaverdikhani}, K., {Henning}, T., \& {Molli{\`e}re}, P. 2020, \apj, 899, 53

\bibitem[{{Molli{\`e}re} {et~al.}(2019){Molli{\`e}re}, {Wardenier}, {van
  Boekel}, {Henning}, {Molaverdikhani}, \& {Snellen}}]{molliere2019}
{Molli{\`e}re}, P., {Wardenier}, J.~P., {van Boekel}, R., {et~al.} 2019, \aap,
  627, A67

\bibitem[{{Montes} {et~al.}(2018){Montes}, {Gonz{\'a}lez-Peinado}, {Tabernero},
  {Caballero}, {Marfil}, {Alonso-Floriano}, {Cort{\'e}s-Contreras},
  {Gonz{\'a}lez Hern{\'a}ndez}, {Klutsch}, \& {Moreno-J{\'o}dar}}]{Montes2018}
{Montes}, D., {Gonz{\'a}lez-Peinado}, R., {Tabernero}, H.~M., {et~al.} 2018,
  \mnras, 479, 1332

\bibitem[{{Morello} {et~al.}(2021){Morello}, {Zingales}, {Martin-Lagarde},
  {Gastaud}, \& {Lagage}}]{morello2021}
{Morello}, G., {Zingales}, T., {Martin-Lagarde}, M., {Gastaud}, R., \&
  {Lagage}, P.-O. 2021, \aj, 161, 174

\bibitem[{{Mortier} \& {Collier Cameron}(2017)}]{sBGLS2017}
{Mortier}, A. \& {Collier Cameron}, A. 2017, \aap, 601, A110

\bibitem[{{Mortier} {et~al.}(2015){Mortier}, {Faria}, {Correia}, {Santerne}, \&
  {Santos}}]{BGLS2015}
{Mortier}, A., {Faria}, J.~P., {Correia}, C.~M., {Santerne}, A., \& {Santos},
  N.~C. 2015, \aap, 573, A101

\bibitem[{{Mugnai} {et~al.}(2021){Mugnai}, {Modirrousta-Galian}, {Edwards},
  {Changeat}, {Bouwman}, {Morello}, {Al-Refaie}, {Baeyens}, {Bieger}, {Blain},
  {Gressier}, {Guilluy}, {Jaziri}, {Kiefer}, {Morvan}, {Pluriel}, {Poveda},
  {Skaf}, {Whiteford}, {Wright}, {Yip}, {Zingales}, {Charnay}, {Drossart},
  {Leconte}, {Venot}, {Waldmann}, \& {Beaulieu}}]{mugnai2021}
{Mugnai}, L.~V., {Modirrousta-Galian}, D., {Edwards}, B., {et~al.} 2021, \aj,
  161, 284

\bibitem[{{Narita} {et~al.}(2019){Narita}, {Fukui}, {Kusakabe}, {Watanabe},
  {Palle}, {Parviainen}, {Monta{\~n}{\'e}s-Rodr{\'\i}guez}, {Murgas},
  {Monelli}, {Aguiar}, {Perez Prieto}, {Oscoz}, {de Leon}, {Mori}, {Tamura},
  {Yamamuro}, {B{\'e}jar}, {Crouzet}, {Hidalgo}, {Klagyivik}, {Luque}, \&
  {Nishiumi}}]{Narita2019}
{Narita}, N., {Fukui}, A., {Kusakabe}, N., {et~al.} 2019, Journal of
  Astronomical Telescopes, Instruments, and Systems, 5, 015001

\bibitem[{{Niraula} {et~al.}(2017){Niraula}, {Redfield}, {Dai}, {Barrag{\'a}n},
  {Gandolfi}, {Cauley}, {Hirano}, {Korth}, {Smith}, {Prieto-Arranz}, {Grziwa},
  {Fridlund}, {Persson}, {Justesen}, {Winn}, {Albrecht}, {Cochran},
  {Csizmadia}, {Duvvuri}, {Endl}, {Hatzes}, {Livingston}, {Narita}, {Nespral},
  {Nowak}, {P{\"a}tzold}, {Palle}, \& {Van Eylen}}]{2017AJ....154..266N}
{Niraula}, P., {Redfield}, S., {Dai}, F., {et~al.} 2017, \aj, 154, 266

\bibitem[{{Nowak} {et~al.}(2020){Nowak}, {Luque}, {Parviainen}, {Pall{\'e}},
  {Molaverdikhani}, {B{\'e}jar}, {Lillo-Box}, {Rodr{\'\i}guez-L{\'o}pez},
  {Caballero}, {Zechmeister}, {Passegger}, {Cifuentes}, {Schweitzer}, {Narita},
  {Cale}, {Espinoza}, {Murgas}, {Hidalgo}, {Zapatero Osorio}, {Pozuelos},
  {Aceituno}, {Amado}, {Barkaoui}, {Barrado}, {Bauer}, {Benkhaldoun},
  {Caldwell}, {Casasayas Barris}, {Chaturvedi}, {Chen}, {Collins}, {Collins},
  {Cort{\'e}s-Contreras}, {Crossfield}, {de Le{\'o}n}, {D{\'\i}ez Alonso},
  {Dreizler}, {El Mufti}, {Esparza-Borges}, {Essack}, {Fukui}, {Gaidos},
  {Gillon}, {Gonzales}, {Guerra}, {Hatzes}, {Henning}, {Herrero}, {Hesse},
  {Hirano}, {Howell}, {Jeffers}, {Jehin}, {Jenkins}, {Kaminski}, {Kemmer},
  {Kielkopf}, {Kossakowski}, {Kotani}, {K{\"u}rster}, {Lafarga}, {Latham},
  {Law}, {Lissauer}, {Lodieu}, {Madrigal-Aguado}, {Mann}, {Massey}, {Matson},
  {Matthews}, {Monta{\~n}{\'e}s-Rodr{\'\i}guez}, {Montes}, {Morales}, {Mori},
  {Nagel}, {Oshagh}, {Pedraz}, {Plavchan}, {Pollacco}, {Quirrenbach},
  {Reffert}, {Reiners}, {Ribas}, {Ricker}, {Rose}, {Schlecker}, {Schlieder},
  {Seager}, {Stangret}, {Stock}, {Tamura}, {Tanner}, {Teske}, {Trifonov},
  {Twicken}, {Vanderspek}, {Watanabe}, {Wittrock}, {Ziegler}, \&
  {Zohrabi}}]{2020A&A...642A.173N}
{Nowak}, G., {Luque}, R., {Parviainen}, H., {et~al.} 2020, \aap, 642, A173

\bibitem[{{Oshagh} {et~al.}(2017){Oshagh}, {Santos}, {Figueira}, {Barros},
  {Donati}, {Adibekyan}, {Faria}, {Watson}, {Cegla}, {Dumusque}, {H{\'e}brard},
  {Demangeon}, {Dreizler}, {Boisse}, {Deleuil}, {Bonfils}, {Pepe}, \&
  {Udry}}]{Oshagh2017}
{Oshagh}, M., {Santos}, N.~C., {Figueira}, P., {et~al.} 2017, \aap, 606, A107

\bibitem[{{Owen} \& {Campos Estrada}(2020)}]{Owen2020}
{Owen}, J.~E. \& {Campos Estrada}, B. 2020, \mnras, 491, 5287

\bibitem[{{Owen} \& {Wu}(2013)}]{Owen2013}
{Owen}, J.~E. \& {Wu}, Y. 2013, \apj, 775, 105

\bibitem[{{Owen} \& {Wu}(2016)}]{Owen2016}
{Owen}, J.~E. \& {Wu}, Y. 2016, \apj, 817, 107

\bibitem[{Parviainen {et~al.}(2019)Parviainen, Tingley, Deeg, Palle, Alonso,
  {Montanes Rodriguez}, Murgas, Narita, Fukui, Watanabe, Kusakabe, Tamura,
  Nishiumi, Prieto-Arranz, Klagyivik, B{\'{e}}jar, Crouzet, Mori, {Hidalgo
  Soto}, {Casasayas Barris}, \& Luque}]{Parviainen2019}
Parviainen, H., Tingley, B., Deeg, H.~J., {et~al.} 2019, Astronomy and
  Astrophysics, 630, A89

\bibitem[{{Quirrenbach} {et~al.}(2014){Quirrenbach}, {Amado}, {Caballero},
  {Mundt}, {Reiners}, {Ribas}, {Seifert}, {Abril}, {Aceituno},
  {Alonso-Floriano}, {Ammler-von Eiff}, {Antona Jim{\'e}nez},
  {Anwand-Heerwart}, {Azzaro}, {Bauer}, {Barrado}, {Becerril}, {B{\'e}jar},
  {Ben{\'\i}tez}, {Berdi{\~n}as}, {C{\'a}rdenas}, {Casal}, {Claret},
  {Colom{\'e}}, {Cort{\'e}s-Contreras}, {Czesla}, {Doellinger}, {Dreizler},
  {Feiz}, {Fern{\'a}ndez}, {Galad{\'\i}}, {G{\'a}lvez-Ortiz},
  {Garc{\'\i}a-Piquer}, {Garc{\'\i}a-Vargas}, {Garrido}, {Gesa}, {G{\'o}mez
  Galera}, {Gonz{\'a}lez {\'A}lvarez}, {Gonz{\'a}lez Hern{\'a}ndez},
  {Gr{\"o}zinger}, {Gu{\`a}rdia}, {Guenther}, {de Guindos},
  {Guti{\'e}rrez-Soto}, {Hagen}, {Hatzes}, {Hauschildt}, {Helmling}, {Henning},
  {Hermann}, {Hern{\'a}ndez Casta{\~n}o}, {Herrero}, {Hidalgo}, {Holgado},
  {Huber}, {Huber}, {Jeffers}, {Joergens}, {de Juan}, {Kehr}, {Klein},
  {K{\"u}rster}, {Lamert}, {Lalitha}, {Laun}, {Lemke}, {Lenzen}, {L{\'o}pez del
  Fresno}, {L{\'o}pez Mart{\'\i}}, {L{\'o}pez-Santiago}, {Mall}, {Mandel},
  {Mart{\'\i}n}, {Mart{\'\i}n-Ruiz}, {Mart{\'\i}nez-Rodr{\'\i}guez}, {Marvin},
  {Mathar}, {Mirabet}, {Montes}, {Morales Mu{\~n}oz}, {Moya}, {Naranjo},
  {Ofir}, {Oreiro}, {Pall{\'e}}, {Panduro}, {Passegger}, {P{\'e}rez-Calpena},
  {P{\'e}rez Medialdea}, {Perger}, {Pluto}, {Ram{\'o}n}, {Rebolo}, {Redondo},
  {Reffert}, {Reinhardt}, {Rhode}, {Rix}, {Rodler}, {Rodr{\'\i}guez},
  {Rodr{\'\i}guez-L{\'o}pez}, {Rodr{\'\i}guez-P{\'e}rez}, {Rohloff}, {Rosich},
  {S{\'a}nchez-Blanco}, {S{\'a}nchez Carrasco}, {Sanz-Forcada}, {Sarmiento},
  {Sch{\"a}fer}, {Schiller}, {Schmidt}, {Schmitt}, {Solano}, {Stahl}, {Storz},
  {St{\"u}rmer}, {Su{\'a}rez}, {Ulbrich}, {Veredas}, {Wagner}, {Winkler},
  {Zapatero Osorio}, {Zechmeister}, {Abell{\'a}n de Paco},
  {Anglada-Escud{\'e}}, {del Burgo}, {Klutsch}, {Lizon}, {L{\'o}pez-Morales},
  {Morales}, {Perryman}, {Tulloch}, \& {Xu}}]{Quirrenbach2014}
{Quirrenbach}, A., {Amado}, P.~J., {Caballero}, J.~A., {et~al.} 2014, in
  Society of Photo-Optical Instrumentation Engineers (SPIE) Conference Series,
  Vol. 9147, Ground-based and Airborne Instrumentation for Astronomy V, ed.
  S.~K. {Ramsay}, I.~S. {McLean}, \& H.~{Takami}, 91471F

\bibitem[{{Quirrenbach} {et~al.}(2022){Quirrenbach}, {Passegger}, {Trifonov},
  {Amado}, {Caballero}, {Reiners}, {Ribas}, {Aceituno}, {B{\'e}jar},
  {Chaturvedi}, {Gonz{\'a}lez-Cuesta}, {Henning}, {Herrero}, {Kaminski},
  {K{\"u}rster}, {Lalitha}, {Lodieu}, {L{\'o}pez-Gonz{\'a}lez}, {Montes},
  {Pall{\'e}}, {Perger}, {Pollacco}, {Reffert}, {Rodr{\'\i}guez}, {L{\'o}pez},
  {Shan}, {Tal-Or}, {Osorio}, \& {Zechmeister}}]{Quirrenbach2022}
{Quirrenbach}, A., {Passegger}, V.~M., {Trifonov}, T., {et~al.} 2022, \aap,
  663, A48

\bibitem[{{Reiners} {et~al.}(2012){Reiners}, {Joshi}, \&
  {Goldman}}]{Reiners2012}
{Reiners}, A., {Joshi}, N., \& {Goldman}, B. 2012, \aj, 143, 93

\bibitem[{{Reiners} {et~al.}(2018){Reiners}, {Ribas}, {Zechmeister},
  {Caballero}, {Trifonov}, {Dreizler}, {Morales}, {Tal-Or}, {Lafarga},
  {Quirrenbach}, {Amado}, {Kaminski}, {Jeffers}, {Aceituno}, {B{\'e}jar},
  {Gu{\`a}rdia}, {Guenther}, {Hagen}, {Montes}, {Passegger}, {Seifert},
  {Schweitzer}, {Cort{\'e}s-Contreras}, {Abril}, {Alonso-Floriano}, {Eiff},
  {Antona}, {Anglada-Escud{\'e}}, {Anwand-Heerwart}, {Arroyo-Torres}, {Azzaro},
  {Baroch}, {Barrado}, {Bauer}, {Becerril}, {Ben{\'{\i}}tez}, {Berdi{\~n}as},
  {Bergond}, {Bl{\"u}mcke}, {Brinkm{\"o}ller}, {del Burgo}, {Cano},
  {C{\'a}rdenas V{\'a}zquez}, {Casal}, {Cifuentes}, {Claret}, {Colom{\'e}},
  {Czesla}, {D{\'{\i}}ez-Alonso}, {Feiz}, {Fern{\'a}ndez}, {Ferro},
  {Fuhrmeister}, {Galad{\'{\i}}-Enr{\'{\i}}quez}, {Garcia-Piquer},
  {Garc{\'{\i}}a Vargas}, {Gesa}, {G{\'o}mez Galera}, {Gonz{\'a}lez
  Hern{\'a}ndez}, {Gonz{\'a}lez-Peinado}, {Gr{\"o}zinger}, {Grohnert},
  {Guijarro}, {de Guindos}, {Guti{\'e}rrez-Soto}, {Hatzes}, {Hauschildt},
  {Hedrosa}, {Helmling}, {Henning}, {Hermelo}, {Hern{\'a}ndez Arab{\'{\i}}},
  {Hern{\'a}ndez Casta{\~n}o}, {Hern{\'a}ndez Hernando}, {Herrero}, {Huber},
  {Huke}, {Johnson}, {de Juan}, {Kim}, {Klein}, {Kl{\"u}ter}, {Klutsch},
  {K{\"u}rster}, {Labarga}, {Lamert}, {Lamp{\'o}n}, {Lara}, {Laun}, {Lemke},
  {Lenzen}, {Launhardt}, {L{\'o}pez del Fresno}, {L{\'o}pez-Gonz{\'a}lez},
  {L{\'o}pez-Puertas}, {L{\'o}pez Salas}, {L{\'o}pez-Santiago}, {Luque},
  {Mag{\'a}n Madinabeitia}, {Mall}, {Mancini}, {Mandel}, {Marfil},
  {Mar{\'{\i}}n Molina}, {Maroto Fern{\'a}ndez}, {Mart{\'{\i}}n},
  {Mart{\'{\i}}n-Ruiz}, {Marvin}, {Mathar}, {Mirabet}, {Moreno-Raya}, {Moya},
  {Mundt}, {Nagel}, {Naranjo}, {Nortmann}, {Nowak}, {Ofir}, {Oreiro},
  {Pall{\'e}}, {Panduro}, {Pascual}, {Pavlov}, {Pedraz}, {P{\'e}rez-Calpena},
  {P{\'e}rez Medialdea}, {Perger}, {Perryman}, {Pluto}, {Rabaza}, {Ram{\'o}n},
  {Rebolo}, {Redondo}, {Reffert}, {Reinhart}, {Rhode}, {Rix}, {Rodler},
  {Rodr{\'{\i}}guez}, {Rodr{\'{\i}}guez-L{\'o}pez}, {Rodr{\'{\i}}guez
  Trinidad}, {Rohloff}, {Rosich}, {Sadegi}, {S{\'a}nchez-Blanco}, {S{\'a}nchez
  Carrasco}, {S{\'a}nchez-L{\'o}pez}, {Sanz-Forcada}, {Sarkis}, {Sarmiento},
  {Sch{\"a}fer}, {Schmitt}, {Schiller}, {Sch{\"o}fer}, {Solano}, {Stahl},
  {Strachan}, {St{\"u}rmer}, {Su{\'a}rez}, {Tabernero}, {Tala}, {Tulloch},
  {Ulbrich}, {Veredas}, {Vico Linares}, {Vilardell}, {Wagner}, {Winkler},
  {Wolthoff}, {Xu}, {Yan}, \& {Zapatero Osorio}}]{Reiners18}
{Reiners}, A., {Ribas}, I., {Zechmeister}, M., {et~al.} 2018, \aap, 609, L5

\bibitem[{{Reyl{\'e}} {et~al.}(2021){Reyl{\'e}}, {Jardine}, {Fouqu{\'e}},
  {Caballero}, {Smart}, \& {Sozzetti}}]{Reyle2021}
{Reyl{\'e}}, C., {Jardine}, K., {Fouqu{\'e}}, P., {et~al.} 2021, \aap, 650,
  A201

\bibitem[{{Ricker} {et~al.}(2015){Ricker}, {Winn}, {Vanderspek}, {Latham},
  {Bakos}, {Bean}, {Berta-Thompson}, {Brown}, {Buchhave}, {Butler}, {Butler},
  {Chaplin}, {Charbonneau}, {Christensen-Dalsgaard}, {Clampin}, {Deming},
  {Doty}, {De Lee}, {Dressing}, {Dunham}, {Endl}, {Fressin}, {Ge}, {Henning},
  {Holman}, {Howard}, {Ida}, {Jenkins}, {Jernigan}, {Johnson}, {Kaltenegger},
  {Kawai}, {Kjeldsen}, {Laughlin}, {Levine}, {Lin}, {Lissauer}, {MacQueen},
  {Marcy}, {McCullough}, {Morton}, {Narita}, {Paegert}, {Palle}, {Pepe},
  {Pepper}, {Quirrenbach}, {Rinehart}, {Sasselov}, {Sato}, {Seager},
  {Sozzetti}, {Stassun}, {Sullivan}, {Szentgyorgyi}, {Torres}, {Udry}, \&
  {Villasenor}}]{Ricker2015}
{Ricker}, G.~R., {Winn}, J.~N., {Vanderspek}, R., {et~al.} 2015, Journal of
  Astronomical Telescopes, Instruments, and Systems, 1, 014003

\bibitem[{{Sabotta} {et~al.}(2021){Sabotta}, {Schlecker}, {Chaturvedi},
  {Guenther}, {Mu{\~n}oz Rodr{\'\i}guez}, {Mu{\~n}oz S{\'a}nchez}, {Caballero},
  {Shan}, {Reffert}, {Ribas}, {Reiners}, {Hatzes}, {Amado}, {Klahr}, {Morales},
  {Quirrenbach}, {Henning}, {Dreizler}, {Pall{\'e}}, {Perger}, {Azzaro},
  {Jeffers}, {Kaminski}, {K{\"u}rster}, {Lafarga}, {Montes}, {Passegger}, \&
  {Zechmeister}}]{Sabotta2021}
{Sabotta}, S., {Schlecker}, M., {Chaturvedi}, P., {et~al.} 2021, \aap, 653,
  A114

\bibitem[{{Sanz-Forcada} {et~al.}(2011){Sanz-Forcada}, {Micela}, {Ribas},
  {Pollock}, {Eiroa}, {Velasco}, {Solano}, \&
  {Garc{\'\i}a-{\'A}lvarez}}]{Sanz-Forcada2011}
{Sanz-Forcada}, J., {Micela}, G., {Ribas}, I., {et~al.} 2011, \aap, 532, A6

\bibitem[{{Schlecker} {et~al.}(2021){Schlecker}, {Mordasini}, {Emsenhuber},
  {Klahr}, {Henning}, {Burn}, {Alibert}, \& {Benz}}]{Schlecker2021}
{Schlecker}, M., {Mordasini}, C., {Emsenhuber}, A., {et~al.} 2021, \aap, 656,
  A71

\bibitem[{{Schmitt} {et~al.}(1995){Schmitt}, {Fleming}, \&
  {Giampapa}}]{Schmitt1995}
{Schmitt}, J. H.~M.~M., {Fleming}, T.~A., \& {Giampapa}, M.~S. 1995, \apj, 450,
  392

\bibitem[{{Sch{\"o}fer} {et~al.}(2019){Sch{\"o}fer}, {Jeffers}, {Reiners},
  {Shulyak}, {Fuhrmeister}, {Johnson}, {Zechmeister}, {Ribas}, {Quirrenbach},
  {Amado}, {Caballero}, {Anglada-Escud{\'e}}, {Bauer}, {B{\'e}jar},
  {Cort{\'e}s-Contreras}, {Dreizler}, {Guenther}, {Kaminski}, {K{\"u}rster},
  {Lafarga}, {Montes}, {Morales}, {Pedraz}, \& {Tal-Or}}]{Schoefer2019}
{Sch{\"o}fer}, P., {Jeffers}, S.~V., {Reiners}, A., {et~al.} 2019, \aap, 623,
  A44

\bibitem[{{Sch{\"o}fer} {et~al.}(2022){Sch{\"o}fer}, {Jeffers}, {Reiners},
  {Zechmeister}, {Fuhrmeister}, {Lafarga}, {Ribas}, {Quirrenbach}, {Amado},
  {Caballero}, {Anglada-Escud{\'e}}, {Bauer}, {B{\'e}jar},
  {Cort{\'e}s-Contreras}, {Alonso}, {Dreizler}, {Guenther}, {Herbort},
  {Johnson}, {Kaminski}, {K{\"u}rster}, {Montes}, {Morales}, {Pedraz}, \&
  {Tal-Or}}]{Schoefer2022}
{Sch{\"o}fer}, P., {Jeffers}, S.~V., {Reiners}, A., {et~al.} 2022, \aap, 663,
  A68

\bibitem[{{Schweitzer} {et~al.}(2019){Schweitzer}, {Passegger}, {Cifuentes},
  {B{\'e}jar}, {Cort{\'e}s-Contreras}, {Caballero}, {del Burgo}, {Czesla},
  {K{\"u}rster}, {Montes}, {Zapatero Osorio}, {Ribas}, {Reiners},
  {Quirrenbach}, {Amado}, {Aceituno}, {Anglada-Escud{\'e}}, {Bauer},
  {Dreizler}, {Jeffers}, {Guenther}, {Henning}, {Kaminski}, {Lafarga},
  {Marfil}, {Morales}, {Schmitt}, {Seifert}, {Solano}, {Tabernero}, \&
  {Zechmeister}}]{Schweitzer2019}
{Schweitzer}, A., {Passegger}, V.~M., {Cifuentes}, C., {et~al.} 2019, \aap,
  625, A68

\bibitem[{{Seifahrt} {et~al.}(2020){Seifahrt}, {Bean}, {St{\"u}rmer}, {Kasper},
  {Gers}, {Schwab}, {Zechmeister}, {Stef{\'a}nsson}, {Montet}, {Dos Santos},
  {Peck}, {White}, \& {Tapia}}]{Seifahrt2020}
{Seifahrt}, A., {Bean}, J.~L., {St{\"u}rmer}, J., {et~al.} 2020, in Society of
  Photo-Optical Instrumentation Engineers (SPIE) Conference Series, Vol. 11447,
  Society of Photo-Optical Instrumentation Engineers (SPIE) Conference Series,
  114471F

\bibitem[{{Seifahrt} {et~al.}(2018){Seifahrt}, {St{\"u}rmer}, {Bean}, \&
  {Schwab}}]{maroonx2018}
{Seifahrt}, A., {St{\"u}rmer}, J., {Bean}, J.~L., \& {Schwab}, C. 2018, in
  Society of Photo-Optical Instrumentation Engineers (SPIE) Conference Series,
  Vol. 10702, Ground-based and Airborne Instrumentation for Astronomy VII, ed.
  C.~J. {Evans}, L.~{Simard}, \& H.~{Takami}, 107026D

\bibitem[{{Shappee} {et~al.}(2014){Shappee}, {Prieto}, {Stanek}, {Kochanek},
  {Holoien}, {Jencson}, {Basu}, {Beacom}, {Szczygiel}, {Pojmanski},
  {Brimacombe}, {Dubberley}, {Elphick}, {Foale}, {Hawkins}, {Mullins},
  {Rosing}, {Ross}, \& {Walker}}]{Shapee2014}
{Shappee}, B., {Prieto}, J., {Stanek}, K.~Z., {et~al.} 2014, in AAS Meeting
  Abstracts, Vol. 223, AAS Meeting Abstracts \#223, 236.03

\bibitem[{{Skrutskie} {et~al.}(2006){Skrutskie}, {Cutri}, {Stiening},
  {Weinberg}, {Schneider}, {Carpenter}, {Beichman}, {Capps}, {Chester},
  {Elias}, {Huchra}, {Liebert}, {Lonsdale}, {Monet}, {Price}, {Seitzer},
  {Jarrett}, {Kirkpatrick}, {Gizis}, {Howard}, {Evans}, {Fowler}, {Fullmer},
  {Hurt}, {Light}, {Kopan}, {Marsh}, {McCallon}, {Tam}, {Van Dyk}, \&
  {Wheelock}}]{2MASS}
{Skrutskie}, M.~F., {Cutri}, R.~M., {Stiening}, R., {et~al.} 2006, \aj, 131,
  1163

\bibitem[{{Smith} {et~al.}(2012){Smith}, {Stumpe}, {Van Cleve}, {Jenkins},
  {Barclay}, {Fanelli}, {Girouard}, {Kolodziejczak}, {McCauliff}, {Morris}, \&
  {Twicken}}]{Smith2012}
{Smith}, J.~C., {Stumpe}, M.~C., {Van Cleve}, J.~E., {et~al.} 2012, \pasp, 124,
  1000

\bibitem[{{Sozzetti} {et~al.}(2007){Sozzetti}, {Torres}, {Charbonneau},
  {Latham}, {Holman}, {Winn}, {Laird}, \& {O'Donovan}}]{Sozzetti2007}
{Sozzetti}, A., {Torres}, G., {Charbonneau}, D., {et~al.} 2007, \apj, 664, 1190

\bibitem[{{Speagle}(2020)}]{Speagle2020}
{Speagle}, J.~S. 2020, \mnras, 493, 3132

\bibitem[{{Stassun} {et~al.}(2019){Stassun}, {Oelkers}, {Paegert}, {Torres},
  {Pepper}, {De Lee}, {Collins}, {Latham}, {Muirhead}, {Chittidi},
  {Rojas-Ayala}, {Fleming}, {Rose}, {Tenenbaum}, {Ting}, {Kane}, {Barclay},
  {Bean}, {Brassuer}, {Charbonneau}, {Ge}, {Lissauer}, {Mann}, {McLean},
  {Mullally}, {Narita}, {Plavchan}, {Ricker}, {Sasselov}, {Seager}, {Sharma},
  {Shiao}, {Sozzetti}, {Stello}, {Vanderspek}, {Wallace}, \&
  {Winn}}]{Stassun2019}
{Stassun}, K.~G., {Oelkers}, R.~J., {Paegert}, M., {et~al.} 2019, \aj, 158, 138

\bibitem[{{Stassun} {et~al.}(2018){Stassun}, {Oelkers}, {Pepper}, {Paegert},
  {De Lee}, {Torres}, {Latham}, {Charpinet}, {Dressing}, {Huber}, {Kane},
  {L{\'e}pine}, {Mann}, {Muirhead}, {Rojas-Ayala}, {Silvotti}, {Fleming},
  {Levine}, \& {Plavchan}}]{Stassun2018}
{Stassun}, K.~G., {Oelkers}, R.~J., {Pepper}, J., {et~al.} 2018, \aj, 156, 102

\bibitem[{{Stef{\'a}nsson} {et~al.}(2020){Stef{\'a}nsson}, {Kopparapu}, {Lin},
  {Mahadevan}, {Ca{\~n}as}, {Kanodia}, {Ninan}, {Cochran}, {Endl}, {Hebb},
  {Wisniewski}, {Gupta}, {Everett}, {Bender}, {Diddams}, {Ford}, {Fredrick},
  {Halverson}, {Hearty}, {Levi}, {Maney}, {Metcalf}, {Monson}, {Ramsey},
  {Robertson}, {Roy}, {Schwab}, {Terrien}, \& {Wright}}]{Stefansson2020}
{Stef{\'a}nsson}, G., {Kopparapu}, R., {Lin}, A., {et~al.} 2020, \aj, 160, 259

\bibitem[{{Stock} {et~al.}(2020){Stock}, {Nagel}, {Kemmer}, {Passegger},
  {Reffert}, {Quirrenbach}, {Caballero}, {Czesla}, {B{\'e}jar}, {Cardona},
  {D{\'\i}ez-Alonso}, {Herrero}, {Lalitha}, {Schlecker}, {Tal-Or},
  {Rodr{\'\i}guez}, {Rodr{\'\i}guez-L{\'o}pez}, {Ribas}, {Reiners}, {Amado},
  {Bauer}, {Bluhm}, {Cort{\'e}s-Contreras}, {Gonz{\'a}lez-Cuesta}, {Dreizler},
  {Hatzes}, {Henning}, {Jeffers}, {Kaminski}, {K{\"u}rster}, {Lafarga},
  {L{\'o}pez-Gonz{\'a}lez}, {Montes}, {Morales}, {Pedraz}, {Sch{\"o}fer},
  {Schweitzer}, {Trifonov}, {Zapatero Osorio}, \& {Zechmeister}}]{Stock2020}
{Stock}, S., {Nagel}, E., {Kemmer}, J., {et~al.} 2020, \aap, 643, A112

\bibitem[{{Stumpe} {et~al.}(2014){Stumpe}, {Smith}, {Catanzarite}, {Van Cleve},
  {Jenkins}, {Twicken}, \& {Girouard}}]{Stumpe2014}
{Stumpe}, M.~C., {Smith}, J.~C., {Catanzarite}, J.~H., {et~al.} 2014, \pasp,
  126, 100

\bibitem[{{Stumpe} {et~al.}(2012){Stumpe}, {Smith}, {Van Cleve}, {Twicken},
  {Barclay}, {Fanelli}, {Girouard}, {Jenkins}, {Kolodziejczak}, {McCauliff}, \&
  {Morris}}]{Stumpe2012}
{Stumpe}, M.~C., {Smith}, J.~C., {Van Cleve}, J.~E., {et~al.} 2012, \pasp, 124,
  985

\bibitem[{{Swain} {et~al.}(2021){Swain}, {Estrela}, {Roudier}, {Sotin},
  {Rimmer}, {Valio}, {West}, {Pearson}, {Huber-Feely}, \& {Zellem}}]{swain2021}
{Swain}, M.~R., {Estrela}, R., {Roudier}, G.~M., {et~al.} 2021, \aj, 161, 213

\bibitem[{{Tal-Or} {et~al.}(2019){Tal-Or}, {Trifonov}, {Zucker}, {Mazeh}, \&
  {Zechmeister}}]{2019MNRAS.484L...8T}
{Tal-Or}, L., {Trifonov}, T., {Zucker}, S., {Mazeh}, T., \& {Zechmeister}, M.
  2019, \mnras, 484, L8

\bibitem[{{Tinetti} {et~al.}(2016){Tinetti}, {Drossart}, {Eccleston},
  {Hartogh}, {Heske}, {Leconte}, {Micela}, {Ollivier}, {Pilbratt}, {Puig},
  {Turrini}, {Vandenbussche}, {Wolkenberg}, {Pascale}, {Beaulieu}, {G{\"u}del},
  {Min}, {Rataj}, {Ray}, {Ribas}, {Barstow}, {Bowles}, {Coustenis}, {Coud{\'e}
  du Foresto}, {Decin}, {Encrenaz}, {Forget}, {Friswell}, {Griffin}, {Lagage},
  {Malaguti}, {Moneti}, {Morales}, {Pace}, {Rocchetto}, {Sarkar}, {Selsis},
  {Taylor}, {Tennyson}, {Venot}, {Waldmann}, {Wright}, {Zingales}, \&
  {Zapatero-Osorio}}]{2016SPIE.9904E..1XT}
{Tinetti}, G., {Drossart}, P., {Eccleston}, P., {et~al.} 2016, in Society of
  Photo-Optical Instrumentation Engineers (SPIE) Conference Series, Vol. 9904,
  Space Telescopes and Instrumentation 2016: Optical, Infrared, and Millimeter
  Wave, ed. H.~A. {MacEwen}, G.~G. {Fazio}, M.~{Lystrup}, N.~{Batalha},
  N.~{Siegler}, \& E.~C. {Tong}, 99041X

\bibitem[{{Tody}(1993)}]{IRAF1993}
{Tody}, D. 1993, in Astronomical Society of the Pacific Conference Series,
  Vol.~52, Astronomical Data Analysis Software and Systems II, ed. R.~J.
  {Hanisch}, R.~J.~V. {Brissenden}, \& J.~{Barnes}, 173

\bibitem[{{Trifonov} {et~al.}(2021){Trifonov}, {Caballero}, {Morales},
  {Seifahrt}, {Ribas}, {Reiners}, {Bean}, {Luque}, {Parviainen}, {Pall{\'e}},
  {Stock}, {Zechmeister}, {Amado}, {Anglada-Escud{\'e}}, {Azzaro}, {Barclay},
  {B{\'e}jar}, {Bluhm}, {Casasayas-Barris}, {Cifuentes}, {Collins}, {Collins},
  {Cort{\'e}s-Contreras}, {de Leon}, {Dreizler}, {Dressing}, {Esparza-Borges},
  {Espinoza}, {Fausnaugh}, {Fukui}, {Hatzes}, {Hellier}, {Henning}, {Henze},
  {Herrero}, {Jeffers}, {Jenkins}, {Jensen}, {Kaminski}, {Kasper},
  {Kossakowski}, {K{\"u}rster}, {Lafarga}, {Latham}, {Mann}, {Molaverdikhani},
  {Montes}, {Montet}, {Murgas}, {Narita}, {Oshagh}, {Passegger}, {Pollacco},
  {Quinn}, {Quirrenbach}, {Ricker}, {Rodr{\'\i}guez L{\'o}pez}, {Sanz-Forcada},
  {Schwarz}, {Schweitzer}, {Seager}, {Shporer}, {Stangret}, {St{\"u}rmer},
  {Tan}, {Tenenbaum}, {Twicken}, {Vanderspek}, \& {Winn}}]{Trifonov2021}
{Trifonov}, T., {Caballero}, J.~A., {Morales}, J.~C., {et~al.} 2021, Science,
  371, 1038

\bibitem[{{Trifonov} {et~al.}(2020){Trifonov}, {Tal-Or}, {Zechmeister},
  {Kaminski}, {Zucker}, \& {Mazeh}}]{2020A&A...636A..74T}
{Trifonov}, T., {Tal-Or}, L., {Zechmeister}, M., {et~al.} 2020, \aap, 636, A74

\bibitem[{{Trotta}(2008)}]{Trotta2008}
{Trotta}, R. 2008, Contemporary Physics, 49, 71

\bibitem[{{Twicken} {et~al.}(2018){Twicken}, {Catanzarite}, {Clarke},
  {Girouard}, {Jenkins}, {Klaus}, {Li}, {McCauliff}, {Seader}, {Tenenbaum},
  {Wohler}, {Bryson}, {Burke}, {Caldwell}, {Haas}, {Henze}, \&
  {Sanderfer}}]{Twicken:DVdiagnostics2018}
{Twicken}, J.~D., {Catanzarite}, J.~H., {Clarke}, B.~D., {et~al.} 2018, \pasp,
  130, 064502

\bibitem[{{Van Eylen} {et~al.}(2018){Van Eylen}, {Agentoft}, {Lundkvist},
  {Kjeldsen}, {Owen}, {Fulton}, {Petigura}, \& {Snellen}}]{vaneylen2018}
{Van Eylen}, V., {Agentoft}, C., {Lundkvist}, M.~S., {et~al.} 2018, \mnras,
  479, 4786

\bibitem[{{Van Eylen} {et~al.}(2021){Van Eylen}, {Astudillo-Defru}, {Bonfils},
  {Livingston}, {Hirano}, {Luque}, {Lam}, {Justesen}, {Winn}, {Gandolfi},
  {Nowak}, {Palle}, {Albrecht}, {Dai}, {Campos Estrada}, {Owen},
  {Foreman-Mackey}, {Fridlund}, {Korth}, {Mathur}, {Forveille}, {Mikal-Evans},
  {Osborne}, {Ho}, {Almenara}, {Artigau}, {Barrag{\'a}n}, {Barros}, {Bouchy},
  {Cabrera}, {Caldwell}, {Charbonneau}, {Chaturvedi}, {Cochran}, {Csizmadia},
  {Damasso}, {Delfosse}, {De Medeiros}, {D{\'\i}az}, {Doyon}, {Esposito},
  {F{\H{u}}r{\'e}sz}, {Figueira}, {Georgieva}, {Goffo}, {Grziwa}, {Guenther},
  {Hatzes}, {Jenkins}, {Kabath}, {Knudstrup}, {Latham}, {Lavie}, {Lovis},
  {Mennickent}, {Mullally}, {Murgas}, {Narita}, {Pepe}, {Persson}, {Redfield},
  {Ricker}, {Santos}, {Seager}, {Serrano}, {Smith}, {Mascare{\~n}o}, {Subjak},
  {Twicken}, {Udry}, {Vanderspek}, \& {Zapatero Osorio}}]{2021MNRAS.507.2154V}
{Van Eylen}, V., {Astudillo-Defru}, N., {Bonfils}, X., {et~al.} 2021, \mnras,
  507, 2154

\bibitem[{{Wandel}(2018)}]{Wandel2018}
{Wandel}, A. 2018, \apj, 856, 165

\bibitem[{{Ward-Duong} {et~al.}(2015){Ward-Duong}, {Patience}, {De Rosa},
  {Bulger}, {Rajan}, {Goodwin}, {Parker}, {McCarthy}, \&
  {Kulesa}}]{Ward-Duong2015}
{Ward-Duong}, K., {Patience}, J., {De Rosa}, R.~J., {et~al.} 2015, \mnras, 449,
  2618

\bibitem[{{Weiner} {et~al.}(2018){Weiner}, {Sand}, {Gabor}, {Johnson},
  {Swindell}, {Kub{\'a}nek}, {Gasho}, {Golota}, {Jannuzi}, {Milne}, {Smith}, \&
  {Zaritsky}}]{2018SPIE10704E..2HW}
{Weiner}, B.~J., {Sand}, D., {Gabor}, P., {et~al.} 2018, in Society of
  Photo-Optical Instrumentation Engineers (SPIE) Conference Series, Vol. 10704,
  Observatory Operations: Strategies, Processes, and Systems VII, 107042H

\bibitem[{{Wo{\'z}niak} {et~al.}(2004){Wo{\'z}niak}, {Vestrand}, {Akerlof},
  {Balsano}, {Bloch}, {Casperson}, {Fletcher}, {Gisler}, {Kehoe}, {Kinemuchi},
  {Lee}, {Marshall}, {McGowan}, {McKay}, {Rykoff}, {Smith}, {Szymanski}, \&
  {Wren}}]{Wozniak2004}
{Wo{\'z}niak}, P.~R., {Vestrand}, W.~T., {Akerlof}, C.~W., {et~al.} 2004, \aj,
  127, 2436

\bibitem[{{Zechmeister} {et~al.}(2014){Zechmeister}, {Anglada-Escud{\'e}}, \&
  {Reiners}}]{2014A&A...561A..59Z}
{Zechmeister}, M., {Anglada-Escud{\'e}}, G., \& {Reiners}, A. 2014, \aap, 561,
  A59

\bibitem[{{Zechmeister} \& {K{\"u}rster}(2009)}]{Zechmeister2009}
{Zechmeister}, M. \& {K{\"u}rster}, M. 2009, \aap, 496, 577

\bibitem[{{Zechmeister} {et~al.}(2018){Zechmeister}, {Reiners}, {Amado},
  {Azzaro}, {Bauer}, {B{\'e}jar}, {Caballero}, {Guenther}, {Hagen}, {Jeffers},
  {Kaminski}, {K{\"u}rster}, {Launhardt}, {Montes}, {Morales}, {Quirrenbach},
  {Reffert}, {Ribas}, {Seifert}, {Tal-Or}, \& {Wolthoff}}]{2018A&A...609A..12Z}
{Zechmeister}, M., {Reiners}, A., {Amado}, P.~J., {et~al.} 2018, \aap, 609, A12

\bibitem[{{Zeng} {et~al.}(2017){Zeng}, {Jacobsen}, \& {Sasselov}}]{Zeng2017}
{Zeng}, L., {Jacobsen}, S.~B., \& {Sasselov}, D.~D. 2017, Research Notes of the
  American Astronomical Society, 1, 32

\bibitem[{{Zeng} {et~al.}(2019){Zeng}, {Jacobsen}, {Sasselov}, {Petaev},
  {Vanderburg}, {Lopez-Morales}, {Perez-Mercader}, {Mattsson}, {Li}, {Heising},
  {Bonomo}, {Damasso}, {Berger}, {Cao}, {Levi}, \& {Wordsworth}}]{Zeng2019}
{Zeng}, L., {Jacobsen}, S.~B., {Sasselov}, D.~D., {et~al.} 2019, PNAS, 116,
  9723

\end{thebibliography}

\begin{appendix}

\section{Data modeling with juliet}\label{appA:data_modelling}

\subsection{Transit-only modeling}\label{appsubsec:phot-juliet}

We fit the {\em TESS} light curve from sectors 17, 42, and 43 to determine more precise transit parameters using the SPOC reported period and ephemeris for each planet.
We used this information as initial input parameters to precisely determine the ephemeris of the transiting planets. As a first step, we allowed the period of the inner planet to vary between 1.6 and 2.1 days, whereas, for the outer planet we kept this range between 15.2 and 15.8 days. We also incorporated the central transit time for both the planets based on the {\em TESS} light curves, analyzed with the SPOC pipelines \citep{SPOC}, namely between BJD 2458765.5 and  2458765.8 for the inner planet, and between 2458766.7 and 2458767.1 for the outer planet. The transit fitting is usually done by fitting the scaled planetary radius, $a/R_\star$, as a free parameter. Since we have a multiple planet system, we instead fitted the stellar density $\rho_\star,$ introduced first by \cite{Sozzetti2007}, which should remain the same for the case of both transiting planets and constrain it in a more robust way. We kept $\rho_\star$ as a free parameter with a normal distribution centered around the stellar density and a width of $3\sigma,$ calculated based on stellar parameters as derived in Table\,\ref{tab:stellar-param}.
Based on \cite{Espinoza2018}, we also parametrized the planet-to-star-ratios, $p_{\rm 1}$, $p_{ \rm 2}$, and the impact parameters, $b_{\rm 1}$, $b_{\rm 2}$ to fit variables $r_{\rm 1,b}$, $r_{\rm 2,b}$, and $r_{\rm 1,c}$, $r_{\rm 2,c}$ respectively for both planets, and set the priors between 0.0 and 1.0 with a uniform distribution. The RVs provide more information in constraining the eccentricity-omega (e\,--\,$\omega$) parameters, and thus these are kept fixed at 0$^\circ$ and 90$^\circ$, respectively, for the transit-only fit. We kept the dilution factor for the {\em TESS} and the ground-based photometry fixed at 1.0 as discussed previously in Section\,\ref{subsec:contamination}. We also chose a quadratic limb darkening law for the {\em TESS} data \citep{Kipping2013} with a uniform distribution between 0.0 and 1.0. We used a linear limb darkening law for the ground-based data due to relatively low photometric precision, but with the same priors as those for the {\em TESS} data. A flux offset was assumed between different photometry data sets, which was allowed to vary between 0.0 and 0.1, and a flux scatter was assumed between $10^{-5}$ to $10^{5}$\,ppm, both with a Jeffreys distribution. We recovered the {\em TESS} derived periods, $P_{\rm b}$ and $P_{\rm c}$ to be 
$1.8805175^{+ 0.0000029}_{- 0.0000030}$\,d and $15.532244^{+ 0.000037}_{- 0.000046}$\,d, and the central transit times, $t_{0b}$,  $t_{0c}$ to be $2458765.67857^{+0.00084}_{- 0.00082}$ and $2458766.9241^{+0.00050}_{- 0.00048}$ (both in BJD), respectively. These initial values with a $3\sigma$ prior width were applied as priors for the joint fit, which is described in \ref{subsec:orbit-model}.

\subsection{RV-only modeling} \label{appsubsec:rv-juliet}

\begin{table}
\centering
\small
\caption{Comparison of RV-only models.}
\label{tab:bayesian}
\begin{tabular}{l c c}
\hline\hline
\noalign{\smallskip}
Models & $\ln{\mathcal{Z}}$  & $|\Delta\ln{\mathcal{Z}}|$\\
\noalign{\smallskip}
\hline
\noalign{\smallskip}
\multicolumn{3}{c}{\it Two-signal models (without activity modeling)}\\[0.1cm]
\noalign{\smallskip}
2cp        & --267.66 & 0.0 \\
1cp\,+\,1kp    & --269.54 & 1.88 \\
1kp\,+\,1cp    & --269.44 & 1.78 \\
2kp        & --271.40 & 3.74 \\
\noalign{\smallskip}
\multicolumn{3}{c}{\it Three-signal models (with activity modeling)}\\
\noalign{\smallskip}
2cp\,+\,sin  & --250.44 &  17.22\\
\textbf{2cp\,+\,GP} & \textbf{--253.81} & \textbf{13.85}\\
1cp\,+1kp+\,GP & --255.10 &  12.56\\
1kp\,+1cp+\,GP & --255.21 &  12.45\\
2kp\,+\,GP     & --256.97 &  10.69\\
\hline
\end{tabular}

\tablefoot{Here, ``cp'', ``kp'', ``sin'', and ``GP'' refer to the circular planet model, the Keplerian (eccentric planet) model, the sinusoidal fit, and Gaussian processes, respectivel. 
$\ln \mathcal{Z}$ and $|\Delta\ln \mathcal{Z}|$ are the log-evidence and relative absolute log-evidence with respect to the simplest model (2cp without activity modeling), respectively.
}
\end{table}

As discussed in \ref{subsubsec:periodogram}, the CARMENES and MAROON-X RVs show the two transiting planet signals in the GLS periodogram at 1.88\,d (and a 2.12\,d alias due to the 1\,d sampling) and 15.53\,d, which both have a nominal false alarm probability (FAP) < 1\%. After subtracting the two planetary signals, another signal is seen close to $\sim$41\,d (FAP < 0.1\%) in the residual periodogram, which is related to the stellar rotation period (see Sect.\,\ref{subsec:Prot}).

To find the best model to reproduce our RV data, we performed an extensive model comparison. To select the final model, we used the criteria described in \cite{Trotta2008}, which consider a difference between models of $|\Delta \ln \mathcal{Z}|> 5$ as ``significant''. In this case, the model with the larger Bayesian log evidence is favored. In the case where $|\Delta \ln\mathcal{Z}|> 2.5,$  the models are ``moderately'' favored one over the other. However, if $|\Delta \ln\mathcal{Z}|\leq 2.5$, then the two models are considered ``indistinguishable'' and a simpler model would be chosen. An overview of the different models and their Bayesian evidence is shown in Table\,\ref{tab:bayesian}. The residual periodogram for our selected model is shown in Fig.\,\ref{fig:spectral_indices} (panel d).

Since both planets are statistically significant, we started simultaneously fitting both planet signals using circular Keplerian orbits (``2cp'' model). For the period and central time priors, we used normal distributions centered in the values determined through the transit fit (Sect.\,\ref{appsubsec:phot-juliet}). The RV amplitude had uniform priors between 0 and 10 m\,s$^{-1}$, the offset parameter of CARMENES and MAROON-X was chosen uniformly between $-10$ and $10$ m\,s$^{-1}$, and the stellar jitter was selected with uniform priors between 0.01 and 10 m\,s$^{-1}$. The residual periodogram of this fit is shown in Fig.\,\ref{fig:spectral_indices} (panel c). 

To include eccentricity in our models, we parameterized it as $\mathcal{S}_1$\,=\,$\sqrt{e}\sin\omega$ and $\mathcal{S}_2$\,=\,$\sqrt{e}\cos\omega$ with uniform priors between $-1$ and $1$ \citep{juliet}. We performed models combining Keplerian planet orbits with fixed eccentricity (``cp'') or kept it free (``kp''). Of these models, those that consider eccentric orbits for one of the two signals are indistinguishable, only the 2kp model is moderately favored ($|\Delta$ln\,Z|\,=\,3.74) compared with the 2cp fit.

To account for the stellar activity, we then investigated whether including the third signal at $\sim$41\,d would improve the log-evidence of the fit. In this case, we modeled it using a sinusoid (sin) or a GP. In the first case, we used uniform priors between 30 and 50 d, which correspond to the suspected region for the stellar rotational period. For the GP selection, we used the QP kernel previously defined in Eq.\,\ref{eq:QP-GP}. As we discuss in Sec\,\ref{subsec:Prot}, GPs were used to model the rotational signal of the star \citep{Angus2018}. The rotation of the star often produces these sinusoidal-like signals \citep{2011A&A...525A.140D, 2014MNRAS.443.2517H}, which are misinterpreted in the RV data as a planetary signal. Thus we used a GP as a better way to model these complex periodic signals. For the GP parameters, we used uniform priors for the GP amplitude ($\sigma_{\rm{GP,RV}}$) between 0 and 100 m\,s$^{-1}$. The inverse length scale of the external parameter ($\alpha_{\rm{GP,RV}}$) and the amplitude of the sine part of the kernel ($\gamma_{\rm{GP,RV}}$) had a Jeffreys distribution between $10^{-10}$ to $10^{-2}$ and between 0.1 to 10, respectively. The rotational period of the GP ($P_{\rm{rot;GP,RV}}$) had uniform priors between 30 and 50\,d. The transiting planets were modeled with both circular and eccentric Keplerians, as done above. These results are summarized in the last five rows of Table\,\ref{tab:bayesian}. Based on the results of the previous table, we noticed that models that include three signals have the largest log-evidence compared with 2cp model. However, the difference between these models are nearly indistinguishable or moderately favored in the case of 2cp\,+\,sin ($|\Delta\,\rm ln\,Z|\sim3.37$). Thus, the first two transiting signals can be best explained with circular planetary orbits, and the third signal close to 41\,d can be equally modeled with a sinusoid or a GP fit. However, it is also important to note that the final amplitude and phase of the transiting planets should remain consistent within methods, despite our choice of model for the third signal. In this scenario, given our previous knowledge and having determined that the stellar rotation period should be around 41\,--\,44\,d, and due to the GP allowing us to account for the effects of rotating spots on the data and, therefore, a better way to model the stellar activity, we chose the 2cp+GP as our fiducial model. 

\subsection{Alternative data modeling with the FCO method}

\begin{figure}
\centering
\includegraphics[width=0.5\textwidth]{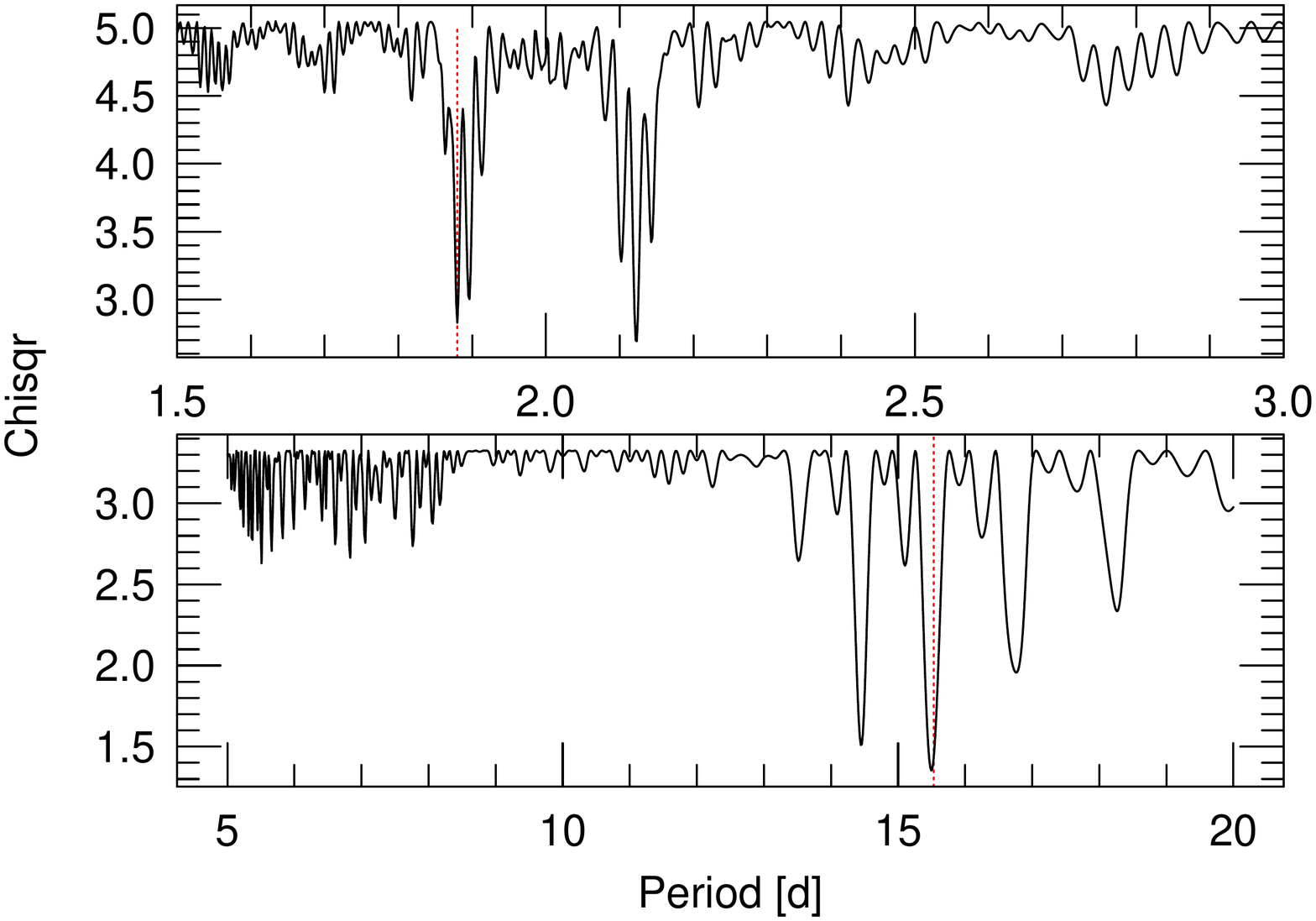}
\includegraphics[width=0.5\textwidth]{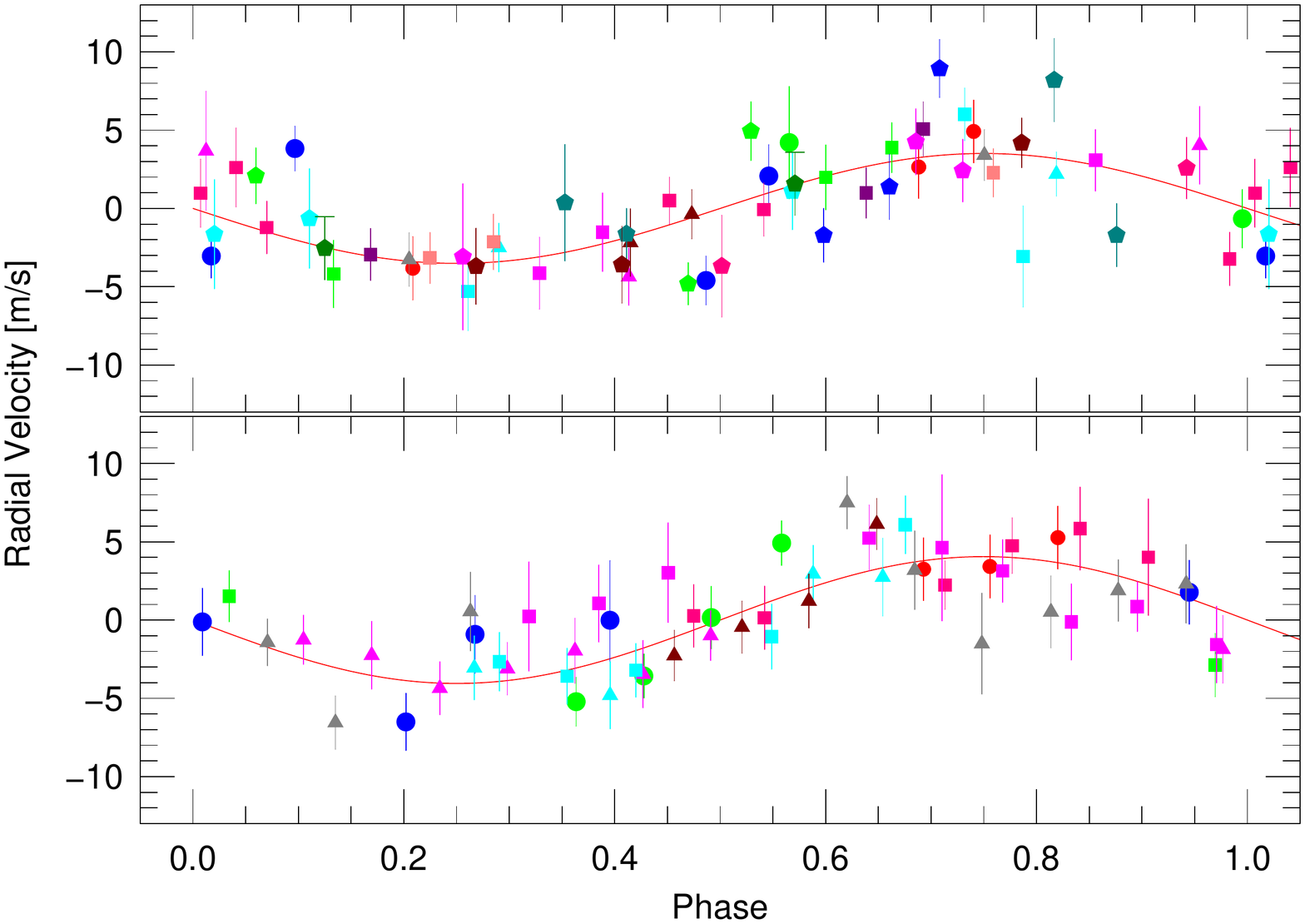}
\caption{Results from the FCO (``floating chunk offset'') method.
{\em Top panels}: Periodograms of reduced $\chi^2$ for CARMENES RVs as a function of input period for the FCO method for planets TOI-1468\,b ({\em top}) and~c ({\em bottom}). Vertical dashed red lines mark the planet orbital periods.
{\em Bottom panel}: Phase-folded CARMENES RVs for planets TOI-1468\,b ({\em top}) and~c ({\em bottom}).
Different colors and symbols represent RV values from individual chunks.} 
\label{fig:fco_periodogram}
\end{figure}

We applied an independent alternative method, a ``floating chunk offset'' (FCO) on CARMENES data to fit the RVs. This method is based on creating an orbital fit to the data by allowing the RV offsets of individual nights to vary with respect to each other \citep{2010A&A...520A..93H,2019dmde.book.....H}. The RV data were divided into 23 chunks, each with 2--4 measurements spanning 2--3\,d. For the most part, the chunks used measurements taken on consecutive nights. As a check we performed a FCO periodogram to ensure that FCO was seeing the planet signal \citep{2010A&A...520A..93H,2019dmde.book.....H}. Fig.\,\ref{fig:fco_periodogram} shows the reduced $\chi^2$ as a function of trial periods for the orbital fit. For this periodogram the phase (i.e., transit ephemeris) was allowed to vary so that the best possible fit was made to the data given the trial period. The best fit (minimum $\chi^2$)  occurs at the transit  period of 1.88\,d and its alias at 2.14\,d. 
FCO detects the correct signal in the RV data.
Applying the FCO method yielded a $K$ amplitude due to planet b of $K_{\rm b}$\,=\,$3.47\pm0.62$\,m\,s$^{-1}$. 
The lower panel of Fig.\,\ref{fig:fco_periodogram} shows the final RV variations phased to the orbital period.

\begin{figure}
\includegraphics[width=0.5\textwidth]{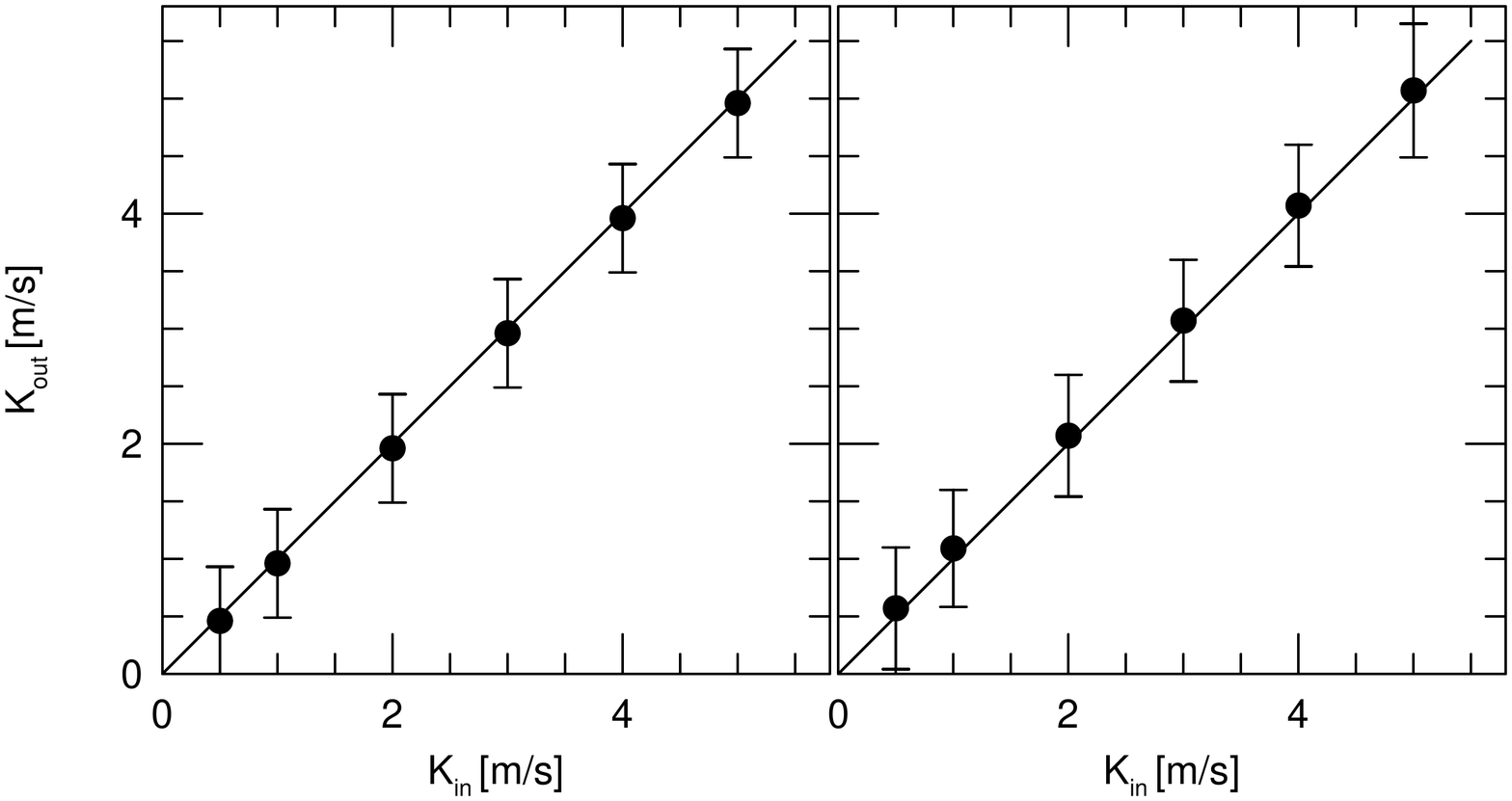}
\caption{Simulations of the FCO method showing the output $K$ amplitude ($K_{\rm out}$)
as a function of the input $K$ amplitude ($K_{\rm in}$) for TOI-1468\,b ({\em left})
and TOI-1468\,c ({\em right}).}
\label{fig:fco_kamp}
\end{figure}

To check whether the FCO method was introducing a systematic error into the $K$-amplitude determination, we created a synthetic data set  using the orbit of TOI-1468\,b with varying $K$-amplitudes. To this we added  the orbital motion  of TOI-1468\,c, a rotational modulation
signal with  a period of 43\,d, and a $K$-amplitude of 2.9\,m\,s$^{-1}$. These periodic signals may cause  the largest systematic errors because they do not introduce a strictly constant RV value to the chunk. Random noise at a level of 2.3\,m\,s$^{-1}$, roughly the rms scatter after removing all  signals from the data, was added and the data divided into chunks
in the same manner as the observations. The left panel of Fig.\,\ref{fig:fco_kamp} shows that the output $K$-amplitude is recovered as a function of the input amplitude
over the range 0.5--5.0\,m\,s$^{-1}$. The error bars represent the standard deviation of ten simulations, with different realizations of the random noise.

Finally, we investigated if we could
extract the signal of the outer planet. For this, we first 
removed the contribution of TOI-1468\,b from the RV data
and created new time chunks so as to sample more of the 15\,d period. We used 12 chunks that had time spans ranging from 2 to 7\,d, shorter than the 41--44\,d region of rotation period. The resulting RV amplitude was $K$\,=\,$4.05\pm0.59$\,m\,s$^{-1}$, which are in excellent agreement with the $K_{\rm b}$ = 3.4 $\pm$ 0.2\,m\,s$^{-1}$ determined with the joint fit. The phased RV variations are shown in the bottom panel of Fig.\,\ref{fig:fco_periodogram}. 
Again, we checked how well FCO could recover known input amplitudes by taking a combination of the orbit of TOI-1468\,c combined with the 41--44\,d rotation signal. The right panel of Fig.\,\ref{fig:fco_kamp} shows that the method recovers the correct amplitude. We indeed see that there is a systematic offset of $+$0.07\,m\,s$^{-1}$ between the input and output $K_{\rm c}$-amplitude, most likely due to the residual effects of the 41--44\,d period. Applying this offset results in a final RV amplitude of $K_{\rm c}$\,=\,$3.9\pm0.59$\,m\,s$^{-1}$, which is consistent within one sigma with the $K_{\rm c}$ = 3.48 $\pm$ 0.35\,m\,s$^{-1}$ previously derived by the joint~fit.

\section{Long tables}

\begin{table*}
    \centering
    \caption{Priors used for TOI-1468\,b and c in the  joint fit with \texttt{juliet}.}
    \label{tab:priors}
    \begin{tabular}{lccr} 
        \hline
        \hline
        \noalign{\smallskip}
        Parameter$^a$  & Prior & Unit & Description \\
        \noalign{\smallskip}
        \hline
        \noalign{\smallskip}
        \multicolumn{4}{c}{\it Stellar and planetary parameters} \\
        \noalign{\smallskip}
        $\rho_\star$ & $\mathcal{U}(11.0,13.0)$ & g\,cm\,$^{-3}$ & Stellar density \\
        $P_{\rm b}$              & $\mathcal{N}(1.880,0.002)$           & d                    & Period of planet b \\
        $t_{0,b}$                & $\mathcal{N}(2458765.681,0.001)$     & d                    & Time of transit center of planet b \\
        $r_{1,b}$                & $\mathcal{U}(0,1)$                 & \dots                & Parameterization for $p$ and $b$ \\
        $r_{2,b}$                & $\mathcal{U}(0,1)$                 & \dots                & Parameterization for $p$ and $b$ \\
        $K_{b}$                  & $\mathcal{U}(0,10)$                & $\mathrm{m\,s^{-1}}$ & RV semi-amplitude of planet b \\
        $e_{b}$                  & 0.0 (fixed)                        & \dots                & Orbital eccentricity of planet b \\
        $\omega_{b}$             & 90.0 (fixed)                       & deg                  & Periastron angle of planet b \\
        \noalign{\smallskip}
         \noalign{\smallskip}
        $P_{\rm c}$              & $\mathcal{N}(15.538,0.005)$           & d                    & Period of planet c \\
        $t_{0,c}$                & $\mathcal{N}(2458766.9236,0.004)$     & d                    & Time of transit center of planet c \\
        $r_{1,c}$                & $\mathcal{U}(0,1)$                  & \dots                & Parameterization for $p$ and $b$ \\
        $r_{2,c}$                & $\mathcal{U}(0,1)$                  & \dots                & Parameterization for $p$ and $b$ \\
        $K_{c}$                  & $\mathcal{U}(0,10)$                 & $\mathrm{m\,s^{-1}}$ & RV semi-amplitude of planet c \\
        $e_{c}$                  & 0.0 (fixed)                         & \dots                & Orbital eccentricity of planet b \\
        $\omega_{c}$             & 90.0 (fixed)                        & deg                  & Periastron angle of planet c \\
        \noalign{\smallskip}
        \multicolumn{4}{c}{\it Photometry parameters} \\
        $D_{\mathrm{TESS\,17,\,42,\,43}}$          & 1.0 (fixed)                  & \dots     & Dilution factor for TESS sectors 17, 42, 43  \\
        $M_{\mathrm{TESS\,17,\,42,\,43}}$          & $\mathcal{N}(0,0.1)$         & \dots     & Relative flux offset for  TESS sectors 17, 42, 43 \\
        $\sigma_{\mathrm{TESS\,17,\,42,\,43}}$     & $\mathcal{LU}(10^{-3},10^{4})$ & ppm       & Extra jitter term for sectors TESS 17, 42, 43 \\
        $q_{1,\mathrm{TESS\,17,\,42,\,43}}$        & $\mathcal{U}(0,1)$           & \dots     & Limb-darkening parameterization for TESS sectors 17, 42, 43\\
        $q_{2,\mathrm{TESS\,17,\,42,\,43}}$        & $\mathcal{U}(0,1)$           & \dots     & Limb-darkening parameterization for TESS sectors 17, 42, 43\\
        $D_{\mathrm{LCO-SSO}}$       & 1.0 (fixed)                  & \dots     & Dilution factor for LCO-SSO \\
        $M_{\mathrm{LCO-SSO}}$       & $\mathcal{N}(0,0.1)$         & \dots     & Relative flux offset for LCO-SSO \\
        $\sigma_{\mathrm{LCO-SSO}}$  & $\mathcal{LU}(10^{-3},10^{4})$ & ppm       & Extra jitter term for LCO-SSO \\
        $q_{1,\mathrm{LCO-SSO}}$     & $\mathcal{U}(0,1)$           & \dots     & Limb-darkening parameterization for LCO-SSO \\
         $D_{\mathrm{LCO-SAAO}}$     & 1.0 (fixed)                  & \dots     & Dilution factor for LCO-SAAO \\
        $M_{\mathrm{LCO-SAAO}}$      & $\mathcal{N}(0,0.1)$         & \dots     & Relative flux offset for LCO-SAAO \\
        $\sigma_{\mathrm{LCO-SAAO}}$ & $\mathcal{LU}(10^{-3},10^{4})$ & ppm       & Extra jitter term for LCO-SAAO \\
        $q_{1,\mathrm{LCO-SAAO}}$    & $\mathcal{U}(0,1)$           & \dots     & Limb-darkening parameterization for LCO-SAAO \\
        $D_{\mathrm{LCO-McD}}$       & 1.0 (fixed)                  & \dots     & Dilution factor for LCO-McD \\
        $M_{\mathrm{LCO-McD}}$       & $\mathcal{N}(0,0.1)$         & \dots     & Relative flux offset for LCO-McD \\
        $\sigma_{\mathrm{LCO-McD}}$  & $\mathcal{LU}(10^{-3},10^{4})$ & ppm       & Extra jitter term for LCO-McD \\
        $q_{1,\mathrm{LCO-McD}}$     & $\mathcal{U}(0,1)$           & \dots     & Limb-darkening parameterization for LCO-McD \\
        $D_{\mathrm{SO-KUIPER}}$     & 1.0 (fixed)                  & \dots     & Dilution factor for SO-KUIPER \\
        $M_{\mathrm{SO-KUIPER}}$     & $\mathcal{N}(0,0.1)$         & \dots     & Relative flux offset for SO-KUIPER \\
        $\sigma_{\mathrm{SO-KUIPER}}$& $\mathcal{LU}(10^{-3},10^{4})$ & ppm       & Extra jitter term for SO-KUIPER \\
        $q_{1,\mathrm{SO-KUIPER}}$   & $\mathcal{U}(0,1)$           & \dots     & Limb-darkening parameterization for SO-KUIPER \\
        \noalign{\smallskip}
        \multicolumn{4}{c}{\it RV parameters} \\
        \noalign{\smallskip}
        $\mu_{\mathrm{CARMENES}}$            & $\mathcal{N}(-10,10)$  & $\mathrm{m\,s^{-1}}$ & RV zero point for CARMENES \\
        $\sigma_{\mathrm{CARMENES}}$         & $\mathcal{LU}(0.01,10)$ & $\mathrm{m\,s^{-1}}$ & Extra jitter term for CARMENES \\
        $\mu_{\mathrm{MAROON-X,Blue,1}}$    & $\mathcal{N}(-10,10)$  & $\mathrm{m\,s^{-1}}$ & RV zero point for MAROON-X Blue 1$^b$ \\
        $\sigma_{\mathrm{MAROON-X,Blue,1}}$   & $\mathcal{LU}(0.01,10)$ & $\mathrm{m\,s^{-1}}$ & Extra jitter term for MAROONX-X Blue 1$^b$ \\
         $\mu_{\mathrm{MAROON-X,Blue,2}}$     & $\mathcal{N}(-10,10)$  & $\mathrm{m\,s^{-1}}$ & RV zero point for MAROONX-X Blue 2$^b$ \\
        $\sigma_{\mathrm{MAROON-X,Blue,2}}$   & $\mathcal{LU}(0.01,10)$ & $\mathrm{m\,s^{-1}}$ & Extra jitter term for MAROONX-X Blue 2$^b$ \\
        $\mu_{\mathrm{MAROON-X,Red,1}}$       & $\mathcal{N}(-10,10)$  & $\mathrm{m\,s^{-1}}$ & RV zero point for MAROON-X Red 1$^b$ \\
        $\sigma_{\mathrm{MAROON-X,Red,1}}$    & $\mathcal{LU}(0.01,10)$ & $\mathrm{m\,s^{-1}}$ & Extra jitter term for MAROON-X Red 1$^b$ \\
         $\mu_{\mathrm{MAROON-X,Red,2}}$      & $\mathcal{N}(-10,10)$  & $\mathrm{m\,s^{-1}}$ & RV zero point for MAROON-X Red 2$^b$ \\
        $\sigma_{\mathrm{MAROON-X,Red,2}}$    & $\mathcal{LU}(0.01,10)$ & $\mathrm{m\,s^{-1}}$ & Extra jitter term for MAROON-X Red 2$^b$ \\
        \noalign{\smallskip}
        \noalign{\smallskip}
        \multicolumn{4}{c}{\it GP hyperparameters} \\
        \noalign{\smallskip}
        $\sigma_\mathrm{GP,RV}$  & $\mathcal{U}(0,100)$        & $\mathrm{m\,s^{-1}}$ & Amplitude of GP component for the RVs \\
        $\alpha_\mathrm{GP,RV}$  & $\mathcal{J}(10^{-10},10^{-6})$ & d$^{-2}$             & Inverse length-scale of GP exponential component for the RVs\\
        $\Gamma_\mathrm{GP,RV}$  & $\mathcal{J}(0.1,10)$       & \dots                & Amplitude of GP sine-squared component for the RVs \\
        $P_\mathrm{rot;GP,RV}$   & $\mathcal{U}(37,45)$        & d                    & Period of the GP quasi-periodic component for the RVs \\
        \noalign{\smallskip}
        \hline
    \end{tabular}
    \tablefoot{
        \tablefoottext{a}{
$\mathcal{N}(\mu,\sigma^2)$ is a normal distribution of mean $\mu$ and variance $\sigma^2$, $\mathcal{U}(a,b)$ and $\mathcal{LU}(a,b)$ are uniform and log-uniform distributions between $a$ and $b$, and $\mathcal{J}$ represents a Jeffreys distribution.}
\tablefoottext{b}{1 and 2 indicate MAROON-X data observed in August 2021 and October-November 2021, respectively, separated by RV offsets as discussed in Sect.~\ref{subsec:rv}.}}
\end{table*}

\begin{table*}
\centering
\small
\caption{Radial velocity measurements and spectroscopic activity indicators for TOI-1468 from CARMENES VIS spectra.} 
\label{tab:RV_table}
\begin{tabular}{l r c r r c c}
\hline
\hline
        \noalign{\smallskip}
        \multicolumn{1}{c}{BJD$^a$} & \multicolumn{1}{c}{RV} & Ca\,{\sc ii}\,IRT$_1$ & \multicolumn{1}{c}{CRX}   & \multicolumn{1}{c}{dLW}   & H$\alpha$ & Na~{\sc i}~D$_2$  \\
         & \multicolumn{1}{c}{(m\,s$^{-1}$)}& & \multicolumn{1}{c}{(m\,s$^{-1}$\,Np$^{-1}$)} & \multicolumn{1}{c}{(m$^{2}$\,s$^{-2}$)} & & \\
        \hline
        \noalign{\smallskip}
2458855.34742   &       $7.8\pm1.58$    &       $0.639\pm0.002$ &       $20.23\pm11.72$ &       $1.72\pm1.60$   &       $0.921\pm0.003$ &       $0.139\pm0.006$ \\
2458856.32553   &       $1.32\pm1.78$   &       $0.638\pm0.002$ &       $-2.05\pm14.30$ &       $-1.38\pm1.60$  &       $0.953\pm0.003$ &       $0.153\pm0.009$ \\
2458857.32583   &       $10.05\pm1.97$  &       $0.639\pm0.002$ &       $20.56\pm16.09$ &       $-3.09\pm1.61$  &       $0.944\pm0.003$ &       $0.167\pm0.009$ \\
2458860.40062   &       $1.36\pm2.69$   &       $0.627\pm0.005$ &       $22.32\pm27.83$ &       $1.10\pm4.47$   &       $0.905\pm0.006$ &       $0.214\pm0.026$ \\
2458865.32617   &       $-4.52\pm1.86$  &       $0.635\pm0.003$ &       $10.80\pm18.08$ &       $-3.18\pm2.35$  &       $0.895\pm0.004$ &       $0.133\pm0.011$ \\
2458866.39842   &       $0.35\pm3.58$   &       $0.619\pm0.005$ &       $-34.97\pm35.48$&       $-1.95\pm4.15$  &       $0.895\pm0.007$ &       $0.099\pm0.036$ \\
2458881.29264   &       $-6.92\pm1.55$  &       $0.632\pm0.003$ &       $13.73\pm15.28$ &       $4.33\pm1.76$   &       $0.896\pm0.003$ &       $0.151\pm0.010$ \\
2458882.29062   &       $-5.37\pm1.40$  &       $0.632\pm0.002$ &       $2.81\pm13.00$  &       $5.72\pm1.63$   &       $0.891\pm0.003$ &       $0.149\pm0.007$ \\
2458883.28485   &       $-0.24\pm2.00$  &       $0.634\pm0.002$ &       $-6.63\pm19.88$ &       $-0.48\pm1.30$  &       $0.902\pm0.003$ &       $0.162\pm0.008$ \\
2458884.32007   &       $1.50\pm1.42$   &       $0.642\pm0.002$ &       $-15.28\pm12.60$        &$43.01\pm1.74$ &       $0.935\pm0.003$ &       $0.163\pm0.007$ \\
2458890.32472   &       $1.88\pm2.04$   &       $0.639\pm0.003$ &       $18.68\pm19.30$ &       $-1.10\pm2.54$  &       $0.960\pm0.004$ &       $0.173\pm0.011$ \\
2458891.31897   &       $6.57\pm2.14$   &       $0.642\pm0.003$ &       $-4.85\pm21.75$ &       $-8.40\pm3.00$  &       $1.002\pm0.004$ &       $0.154\pm0.013$ \\
2458894.31705   &       $-4.81\pm1.82$  &       $0.655\pm0.002$ &       $-6.12\pm16.12$ &       $-4.26\pm1.74$  &       $1.046\pm0.003$ &       $0.182\pm0.008$ \\
2458895.33504   &       $3.57\pm2.48$   &       $0.633\pm0.003$ &       $0.65\pm25.80$  &       $-11.16\pm2.75$ &       $0.933\pm0.005$ &       $0.136\pm0.021$ \\
2458897.32369   &       $3.20\pm3.81$   &       $0.641\pm0.005$ &       $11.32\pm42.30$ &       $-6.29\pm5.11$  &       $1.164\pm0.009$ &       $0.185\pm0.039$ \\
2459050.64211   &       $-1.85\pm2.16$  &       $0.640\pm0.002$ &       $-17.50\pm17.46$&   $-8.74\pm2.37$        &       $0.923\pm0.003$ &       $0.121\pm0.008$ \\
2459052.64295   &       $-2.53\pm2.95$  &       $0.649\pm0.003$ &       $15.59\pm19.02$ &       $-19.89\pm3.35$ &       $1.008\pm0.005$ &       $0.155\pm0.015$ \\
2459055.63224   &       $-0.01\pm1.68$  &       $0.642\pm0.002$ &   $7.25\pm13.24$      &       $-6.29\pm1.82$  &       $0.942\pm0.003$ &       $0.153\pm0.008$ \\
2459056.65812   &       $6.84\pm1.52$   &       $0.635\pm0.002$ &       $1.69\pm13.58$  &       $-5.71\pm1.93$  &       $0.975\pm0.003$ &       $0.161\pm0.007$ \\
2459061.66776   &       $-0.82\pm2.16$  &       $0.639\pm0.002$ &       $21.72\pm19.19$ &       $-3.45\pm1.72$  &       $0.955\pm0.003$ &       $0.147\pm0.009$ \\
2459063.65749   &       $0.74\pm1.56$   &       $0.647\pm0.003$ &       $1.09\pm14.97$  &       $-2.10\pm1.91$  &       $0.995\pm0.004$ &       $0.194\pm0.010$ \\
2459064.66176   &       $0.18\pm2.15$   &       $0.634\pm0.003$ &       $9.29\pm21.15$  &       $10.98\pm3.19$  &       $0.943\pm0.005$ &       $0.163\pm0.015$ \\
2459065.66634   &       $-0.82\pm1.68$  &       $0.637\pm0.003$ &       $5.62\pm04.57$  &       $0.77\pm1.98$   &       $0.971\pm0.004$ &       $0.203\pm0.010$ \\
2459066.66004   &       $-2.04\pm1.66$  &       $0.643\pm0.002$ &       $-9.35\pm14.47$ &       $0.02\pm1.73$   &       $1.028\pm0.003$ &       $0.187\pm0.006$ \\
2459067.65698   &       $2.70\pm2.07$   &       $0.645\pm0.003$ &       $34.40\pm17.28$ &       $4.76\pm2.08$   &       $0.973\pm0.003$ &       $0.179\pm0.009$ \\
2459068.65963   &       $-3.45\pm2.15$  &       $0.640\pm0.003$ &       $19.47\pm20.63$ &       $25.04\pm3.15$  &       $0.959\pm0.005$ &       $0.204\pm0.015$ \\
2459069.65498   &       $4.69\pm1.58$   &       $0.651\pm0.002$ &       $13.88\pm14.48$ &       $13.40\pm1.73$  &       $1.012\pm0.003$ &       $0.206\pm0.007$ \\
2459071.66532   &       $7.02\pm1.67$   &       $0.639\pm0.003$ &       $3.05\pm14.58$  &       $22.54\pm2.32$  &       $0.933\pm0.003$ &       $0.217\pm0.009$ \\
2459072.66008   &       $-4.38\pm2.53$  &       $0.639\pm0.003$ &       $14.10\pm17.62$ &       $2.87\pm1.56$   &       $1.011\pm0.003$ &       $0.212\pm0.009$ \\
2459073.64987   &       $-2.11\pm3.23$  &       $0.645\pm0.005$ &       $47.67\pm29.75$ &       $-0.64\pm4.40$  &       $0.954\pm0.007$ &       $0.285\pm0.027$ \\
2459074.66769   &       $-6.58\pm2.30$  &       $0.643\pm0.003$ &       $23.07\pm17.63$ &       $-4.53\pm2.71$  &       $0.974\pm0.004$ &       $0.204\pm0.012$ \\
2459075.65938   &       $0.64\pm2.00$   &       $0.634\pm0.002$ &       $7.33\pm13.53$  &       $-0.27\pm1.99$  &       $0.921\pm0.003$ &       $0.159\pm0.007$ \\
2459076.66055   &       $-4.03\pm2.52$  &       $0.641\pm0.004$ &       $-9.75\pm23.32$ &       $-10.87\pm3.24$ &       $0.973\pm0.005$ &       $0.162\pm0.017$ \\
2459078.6601    &       $-6.64\pm1.54$  &       $0.633\pm0.002$ &       $1.50\pm12.09$  &       $2.79\pm1.74$   &       $0.942\pm0.003$ &       $0.174\pm0.007$ \\
2459079.65943   &       $-10.23\pm1.72$ &       $0.638\pm0.002$ &       $0.55\pm15.42$  &       $-5.40\pm1.78$  &       $0.909\pm0.003$ &       $0.171\pm0.007$ \\
2459081.64765   &       $-4.83\pm2.53$  &       $0.636\pm0.004$ &       $-31.10\pm24.79$&   $-6.97\pm2.88$        &       $0.919\pm0.005$ &       $0.231\pm0.017$ \\
2459084.65177   &       $-1.40\pm1.58$  &       $0.637\pm0.002$ &       $12.55\pm14.46$ &       $-0.10\pm2.05$  &       $0.922\pm0.003$ &       $0.170\pm0.007$ \\
2459085.64853   &       $-5.17\pm1.66$  &       $0.636\pm0.002$ &       $-20.04\pm14.73$&       $-3.90\pm1.87$  &       $0.915\pm0.003$ &       $0.168\pm0.007$ \\
2459086.63404   &       $2.85\pm1.72$   &       $0.642\pm0.002$ &       $-24.41\pm16.80$&       $0.73\pm1.92$   &       $0.902\pm0.003$ &       $0.168\pm0.007$ \\
2459087.63436   &       $0.84\pm1.66$   &       $0.628\pm0.002$ &       $-21.26\pm10.61$&       $41.74\pm1.28$  &       $0.893\pm0.003$ &       $0.157\pm0.007$ \\
2459088.63908   &       $6.33\pm1.54$   &       $0.643\pm0.002$ &       $5.77\pm13.66$  &      $-13.00\pm3.13$ &       $0.906\pm0.003$ &       $0.127\pm0.010$         \\
2459089.62835   &       $1.65\pm1.69$   &       $0.630\pm0.003$ &       $-18.55\pm15.33$&       $-6.01\pm2.15$  &       $0.893\pm0.003$ &       $0.159\pm0.010$ \\
2459090.62817   &       $9.09\pm2.63$   &       $0.638\pm0.004$ &       $-43.42\pm25.50$&       $-5.52\pm3.29$  &       $0.931\pm0.005$ &       $0.224\pm0.020$ \\
2459091.63591   &       $1.65\pm3.66$   &       $0.633\pm0.005$ &       $-11.63\pm36.78$&       $-16.21\pm4.61$ &       $0.918\pm0.008$ &       $0.214\pm0.038$ \\
2459092.61955   &       $-0.24\pm2.01$  &       $0.632\pm0.003$ &       $-15.10\pm19.79$&       $-1.59\pm1.80$  &       $0.918\pm0.004$ &       $0.214\pm0.038$ \\
2459093.62589   &       $-0.40\pm1.60$  &       $0.634\pm0.002$ &       $-20.78\pm14.00$&       $1.14\pm1.54$   &       $0.914\pm0.003$ &       $0.187\pm0.007$ \\
2459095.61627   &       $-9.75\pm1.35$  &       $0.637\pm0.002$ &       $-10.95\pm11.26$&       $5.60\pm1.54$   &       $0.935\pm0.003$ &       $0.195\pm0.007$ \\
2459097.60844   &       $-0.26\pm1.84$  &       $0.646\pm0.002$ &       $-12.96\pm15.74$&       $20.61\pm2.34$  &       $1.011\pm0.003$ &       $0.210\pm0.007$ \\
2459098.60592   &       $-3.00\pm1.76$  &       $0.631\pm0.002$ &       $-13.44\pm13.67$&       $6.50\pm1.84$   &       $0.931\pm0.003$ &       $0.160\pm0.006$ \\
2459099.61862   &       $0.71\pm1.72$   &       $0.644\pm0.002$ &       $-36.81\pm15.52$&       $0.95\pm2.03$   &       $0.946\pm0.003$ &       $0.179\pm0.008$ \\
2459101.61623   &       $3.91\pm2.08$   &       $0.632\pm0.003$ &       $-10.63\pm16.85$&       $-2.47\pm2.08$  &       $0.997\pm0.004$ &       $0.183\pm0.010$ \\
2459103.58597   &       $11.35\pm1.86$  &       $0.644\pm0.002$ &       $9.03\pm14.86$  &       $-1.77\pm1.49$  &       $0.987\pm0.003$ &       $0.192\pm0.007$ \\
2459113.57451   &       $-2.97\pm3.47$  &       $0.638\pm0.006$ &       $4.27\pm35.40$  &       $-15.21\pm5.12$ &       $0.957\pm0.009$ &       $0.500\pm0.045$ \\
2459114.60508   &       $-0.31\pm2.44$  &       $0.645\pm0.003$ &       $16.01\pm23.61$ &       $-16.69\pm3.48$ &       $0.965\pm0.004$ &       $0.217\pm0.014$ \\
2459115.62434   &       $-1.80\pm3.15$  &       $0.637\pm0.004$ &       $4.49\pm30.76$  &       $-7.75\pm3.38$  &       $0.995\pm0.006$ &       $0.384\pm0.021$ \\
2459118.58592   &       $5.63\pm1.98$   &       $0.636\pm0.003$ &       $10.15\pm17.00$ &       $-18.02\pm3.60$ &       $0.936\pm0.004$ &       $0.169\pm0.013$ \\
2459119.6587    &       $-1.90\pm4.66$  &       $0.637\pm0.008$ &       $55.21\pm49.47$ &       $-57.30\pm13.3$ &       $0.932\pm0.013$ &       $0.836\pm0.124$ \\
2459120.54989   &       $4.02\pm2.01$   &       $0.640\pm0.002$ &       $-13.88\pm18.44$&       $-4.23\pm1.79$  &       $0.938\pm0.003$ &       $0.241\pm0.009$ \\
2459121.56249   &       $-6.52\pm2.42$  &       $0.627\pm0.003$ &       $12.26\pm22.77$ &       $23.66\pm3.94$  &       $0.909\pm0.004$ &       $0.287\pm0.014$ \\
2459122.53527   &       $1.55\pm1.58$   &       $0.644\pm0.002$ &       $16.50\pm15.08$ &       $13.23\pm2.47$  &       $0.982\pm0.003$ &       $0.251\pm0.010$ \\
2459123.70277   &       $-6.05\pm2.40$  &       $0.638\pm0.003$ &       $14.75\pm23.50$ &       $34.71\pm4.57$  &       $0.950\pm0.005$ &       $0.406\pm0.018$ \\
2459127.64162   &       $-9.39\pm3.28$  &       $0.634\pm0.004$ &       $10.69\pm33.42$ &       $10.43\pm3.84$  &       $0.906\pm0.006$ &       $0.406\pm0.018$ \\
2459128.47056   &       $-3.93\pm1.95$  &       $0.634\pm0.002$ &       $-44.58\pm16.16$&       $0.71\pm1.88$   &       $0.910\pm0.003$ &       $0.231\pm0.009$ \\
2459131.53261   &       $2.38\pm2.00$   &       $0.630\pm0.002$ &       $-19.23\pm18.37$&       $-1.79\pm1.92$  &       $0.897\pm0.003$ &       $0.220\pm0.008$ \\
2459132.57457   &       $-2.38\pm2.08$  &       $0.643\pm0.003$ &       $8.29\pm17.19$  &       $-3.08\pm2.03$  &       $0.946\pm0.003$ &       $0.265\pm0.011$ \\
\hline
\end{tabular}
\tablefoot{
\tablefoottext{a}{Barycentric Julian date in the barycentric dynamical time standard.}}
\end{table*}

\begin{table*}
\centering
\small
\caption{Radial velocity measurements and spectroscopic activity indicators for TOI-1468 from MAROON-X spectra.} 
\label{tab:RV_table_MAROONX}
\begin{tabular}{l r c r r c c}
\hline
\hline
        \noalign{\smallskip}
        \multicolumn{1}{c}{BJD$^a$} & \multicolumn{1}{c}{RV} & Ca~{\sc ii}\,IRT$_1$ & \multicolumn{1}{c}{CRX}   & \multicolumn{1}{c}{dLW}   & H$\alpha$ & Na~{\sc i}~D$_2$  \\
         & \multicolumn{1}{c}{(m\,s$^{-1}$)}& & \multicolumn{1}{c}{(m\,s$^{-1}$\,Np$^{-1}$)} & \multicolumn{1}{c}{(m$^{2}$\,s$^{-2}$)} & &\\
        \hline
        \noalign{\smallskip}
        \textit{Blue arm}: & & & & & \\
                \noalign{\smallskip}
2459439.93808 & $-2.21 \pm 2.11$ &...& $40.76 \pm 39.32$ & $24.00 \pm 2.74$ & $0.873 \pm 0.006$ & $0.226 \pm 0.007$ \\
2459441.04179 & $-2.55 \pm 1.61$ &...& $2.73 \pm 21.52$ & $7.88 \pm 2.12$ & $0.829 \pm 0.006$ & $0.195 \pm 0.005$ \\
2459442.11303 & $1.35 \pm 1.32$ &...& $19.62 \pm 14.50$ & $2.93 \pm 1.74$ & $0.866 \pm 0.005$ & $0.195 \pm 0.004$ \\
2459443.95968 & $8.64 \pm 2.08$ &...& $10.90 \pm 31.71$ & $21.90 \pm 2.71$ & $0.857 \pm 0.007$ & $0.217 \pm 0.007$ \\
2459444.97751 & $0.57 \pm 1.79$ &...& $-6.56 \pm 28.49$ & $16.27 \pm 2.33$ & $1.039 \pm 0.006$ & $0.255 \pm 0.005$ \\
2459445.97247 & $4.60 \pm 1.77$ &...& $-17.28 \pm 28.16$ & $17.44 \pm 2.31$ & $0.840 \pm 0.006$ & $0.208 \pm 0.005$ \\
2459447.11676 & $-5.15 \pm 2.44$ &...& $41.35 \pm 30.49$ & $20.57 \pm 3.20$ & $0.911 \pm 0.009$ & $0.229 \pm 0.009$ \\
2459448.03916 & $5.93 \pm 1.48$ &...& $29.41 \pm 19.83$ & $13.31 \pm 1.94$ & $0.863 \pm 0.006$ & $0.204 \pm 0.004$ \\
2459449.02174 & $-4.56 \pm 1.15$ &...& $-35.46 \pm 16.07$ & $12.38 \pm 1.50$ & $0.831 \pm 0.004$ & $0.195 \pm 0.003$ \\
2459449.99017 & $-0.56 \pm 1.52$ &...& $-24.76 \pm 19.96$ & $16.91 \pm 1.99$ & $0.852 \pm 0.005$ & $0.207 \pm 0.004$ \\
2459514.74231 & $-0.14 \pm 1.97$ &...& $25.67 \pm 24.13$ & $22.17 \pm 2.59$ & $0.829 \pm 0.006$ & $0.205 \pm 0.006$ \\
2459516.02101 & $3.92 \pm 2.45$ &...& $47.95 \pm 34.45$ & $33.00 \pm 3.22$ & $0.822 \pm 0.008$ & $0.202 \pm 0.008$ \\
2459527.71072 & $0.05 \pm 1.74$ &...& $-38.69 \pm 27.41$ & $14.57 \pm 2.30$ & $0.851 \pm 0.006$ & $0.202 \pm 0.005$ \\
2459529.90752 & $-2.59 \pm 2.23$ &...& $16.73 \pm 34.69$ & $25.33 \pm 2.95$ & $0.926 \pm 0.008$ & $0.208 \pm 0.007$ \\
2459537.89012 & $1.02 \pm 1.45$ &...& $-10.29 \pm 20.38$ & $11.79 \pm 1.91$ & $0.875 \pm 0.005$ & $0.196 \pm 0.004$ \\
2459541.76327 & $3.10 \pm 1.65$ &...& $-36.66 \pm 25.24$ & $6.00 \pm 2.20$ & $0.965 \pm 0.006$ & $0.201 \pm 0.005$ \\
        \noalign{\smallskip}
\textit{Red arm}: & & & & & \\
        \noalign{\smallskip}
2459439.93808 & $-1.82 \pm 0.93$ & $0.650 \pm 0.001$ & $12.57 \pm 14.57$ & $3.21 \pm 1.19$ & $0.790 \pm 0.631$ &...\\
2459441.04179 & $-3.57 \pm 0.91$ & $0.642 \pm 0.001$ & $-1.47 \pm 13.92$ & $4.14 \pm 1.16$ & $0.708 \pm 0.596$ &...\\
2459442.11303 & $3.60 \pm 0.76$ & $0.640 \pm 0.001$ & $0.90 \pm 11.42$ & $3.84 \pm 0.96$ & $0.710 \pm 0.430$ &...\\
2459443.95968 & $4.91 \pm 1.01$ & $0.638 \pm 0.001$ & $-3.02 \pm 12.89$ & $8.16 \pm 1.29$ & $0.728 \pm 0.735$ &...\\
2459444.97751 & $-2.47 \pm 0.84$ & $0.670 \pm 0.001$ & $-10.44 \pm 12.22$ & $5.61 \pm 1.07$ & $0.893 \pm 0.535$ &...\\
2459445.97247 & $3.67 \pm 0.83$ & $0.644 \pm 0.001$ & $-17.87 \pm 13.65$ & $7.39 \pm 1.06$ & $0.731 \pm 0.551$ &...\\
2459447.11676 & $-2.47 \pm 1.25$ & $0.644 \pm 0.002$ & $17.63 \pm 20.05$ & $11.69 \pm 1.60$ & $0.858 \pm 1.039$ &...\\
2459448.03916 & $3.85 \pm 0.83$ & $0.646 \pm 0.001$ & $-6.73 \pm 10.31$ & $6.88 \pm 1.05$ & $0.776 \pm 0.499$ &...\\
2459449.02174 & $-3.16 \pm 0.65$ & $0.644 \pm 0.001$ & $8.89 \pm 09.22$ & $5.63 \pm 0.82$ & $0.758 \pm 0.345$ &...\\
2459449.99017 & $-0.28 \pm 0.82$ & $0.648 \pm 0.001$ & $-4.37 \pm 08.33$ & $6.96 \pm 1.04$ & $0.738 \pm 0.513$ &...\\
2459514.74231 & $-0.46 \pm 0.97$ & $0.644 \pm 0.001$ & $-5.30 \pm 12.92$ & $9.96 \pm 1.24$ & $0.733 \pm 0.007$ &...\\
2459516.02101 & $1.84 \pm 1.19$ & $0.643 \pm 0.001$ & $6.76 \pm 17.23$ & $13.24 \pm 1.52$ & $0.690 \pm 0.009$ &...\\
2459527.71072 & $-0.56 \pm 0.87$ & $0.645 \pm 0.001$ & $-3.88 \pm 12.29$ & $3.37 \pm 1.11$ & $0.742 \pm 0.006$ &...\\
2459529.90752 & $-8.02 \pm 1.18$ & $0.649 \pm 0.002$ & $-3.02 \pm 15.57$ & $5.64 \pm 1.51$ & $0.814 \pm 0.009$ &...\\
2459537.89012 & $0.76 \pm 0.79$ & $0.647 \pm 0.001$ & $18.13 \pm 08.68$ & $11.10 \pm 0.99$ & $0.765 \pm 0.006$ &...\\
2459541.76327 & $3.98 \pm 0.91$ & $0.648 \pm 0.001$ & $3.49 \pm 10.61$ & $5.39 \pm 1.16$ & $0.847 \pm 0.007$ &...\\
    \hline
\end{tabular}
\tablefoot{\tablefoottext{a}{Barycentric Julian date in the barycentric dynamical time standard.}}
\end{table*}

\section{Figures}

\begin{figure*}
   \centering
   \includegraphics[width=\hsize]{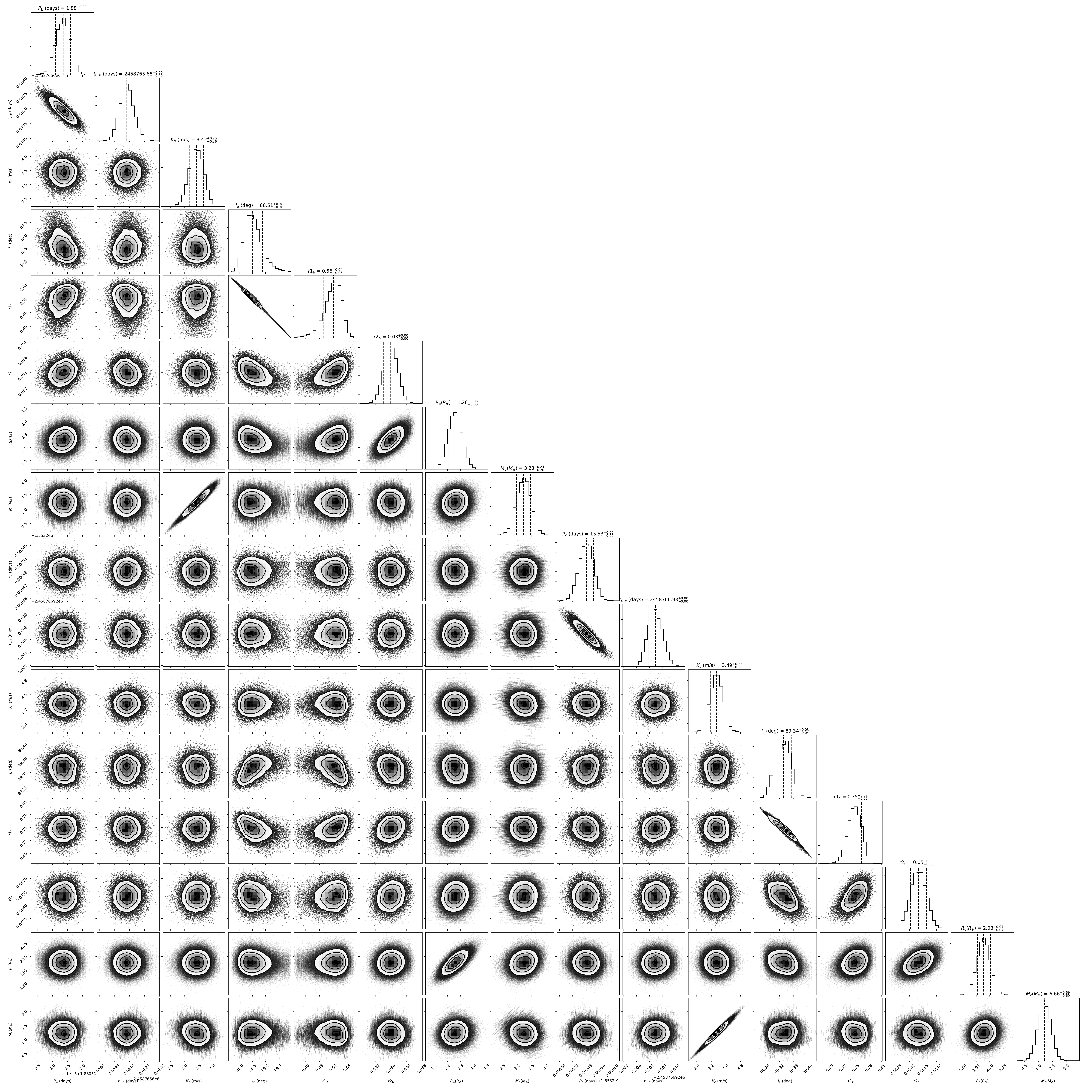}
   \caption{Posterior distribution for the joint model parameters (2cp+GP) derived with {\tt juliet}.}
    \label{Fig:corner_plot-2pl}
\end{figure*}

\end{appendix}

\end{document}